\documentclass[11pt]{article}
\pdfoutput=1
\usepackage{epsfig}
\usepackage{amssymb}
\usepackage{amsfonts}
\usepackage{subfigure}
\def\beq{\begin{equation}}
\def\eeq{\end{equation}}
\def\bea{\begin{eqnarray}}
\def\eea{\end{eqnarray}}
\def\ksl{\hbox{\hbox{${k}$}}\kern-1.9mm{\hbox{${/}$}}}
\def\a{\alpha}
\def\b{\beta}

\newcommand{\eps}{\varepsilon}

\newcommand{\nn}{\nonumber}
\newcommand{\text}{\rm}

\def\lsim{\raise0.3ex\hbox{$\;<$\kern-0.75em\raise-1.1ex\hbox{$\sim\;$}}} 

\def\gsim{\raise0.3ex\hbox{$\;>$\kern-0.75em\raise-1.1ex\hbox{$\sim\;$}}}

\begin{document}

\begin{center} 
{\bf \large  Superconformal Sum Rules and the  Spectral Density Flow \\ of the Composite 
 Dilaton (ADD) Multiplet in $\mathcal{N}=1$ Theories}
\vspace{0.2cm}
{\bf \Large } 

\vspace{1.5cm}
{\bf Claudio Corian\`{o}, Antonio Costantini, Luigi Delle Rose and Mirko Serino}
\vspace{1cm}

{\it Dipartimento di Matematica e Fisica "Ennio De Giorgi", 
Universit\`{a} del Salento and \\ INFN-Lecce, Via Arnesano, 73100 Lecce, Italy\footnote{claudio.coriano@le.infn.it, antonio.costantini@le.infn.it, luigi.dellerose@le.infn.it, mirko.serino@le.infn.it}}\\

\vspace{.5cm}
\begin{abstract} 
We discuss the signature of the anomalous breaking of the superconformal symmetry in $\mathcal{N}=1$ super Yang Mills theory, mediated by the Ferrara-Zumino hypercurrent ($\mathcal{J}$) with two vector ($\mathcal V$) supercurrents $(\mathcal{JVV})$ and its manifestation in the anomaly action, in the form of anomaly poles.  
This allows to investigate in a unified way both conformal and chiral anomalies. The analysis is performed in parallel to the Standard Model, for comparison. 
We investigate, in particular, massive deformations of the $\mathcal{N}=1$ theory and the spectral densities of the anomaly form factors which are extracted from the components of this correlator.  
In this extended framework it is shown that all the anomaly form factors are characterized by spectral densities which flow with the mass deformation. In particular, the continuum contributions from the two-particle cuts of the intermediate states turn into into poles in the zero mass limit, with a single sum rule satisfied by each component. 
Non anomalous form factors, instead, in the same anomalous correlators, are characterized by non-integrable spectral densities. These tend to uniform distributions as one moves towards the conformal point, with a clear dual behaviour. As in a previous analysis of the dilaton pole of the Standard Model, also in this case the poles can be interpreted as signaling the exchange of a composite dilaton/axion/dilatino (ADD) multiplet in the effective Lagrangian. The pole-like behaviour of the anomaly form factors is shown to be a global feature of the correlators, present at all energy scales, due to the sum rules. A similar behaviour is shown to be present in the Konishi current, which identifies additional composite states. We conclude that global anomalous currents characterized by a single flow in the perturbative picture always predict the existence of composite interpolating fields. In case of gauging of these currents, as in superconformal theories coupled to gravity, we show that the cancellation of the corresponding anomalies requires either a vanishing $\beta$ function or the inclusion of an extra gravitational sector which effectively sets the residue at the anomaly poles of the gauged currents to vanish.

\end{abstract}
\end{center}

\newpage
\section{Introduction}
Dilaton fields are expected to play a very important role in the dynamics of the early universe and are present in almost any model which attempts to unify gravity with the ordinary gauge interactions (see for instance \cite{Gasperini:2007ar}). Important examples of these constructions are effective field theories derived from strings, describing their massless spectra, but also theories of gravity compactified on extra dimensional spaces, where the dilaton (graviscalar) emerges in 4 spacetime dimensions from the extra components of the higher dimensional metric (see for instance \cite{LopesCardoso:1991zt,LopesCardoso:1992yd,Derendinger:1991kr,Derendinger:1991hq,Derendinger:1985cv}). In these formulations, due to the geometrical origin of these fields, the dilaton is, in general, a fundamental (i.e. not a composite) field. 
Other extensions, also of significant interest,
in which a fundamental dilaton induces a gauge connection for the abelian Weyl symmetry in a curved spacetime, have been considered (see the discussion in \cite{Codello:2012sn,Buchmuller:1988cj, Coriano:2013nja}). However, also in this case, the link of this fundamental particle to gravity renders it a crucial player in the physics of the early universe, and not a particle to be searched for at colliders. In fact, its interaction with ordinary matter should be suppressed by the Planck scale, except if one entails scenarios with large extra dimensions.

More recently, following an independent route, several extensions of the Standard Model with an {\em effective} dilaton have been considered. They conjecture the existence of a scale-invariant extension of the Higgs sector \cite{Goldberger:2007zk, Coriano:2012nm,Coriano:2012dg}. In this case the breaking of the underlying conformal dynamics, in combination with the spontaneous breaking of the electroweak symmetry \cite{Coriano:2013nja}, suggests, in fact, that the dilaton could emerge as a composite field, appearing as a Nambu-Goldstone mode of the broken conformal symmetry. 
A massless dilaton of this type could acquire a mass by some explicit potential and could mix with the Higgs of the Standard Model.\\
By reasoning in terms of the conformal symmetry of the Standard Model, which should play a role at high energy, the dilaton would be the physical manifestation of the trace anomaly in the Standard Model, in analogy to the pion, which is interpolated by the $U(1)_A$ chiral current and the corresponding $\langle AVV \rangle$ (axial-vector/vector/vector) interaction in QCD. 
As in the $\langle AVV \rangle$ case,  this composite state should be identified with the anomaly 
pole of the related anomaly correlator (the $\langle TVV \rangle$ diagram, with $T$ the energy momentum tensor (EMT)), at least at the level of the 1-particle irreducible (1PI) anomaly effective action \cite{Coriano:2012nm}. Considerations of this nature brings us to the conclusion that the effective massless Nambu-Goldstone modes which should appear as a result of the existence of global anomalies, should be looked for in specific perturbative form factors under special kinematical limits. For this reason they are easier to investigate in the on-shell anomaly effective action, with a single mass parameter which drives the conformal/superconformal deformation. This action has the advantage of being gauge invariant and easier to compute than its off-shell relative. We remark, however, that this picture remains limited to perturbation theory and may be drastically modified by the non perturbative dynamics. The radiative nature of the breaking of a certain global symmetry, as in the case of the anomaly, does not guarantee the massless nature of these modes, which could acquire a nonzero mass. 

The extension of this analysis to the superconformal case is particularly interesting in view of recent results  concerning the derivation of the superconformal anomaly action for the Goldstone supermultiplet in a theory where conformal symmetry is spontaneously broken \cite{Schwimmer:2010za}. In this case it has been argued in favour of the existence of a conformal anomaly matching between the broken and the unbroken phases of the superconformal theory. Our results are in line with these previous elaborations. 

\subsection{Anomalies and anomaly poles}
To take advantage at full scale of the analogy between chiral and conformal anomalies, one should turn to supersymmetry, where the correlation between poles and anomalies should be more direct. In fact, in an ordinary quantum field theory, the $\langle TVV \rangle$ diagram (and the corresponding anomaly action) is characterized, as we are going to show, by pole structures both in those form factors that contribute to the trace anomaly and in those that don't. For this reason we turn our attention to the effective action of the superconformal (the Ferrara-Zumino, FZ) multiplet, where chiral and conformal anomalies share similar signatures, being part of the same multiplet. \\
We are going to prove rigorously in perturbation theory that the anomaly of the FZ multiplet is associated with the exchange of three composite states in the 1PI superconformal anomaly action. These have been discussed in the past, in the context of the spontaneous breaking of the superconformal symmetry \cite{Dudas:1993mm}. They are identified with 
the anomaly poles present in the effective action, extracted from a supersymmetric correlator containing the superconformal hypercurrent and two vector currents, and correspond to the dilaton, the dilatino and the axion. 
 This exchange is identified by a direct analysis of the anomalous correlators in perturbation theory or by the study of the flow of their spectral densities under massive deformations.
The flow describes a 1-parameter family of spectral densities - one family for each component of the correlator - which satisfy mass independent sum rules, and are, therefore, independent of the superpotential. This behaviour turns a dispersive cut of the spectral density $\rho(s,m^2)$ into a pole (i.e. a $\delta(s)$ contribution) as the deformation parameter $m$ goes to zero. 
Moreover, denoting with $k^2$ the momentum square of the anomaly vertex, each of the spectral densities induces on the corresponding form factor a $1/k^2$ behaviour also at large $k^2$, as a consequence of the sum rule. \\
We also recall that the partnership between dilatons and axions is not new in the context of anomalies, and it has been studied in the past - for abelian gauge anomalies - in the case of the supersymmetric St\"uckelberg multiplet \cite{Kors:2004ri, Coriano:2008xa, Coriano:2008aw, Coriano:2010ws}. \\
The three states associated to the three anomaly poles mentioned above are described - in the perturbative picture - by the exchange of two collinear particles. These correspond to a fermion/antifermion pair in the axion case, a fermion/antifermion pair and a pair of scalar particles in the dilaton case, and a collinear scalar/fermion pair for the dilatino. 
The Konishi current will be shown to follow an identical pattern and allows the identification of extra states, 
one for each fermion flavour present in the theory. \\ 
This pattern appears to be general in the context of anomalies, and unique in the case of supersymmetry. In fact, we are going to show that in a supersymmetric theory anomaly correlators have a single pole in each component of the anomaly multiplet, a single spectral flow and a single sum rule, proving the existence of a one-to-one correspondence between anomalies and poles in these correlators. 

Our work is organized as follows. We first illustrate the motivations of our study by overviewing the analysis of the spectral densities performed in the investigations of the conformal anomaly in the $\langle TVV \rangle$ vertex at nonzero momentum transfers in QED \cite{Giannotti:2008cv}\cite{Armillis:2009pq} and QCD \cite{Armillis:2010qk}. The case of a general non-abelian theory, with the inclusion of scalars, is new and is discussed - in the massless case - in Section \ref{TVVsection}. This may serve to highlight some specific properties of these types of vertices which have not received sufficient attention in the past. 

Then we turn to a perturbative study of the effective action in the case of supersymmetric $\mathcal{N}=1$ theories, focusing on the components of the FZ multiplet and on the corresponding anomalies. This is followed by a study of the spectral densities of the relevant diagrams which are responsible for the superconformal anomaly. We show the existence of a unique sum rule for each component of the multiplet, and of a spectral density flow driven by the mass perturbations. As the deformation (mass) parameter turns to zero, restoring the superconformal symmetry, the flow gets localized at zero 
invariant mass, signalling the exchange of a massless pole in the anomaly effective action. We will compare  non supersymmetric and supersymmetric realizations, highlighting the differences between the two cases. In particular we will show in detail how the cancellation of the extra poles of non anomalous form factors is realized in supersymmetric theories. Finally, we present the structure of the anomaly action as a combination of the pole contributions, plus the non anomalous (logarithmic) terms. In superspace, the first had been identified in the past relying on supersymmetric arguments \cite{LopesCardoso:1991zt}. We will conclude our analysis with some comments on the possible implications of our results about the physical manifestation of anomaly poles - for global anomalous symmetries - and their cancellations in the case of their superconformal gaugings.

 \section{Sum rules}
\begin{figure}[t]
\centering
\subfigure[]{\includegraphics[scale=0.8]{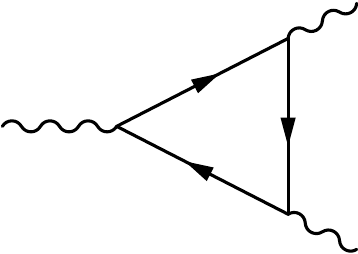}} \hspace{2cm}
\subfigure[]{\includegraphics[scale=0.8]{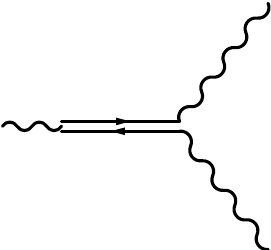}} \hspace{2cm}
\subfigure[]{\includegraphics[scale=0.8]{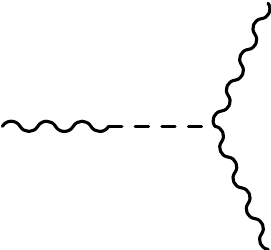}}
\caption{The triangle diagram in the fermion case (a), the collinear fermion configuration responsible for the anomaly (b) and a diagrammatic representation of the exchange via an intermediate state (dashed line) (c). \label{fig1}}
\end{figure}
As pointed out long ago in the literature on the chiral anomaly \cite{Dolgov:1971ri}, its perturbative signature is in the appearance of a massless pole 
(an anomaly pole) in the spectrum of the $\langle AVV \rangle$ diagram. The pole is present, in perturbation theory, only in a specific kinematic configuration, namely at zero fermion mass and with on-shell vector lines. The intermediate state which is exchanged in the effective action, see Fig. (\ref{fig1}), and which is mediated by the $\langle AVV \rangle$ diagram, is characterized by two massless and collinear fermions moving on the light cone. It is rather compelling to interpret the appearance of this intermediate configuration - within the obvious limitations of the perturbative picture - as signalling the possible exchange of a bound state in the quantum effective action. 

In a phenomenological context, what gives a far broader significance to this kinematical mechanism is the appearance of a certain kinematic {\em duality}, which accompanies any perturbative anomaly. In this case it is better known as $Q^2$-{\em duality}, relating the resonance and the asymptotic region of a certain correlator in a nontrivial way \cite{Bertlmann:1981by}. 
This property, in general, finds its justification in the existence of a sum rule for the spectral density $\rho(s,m^2)$. Generically, it is given in the form 
\beq
\label{sr}
\frac{1}{\pi} \int_0^{\infty}\rho(s, m^2) ds =  f,
\eeq
with the constant $f$ independent of any mass (or other) parameter which characterizes the thresholds or the strengths of the resonant states eventually present in the integration region $(s>0)$.
It should be stressed that sum rules formulated for the study of the structure of the resonances, i.e. their strengths and masses, as in the QCD case, involve a parameterization of the resonant behaviour 
of $\rho(s,m)$ at low $s$ values, via a phenomenological approach, with the inclusion of the asymptotic behaviour of the correlator, amenable to perturbation theory, for larger $s$. For this significant interplay between the infrared (IR) and the ultraviolet (UV) regions, the term {\em duality} is indeed quite appropriate to qualify the implications of a given sum rule. 

It was pointed out, some time ago, that one specific tract of the chiral anomaly is in the existence of a sum rule for the $\langle AVV \rangle$ diagram \cite{Horejsi:1985qu}, later extended to a similar study of the trace of the energy momentum tensor ($\textrm{tr}T$) for the $\langle \textrm{tr}TVV \rangle$ in QED (with $V$ a vector current), at zero momentum transfers \cite{Horejsi:1997yn,Horejsi:1994aj}. Similar analysis were performed on the $\langle TT \rangle$ correlator in 2-dimensional gravity \cite{Bertlmann:2000da}, which is affected by the trace anomaly. 
The analysis brought substantial evidence that the sum rule, combined with the original identification of the anomaly pole from the perturbative spectral density \cite{Dolgov:1971ri}, were two important and related aspects of the anomaly phenomenon. We recall that the study of these types of correlators has a quite long history \cite{Freedman:1974ze, Adler:1976zt,Chanowitz:1972da}.

More recently, very general perturbative analysis of the $\langle TVV \rangle$ correlator (or graviton-gauge-gauge vertex), performed off-shell and at nonzero momentum transfer, have shown that the general features observed 
in the $\langle AVV \rangle$ and $\langle \textrm{tr}TVV \rangle$ cases where preserved \cite{Giannotti:2008cv,Armillis:2009pq, Armillis:2010qk, Armillis:2009im}. \\ 
A specific feature of the spectral density of the chiral and conformal anomalies is that the pole is introduced in the spectrum in a specific kinematical limit, as a degeneracy of the two-particle 
cut when any second scale (for instance the fermion mass) is sent to zero. The property of the {\em cut turning into a pole} is peculiar to finite (non superconvergent) sum rules. It is related to a spectral density which is normalized by the sum rule just like an ordinary weighted distribution, and whose support is located at the edge of the allowed phase space ($s=0$) as the conformal deformation turns to zero. This allows to single out a unique interpolating state out of all the possible exchanges permitted in the continuum, i.e. for $s> 4 m^2$, as the theory flows towards its conformal/superconformal point.

\subsection{ Sum rule and the UV/IR conspiracy of the anomaly} 
As we have just mentioned, the existence of a sum rule for the form factor responsible for a certain anomaly indicates a UV/IR connection manifested by the corresponding spectral density, but it is not exclusively related, obviously, to the anomaly phenomenon. In fact, non anomalous form factors, in some cases, as we are going to show next, share a similar behaviour. We would expect, though, that the breaking of a symmetry should manifest in the apperance of a massless state in the spectrum of the effective theory, and in this respect the saturation of the spectral density with a single resonance 
in an anomaly form factor acquires a special status. 
Elaborating on Eq. (\ref{sr}), one can show that 
the effect of the anomaly is, in general, related to the behaviour of the spectral density at any values of $s$, although, in some kinematical limits, it is the region around the light cone ($s\sim 0$) which dominates the sum rule, and amounts to a resonant contribution.
In fact, the combination of the scaling behaviour of the corresponding form factor $F(Q^2)$ (equivalently of its density $\rho$) with the requirement of integrability of the spectral density, essentially fix $f$ to be a constant and the sum rule (\ref{sr}) to be saturated by a single massless resonance. Obviously, a superconvergent sum rule, obtained for $f=0$, would not share this behaviour. At the same time, {\em the absence of subtractions} in the dispersion relations guarantees the significance of the sum rule, being this independent of any ultraviolet cutoff. 

It is quite straightforward to show that Eq. (\ref{sr}) is a constraint on  asymptotic behaviour of the related form factor.
The proof is obtained by observing that the dispersion relation for a form factor in the spacelike region ($Q^2=-k^2> 0$)
\beq
F(Q^2,m^2)=\frac{1}{\pi}\int_0^{\infty} ds\frac{ \rho(s,m^2) }{s + Q^2} \,,
\eeq
once we expand the denominator in $Q^2$ as $\frac{1}{s + Q^2}=\frac{1}{Q^2} - \frac{1}{Q^2} s\frac{1}{Q^2} +\ldots$ and make use of Eq. (\ref{sr}), induces the following asymptotic behaviour on $F(Q^2,m^2)$
\beq
\lim_{Q^2\to \infty} Q^2 F(Q^2, m^2)=f.
\label{asym}
\eeq
The $F\sim f/Q^2$ behaviour at large $Q^2$, with $f$ independent of $m$, shows the pole dominance of $F$ for $Q^2 \rightarrow \infty$. The UV/IR conspiracy of the anomaly,  discussed in \cite{Armillis:2009im, Armillis:2009sm,Armillis:2009pq}, is in the reappearance of the pole contribution at very large value of the invariant $Q^2$, even for a nonzero mass $m$. In fact, as we are going to show in the following sections, the spectral density has support around the $s=0$ region $(\rho(s)\sim \delta(s))$, as in the massless $(m=0)$ case.
This point is quite subtle, since the flows of the spectral densities with $m$ show the decoupling of the anomaly pole for a nonzero mass. Here, the term {\em decoupling} will be used to refer to the non resonant behaviour of $\rho$.
Therefore, the presence of a $1/Q^2$ term in the anomaly form factors is a property of the entire flow which a) converges to a localized massless state (i.e. $\rho(s)\sim\delta(s)$) as 
$m\to 0$, while  
b) the presence of a non vanishing sum rule guarantees the validity of the asymptotic constraint illustrated in Eq. (\ref{asym}). Notice that although for conformal deformations driven by a single mass parameter the independence of the asymptotic value $f$ on $m$ is a simple consequence of the scaling behaviour of $F(Q^2,m^2)$, it holds quite generally even for a completely off-shell kinematics \cite{Giannotti:2008cv}.

In summary, in complete agreement with a previous analysis by Giannotti and Mottola \cite{Giannotti:2008cv}, we are going to verify that for a generic supersymmetric $\mathcal{N}=1$ theory, the two basic features of the anomalous behaviour of a certain form factor responsible for chiral or conformal anomalies are: \,\,1)  the existence of a spectral flow which turns a dispersive cut into a pole as $m$ goes to zero and 2) the existence of a sum rule which relates the asymptotic behaviour of the anomaly form factor to the strength of the pole resonance.\\
 In a supersymmetric theory this correspondence, as we are going to show, is unique, since the only poles present in the explicit expressions are those 
 of the anomaly form factors. This feature is shared also by the $\langle AVV \rangle$ in non supersymmetric theories, where one can identify 
 a single pole in the related form factor, a single sum rule and a single spectral density flow. In the $\langle TVV \rangle$ diagram, for a general field theory, instead, this feature is absent. 
 The appearance of extra poles in the form factors of the traceless parts of this second correlators leaves unanswered the question about the physical meaning of these additional singularities \cite{Giannotti:2008cv,Armillis:2009pq, Armillis:2010qk, Armillis:2009im}. On the other end, the effective massless states emerging from the anomaly sectors should be identified with the Nambu-Goldstone modes of the corresponding broken symmetries, which are such because of non conserved dilatation and chiral currents.  
 
 The reason for turning to supersymmetry should be obvious. One expects, in general, that the perturbative structure of the chiral and conformal \cite{Capper:1975ig} anomalies, in this case, should be similar. This should occur for a supersymmetric anomaly multiplet, where both the $\langle TVV \rangle$ and the $\langle AVV \rangle$-like diagrams are components of the same anomalous correlator. At the same time, we expect to recover, for each single component, the properties found in the past, separately in the chiral and in the conformal cases \cite{Horejsi:1985qu,Horejsi:1997yn,Horejsi:1994aj}, but, hopefully, without the extra poles present in the non anomalous tensor structures and form factors of $\langle TVV \rangle$. \\ 
We are going to prove, by an explicit computation, that this is indeed the case. We also stress the fact that our analysis, in particular in the $\langle TVV \rangle$ case, is entirely performed at nonzero momentum transfer, working with the insertion of the uncontracted $T$, rather than with its trace, as done in \cite{Horejsi:1997yn,Horejsi:1994aj}. 
The study is centered around the $\langle \mathcal{J V V} \rangle$ correlator,  in $\mathcal N=1$ and $\mathcal N=4$ theories, containing the hypercurrent $\mathcal{J}$ (the Ferrara-Zumino multiplet \cite{Ferrara:1974pz}) and two vector supercurrents $\mathcal V$. We will show that the two requirements enunciated above are satisfied by every component of this multiplet.
However, before moving to the supersymmetric case, we briefly review the results of the computation of the $\langle TVV\rangle$ in an ordinary non-abelian gauge theory. This may serve to illustrate the differences between an ordinary gauge theory and its supersymmetric version, which has triggered the current analysis.

 \section{The $\langle TVV \rangle$ and $\langle AVV \rangle$ vertices in an ordinary gauge theory}

 \label{TVVsection}
 We consider a non-abelian gauge theory containing massless scalars, fermions and gauge fields,  with fermions and scalars assigned to the representations $R_f$ and $R_s$ respectively, and define a correlator with a symmetric EMT ($T$) and two vector (gauge) currents ($V$). The 
correlator can be interpreted, in a weak gravitational background, as describing the one graviton-two gauge fields vertex, which is affected by the trace anomaly. The analysis in QED is contained in \cite{Giannotti:2008cv, Armillis:2009pq}, while generalization to QCD and to the Standard Model can be found in \cite{Armillis:2010qk, Coriano:2011zk}. The general tensor structure of this type of vertex has been given in QED by Giannotti and Mottola \cite{Giannotti:2008cv} in terms of a non minimal basis of $13$ form factors $(t_1, t_2,\ldots,t_{13})$ in the course of their studies on the 1PI conformal anomaly effective action for gravity. Here we present the tensor expansion of the one-loop $\langle TVV \rangle$ vertex with on-shell vector lines in a non-abelian gauge theory. Details can be found in  \cite{Armillis:2010qk}. \\
The on-shell expansion of the $\langle TVV \rangle$ correlator in a non-abelian gauge theory is expressed in terms of just 3 independent form factors \cite{Armillis:2010qk}
\bea
\Gamma^{\mu\nu\alpha\beta}_{(T)}(p,q) = f_1(k^2) \, \phi_1^{\mu\nu\alpha\beta}(p,q) + f_2(k^2) \, \phi_2^{\mu\nu\alpha\beta}(p,q) + f_3(k^2) \, \phi_3^{\mu\nu\alpha\beta}(p,q) \,,
\eea
where the tensor structures are defined by
\bea 
\label{phitensors}
\phi_1^{\mu\nu\alpha\beta}(p,q) &\equiv& t_1^{\mu\nu\alpha\beta}(p,q) = (k^2 \eta^{\mu\nu} - k^\mu k^\nu) u^{\alpha\beta}(p,q)\,, \nn \\
\phi_2^{\mu\nu\alpha\beta}(p,q) &\equiv& t_3^{\mu\nu\alpha\beta}(p,q) + t_5^{\mu\nu\alpha\beta}(p,q) -4 t_7^{\mu\nu\alpha\beta}(p,q) = - 2 u^{\alpha\beta}(p,q) [k^2 \eta^{\mu\nu} + 2 (p^\mu p^\nu + q^\mu q^\nu) \nn \\
&-& 4 (p^\mu q^\nu + q^\mu p^\nu)] \,, \nn \\
\phi_3^{\mu\nu\alpha\beta}(p,q) &\equiv& t_{13}^{\mu\nu\alpha\beta}(p,q) = (p^\mu q^\nu + p^\nu q^\mu) \eta^{\alpha\beta} + p \cdot q (\eta^{\alpha\nu} \eta^{\beta\mu} + \eta^{\alpha\mu} \eta^{\beta\nu}) - \eta^{\mu\nu} u^{\alpha \beta}(p,q) \nn \\
&-&  (\eta^{\beta\nu}p^\mu + \eta^{\beta\mu}p^\nu)q^\alpha - (\eta^{\alpha\nu}q^\mu + \eta^{\alpha\mu}q^\nu)p^\beta,
\eea
with 
\bea
\label{utensor}
u^{\alpha\beta}(p,q) = \eta^{\alpha\beta} p \cdot q - p^\beta q^\alpha \,.
\eea
Here $k=p+q$ is the incoming momentum in the EMT line, while $p^\alpha$ and $q^\beta$ are the two outgoing momenta from the two vector currents.\\
For massless fields running in the loops, of these 3 tensor structures only $\phi_1$ is traceful, contributing to the trace anomaly, the remaining ones being traceless. Fermions, scalars and gauge fields give contributions which are separately gauge invariant. Both $f_1$, the anomaly form factor, and the form factor $f_2$ of the traceless tensor $\phi_2$ are found to be finite, while $f_3$, the form factor of the traceless $\phi_3$, gets renormalized. In the most general paremeterization of the 
vertex - assuming nonzero virtualities of all the external lines and an internal mass ($m$) for the field in the loops - 
$t_1^{\mu\nu\alpha\beta}$ is still the only traceful and anomalous tensor structure.\\ $t_2$ is also traceful, but describes the explicit breaking of the conformal symmetry (its form factor is proportional to $m$ and therefore it is non anomalous) and $t_{13}$ is the only tensor structure affected by renormalization. The remaining form factors, corresponding to the contributions $(t_1,t_2,t_3,\ldots, t_{12})$ are finite. \\ 
In the on-shell and massless case, for a Dirac fermion $(f)$ in the representation $R_f$ running in the loops, the form factors are given by
\bea
\label{FFfermions}
f_1^{(f)}(k^2) &=& - \frac{g^2 \, T(R_f)}{18 \pi^2 \, k^2} \,, \qquad
f_2^{(f)}(k^2) = - \frac{g^2 \, T(R_f)}{144 \pi^2 \, k^2} \,, \nn \\
f_3^{(f)}(k^2) &=& \frac{g^2 \, T(R_f)}{144 \pi^2} \left\{ 11 + 12 \, \mathcal B_0(k^2,0)\right\}
\eea
where $\textrm{Tr} \, T^a T^b = T(R) \delta^{ab}$ is the Dynkin index of the representation $R$.\\  
Analogous results hold for a conformally coupled complex scalar $(s)$ in the representation $R_s$
\bea
\label{FFscalars}
f_1^{(s)}(k^2) &=&  - \frac{g^2 \, T(R_s)}{72 \pi^2 \, k^2} \,, \qquad
f_2^{(s)}(k^2) =  \frac{g^2 \, T(R_s)}{288 \pi^2 \, k^2} \,, \nn \\
f_3^{(s)}(k^2) &=& \frac{g^2 \, T(R_s)}{288 \pi^2} \left\{ 7 + 6 \, \mathcal B_0(k^2,0)\right\}, \,
\eea
while for a gauge field ($A$) in the adjoint representation one obtains
\bea
\label{FFgauge}
f_1^{(A)}(k^2) &=&  \frac{11 g^2 \, T(A)}{72 \pi^2 \, k^2} \,, \qquad
f_2^{(A)}(k^2) =  \frac{g^2 \, T(A)}{288 \pi^2 \, k^2} \,, \nn \\
f_3^{(A)}(k^2) &=& - \frac{g^2 \, T(A)}{8 \pi^2} \left\{ \frac{65}{36} - \mathcal B_0(0,0) + \frac{11}{6} \mathcal B_0(k^2,0) + k^2 \, \mathcal C_0(k^2,0) \right\}.
\eea
A discussion of the scalar integrals is given in Appendix \ref{AppScalarIntegrals}. It is a common lore to denote with $\mathcal B_0$ and $\mathcal C_0$ the scalar 2- and 3- point functions. 
Note that in the expression of $\mathcal C_0(k^2,m^2) $, the scalar triangle integral, the first entry is the only nonzero external invariant, while $m$ is the mass of the virtual particles. Also note that $f_1$ and $f_2$ are both finite, while $f_3$ needs renormalization, as we have just mentioned. \\
Each contribution is separately gauge invariant and it is characterized by an anomaly pole in the corresponding form factor $f_1$, described by a spectral density which is a Dirac delta ($\sim \delta(k^2)$). However, additional poles are present also in $f_2$, which multiplies a traceless structure, and are not directly linked to the conformal anomaly. These extra poles have demised, so far, any interpretation, but they seem to share the same properties of  the anomaly poles of the correlator. It is then clear that both $f_1$ and $f_2$, in this case, should be treated on the same footing since, as we are going to show, they are both characterized by spectral densities satisfying the conditions enunciated in the previous sections.
We will see, however, that supersymmetry gives a surprisingly simple answer on this issue, since the extra, non anomalous poles in the supersymmetric case are simply not present.\\ 
The existence of extra poles is a characteristic of the $\langle TVV \rangle$ correlator in ordinary gauge theories, but not of the $\langle AVV \rangle$, where $A$ is an axial-vector current. We recall that for an axial anomaly, the usual Rosenberg parameterization in terms of six form factors $(A_1,\ldots, A_6)$, and the use of the Ward identities and on-shellness conditions on the vector lines, reduce the anomaly amplitude $\Delta^{\lambda\mu\nu}$ to the simple form \cite{Armillis:2009sm}
\beq
\Delta^{\lambda\mu\nu}=A_6(k^2,m^2) k^{\lambda}\epsilon[p,q,\nu,\mu] + (A_4(k^2,m^2) + A_6(k^2,m^2)) ( q^\nu\eps[p,q,\mu,\lambda] - p^{\mu} \epsilon[p,q,\nu,\lambda]), 
\label{anom1}
\eeq
with $k$ denoting the incoming momentum of the axial-vector line (of Lorentz index $\lambda$), and with $p$ and $q$ denoting the outgoing momenta of the $(\mu,\nu)$ vector lines. Note that in this case the transversality condition for the vector currents removes the second combination of form factors, leaving only a nonzero $A_6$, which is given by 
\beq
A_6(k^2,m^2)=\frac{1}{2 \pi^2 k^2}\left(1 +\frac{m^2}{k^2}\log^2\left(\frac{\sqrt{\tau(k^2,m^2)}+1}{\sqrt{\tau(k^2,m^2)}-1}\right)\right) \qquad k^2< 0
\label{a6}
\eeq
with $\tau(k^2,m^2)=1- 4 m^2/k^2$. In the massless limit, the spectral density of this form factor, for $k^2>0$, is proportional to a Dirac $\delta$-function, since the logarithmic term vanishes, and is accompanied by a sum rule. In any case, the spectral density of the $A_4 + A_6$ form factor is not integrable, and the link between the chiral anomaly and the corresponding pole is again unique. We will illustrate this point in a following section. We now turn to discuss the structure of the correlator which is responsible for the superconformal anomaly, proceeding with a perturbative analysis of its components and of its related spectral densities.     

\section{Theoretical framework}
In this section we review the definition and some basic properties of the Ferrara-Zumino supercurrent multiplet, which from now on we will denote also as the \emph{hypercurrent}, in order to distinguish it from its fermionic component, usually called the {\em supercurrent}.  \\
We consider a $\mathcal N=1$ supersymmetric Yang-Mills theory with a chiral supermultiplet in the matter sector. In the superfield formalism the action is given by
\bea
\label{SUSYactionSF}
S = \left( \frac{1}{16 g^2 T(R)} \int d^4 x \, d^2 \theta \, \textrm{Tr} W^2 + h.c.  \right) 
+ \int d^4 x \, d^4 \theta \, \bar \Phi e^V \Phi 
+ \left( \int d^4 x \, d^2 \theta \, \mathcal W(\Phi) +h.c. \right)
\eea
where the supersymmetric field strength $W_A$ and gauge vector field $V$ are contracted with the hermitian generators $T^a$ of the gauge group to which the chiral superfield $\Phi$ belongs. 
In particular
\bea
V =2 g  V^a T^a \,, \qquad \mbox{and} \qquad W_A = 2 g  W^a_A T^a = -\frac{1}{4} \bar D^2 e^{- V} D_A \, e^V. 
\eea
In order to clarify our conventions we give the component expansion of the chiral superfield $\Phi$
\bea
\label{PHIexpansion}
\Phi_i = \phi_i + \sqrt{2} \theta \chi_i + \theta^2 F_i \,,
\eea
and of the superfields $W_A^a$ and $V^a$ in the Wess-Zumino gauge
\bea
\label{GAUGEexpansion}
W^a_A &=&  \lambda^a_A +\theta_A \, D^a - (\sigma^{\mu \nu} \theta)_A F^a_{\mu\nu} + i \theta^2 \, \sigma^{\mu}_{ A \dot B} \mathcal D_\mu \bar \lambda^{a \, \dot B} \,, \\
V^a &=&  \theta \sigma^\mu \bar \theta A^a_\mu + \theta^2 \bar \theta \bar \lambda^a + \bar \theta^2 \theta \lambda^a + \frac{1}{2} \theta^2 \bar \theta^2 \left( D^a + i \partial_\mu A^{a \, \mu} \right) \,,
\eea
where $\phi_i$ is a complex scalar and $\chi_i$ its superpartner, a left-handed Weyl fermion, $A^a_\mu$ and $\lambda^a$ are the gauge vector field and the gaugino respectively, $F^a_{\mu\nu}$ is the gauge field strength while $F_i$ and $D^a$ correspond to the $F$- and $D$-terms. Moreover, we have defined $\sigma^{\mu\nu}=(i/4)(\sigma^\mu \bar \sigma^\nu -\sigma^\nu \bar \sigma^\mu )$. \\
Using the component expansions introduced in Eq.(\ref{PHIexpansion}) and (\ref{GAUGEexpansion}) we obtain the supersymmetric lagrangian in the component formalism, which we report for convenience
\bea
\label{SUSYlagrangianCF}
\mathcal L &=& - \frac{1}{4} F^a_{\mu\nu} F^{a \, \mu\nu} + i \lambda^a \sigma^\mu \mathcal D_\mu^{ab} \bar \lambda^b
+ ( \mathcal D_{ij}^\mu \phi_j )^\dag (\mathcal D_{ik \, \mu} \phi_k) + i \chi_j \sigma_\mu \mathcal D_{ij}^{\mu \, \dag} \bar \chi_i \nn \\
&& - \sqrt{2} g \left( \bar \lambda^a \bar \chi_i T^a_{i j} \phi_j  + \phi_i^\dag T^a_{ij} \lambda^a \chi_j \right) - V(\phi, \phi^\dag) - \frac{1}{2} \left( \chi_i \chi_j \mathcal W_{ij}(\phi) + h.c.  \right) \,,
\eea
where the gauge covariant derivatives on the matter fields and on the gaugino are defined respectively as
\bea
\mathcal D^\mu_{ij} = \delta_{ij} \partial^\mu + i g A^{a \, \mu} T^a_{ij} \,, \qquad
\mathcal D_\mu^{ac} = \delta^{ac} \partial^\mu -g \, t^{abc} A^b_\mu \,,
\eea
with $t^{abc}$ the structure constants of the adjoint representation, and the scalar potential is given by
\bea
V(\phi, \phi^\dag) = \mathcal W^\dag_i(\phi^\dag) \mathcal W_i(\phi) + \frac{1}{2} g^2 \left( \phi_i^\dag T^a_{ij} \phi_j \right)^2 \,.
\eea
For the derivatives of the superpotential we have been used the following definitions
\bea
\mathcal W_i(\phi) = \frac{\partial \mathcal W(\Phi)}{\partial \Phi_i} \bigg|  \,, \qquad \mathcal W_{ij}(\phi) = \frac{\partial^2 \mathcal W(\Phi)}{\partial \Phi_i \partial \Phi_j} \bigg| \,,
\eea
where the symbol $|$ on the right indicates that the quantity is evaluated at $\theta = \bar \theta = 0$. \\
Notice that in the above equations the $F$- and $D$-terms have been removed exploiting their equations of motion.
Having defined the model, we can introduce the Ferrara-Zumino hypercurrent
\bea
\label{Hypercurrent}
\mathcal J_{A \dot A} = \textrm{Tr}\left[ \bar W_{\dot A} e^V W_A e^{- V}\right]
- \frac{1}{3} \bar \Phi \left[  \stackrel{\leftarrow}{\bar \nabla}_{\dot A}  e^V \nabla_A - e^V \bar D_{\dot A} \nabla_A +  \stackrel{\leftarrow}{\bar \nabla}_{\dot A} \stackrel{\leftarrow}{D_A} e^V \right] \Phi \,,
\eea
where $\nabla_A$ is the gauge-covariant derivative in the superfield formalism whose action on chiral superfields is given by
\bea
\nabla_A \Phi = e^{-V} D_A \left( e^V \Phi \right)\,, \qquad  \bar \nabla_{\dot A} \bar \Phi = e^{V} \bar D_{\dot A} \left( e^{-V} \bar \Phi \right)\,.
\eea
The conservation equation for the hypercurrent $\mathcal J_{A \dot A}$ is
\bea
\label{HyperAnomaly}
\bar D^{\dot A} \mathcal J_{A \dot A} = \frac{2}{3} D_A \left[ - \frac{g^2}{16 \pi^2} \left( 3 T(A) - T(R)\right) \textrm{Tr}W^2 - \frac{1}{8} \gamma \, \bar D^2 (\bar \Phi e^V \Phi)+ \left( 3 \mathcal W(\Phi) - \Phi \frac{\partial \mathcal W(\Phi)}{\partial \Phi} \right) \right] \,,
\eea
where $\gamma$ is the anomalous dimension of the chiral superfield. \\
The first two terms in Eq. (\ref{HyperAnomaly}) describe the quantum anomaly of the hypercurrent, while the last is of classical origin and it is entirely given by the superpotential. In particular, for a classical scale invariant theory, in which $\mathcal W$ is cubic in the superfields or identically zero, this term identically vanishes. If, on the other hand, the superpotential is quadratic the conservation equation of the hypercurrent acquires a non-zero contribution even at classical level. This describes the explicit breaking of the conformal symmetry.

We can now project the hypercurrent $\mathcal J_{A \dot A}$ defined in Eq.(\ref{Hypercurrent}) onto its components. The lowest component is given by the $R^\mu$ current, the $\theta$ term is associated with the supercurrent $S^\mu_A$, while the $\theta \bar \theta$ component contains the  energy-momentum tensor $T^{\mu\nu}$. In the $\mathcal N=1$ super Yang-Mills theory described by the Lagrangian in Eq. (\ref{SUSYlagrangianCF}), these three currents are defined as
\bea
\label{Rcurrent}
R^\mu &=& \bar \lambda^a \bar \sigma^\mu \lambda^a 
+ \frac{1}{3} \left( - \bar \chi_i \bar \sigma^\mu \chi_i + 2 i \phi_i^\dag \mathcal D^\mu_{ij} \phi_j - 2 i (\mathcal D^\mu_{ij} \phi_j)^\dag \phi_i \right) \,, \\
\label{Scurrent}
S^\mu_A &=& i (\sigma^{\nu \rho} \sigma^\mu \bar \lambda^a)_A F^a_{\nu\rho}
 - \sqrt{2} ( \sigma_\nu \bar \sigma^\mu \chi_i)_A (\mathcal D^{\nu}_{ij} \phi_j)^\dag - i \sqrt{2} (\sigma^\mu \bar \chi_i) \mathcal W_i^\dag(\phi^\dag) \nn \\
&-&  i g (\phi^\dag_i T^a_{ij} \phi_j) (\sigma^\mu \bar \lambda^a)_A + S^\mu_{I \, A}\,, \\
\label{EMT}
T^{\mu\nu} &=&  - F^{a \, \mu \rho} {F^{a \, \nu}}_\rho 
+ \frac{i}{4} \left[ \bar \lambda^a \bar \sigma^\mu (\delta^{ac} \stackrel{\rightarrow}{\partial^\nu} - g \, t^{abc} A^{b \, \nu} ) \lambda^c + 
\bar \lambda^a \bar \sigma^\mu (- \delta^{ac} \stackrel{\leftarrow}{\partial^\nu} - g \, t^{abc} A^{b \, \nu} ) \lambda^c + (\mu \leftrightarrow \nu) \right] \nn \\
&+&  ( \mathcal D_{ij}^\mu \phi_j )^\dag (\mathcal D_{ik}^\nu \phi_k)  +   ( \mathcal D_{ij}^\nu \phi_j )^\dag (\mathcal D_{ik}^\mu \phi_k) +
\frac{i}{4} \left[ \bar \chi_i \bar \sigma^\mu ( \delta_{ij} \stackrel{\rightarrow}{\partial^\nu} + i g T^a_{ij} A^{a \, \nu} ) \chi_j \right. \nn \\
&+& \left.  \bar \chi_i \bar \sigma^\mu ( - \delta_{ij} \stackrel{\leftarrow}{\partial^\nu} + i g T^a_{ij} A^{a \, \nu} ) \chi_j + (\mu \leftrightarrow \nu) \right]  - \eta^{\mu\nu} \mathcal L + T^{\mu\nu}_I \,, 
\eea
where $\mathcal L$ is given in Eq.(\ref{SUSYlagrangianCF}) and $S^\mu_I$ and $T^{\mu\nu}_I$ are the terms of improvement in $d=4$ of the supercurrent and of the EMT respectively. As in the non-supersymmetric case, these terms are necessary only for a scalar field and, therefore, receive contributions only from the chiral multiplet. They are explicitly given by
\bea
S^\mu_{I \, A} &=& \frac{4 \sqrt{2}}{3} i \left[ \sigma^{\mu\nu} \partial_\nu (\chi_i \phi_i^\dag) \right]_A \,, \\
T^{\mu\nu}_I &=& \frac{1}{3} \left( \eta^{\mu \nu} \partial^2 - \partial^\mu \partial^\nu \right) \phi^\dag_i \phi_i \,.
\eea
The terms of improvement are automatically conserved and guarantee, for $\mathcal W(\Phi) = 0$, upon using the equations of motion, the vanishing of the classical trace of $T^{\mu\nu}$ and of the classical gamma-trace of the supercurrent $S^\mu_A$. %

The anomaly equations in the component formalism, which can be projected out from Eq. (\ref{HyperAnomaly}), are
\bea
\label{AnomalyR}
\partial_\mu R^\mu &=& \frac{g^2}{16 \pi^2} \left( T(A) - \frac{1}{3} T(R) \right) F^{a \, \mu\nu} \tilde F^a_{\mu\nu} \,, \\
\label{AnomalyS}
\bar \sigma_\mu S^\mu_A &=&  - i \frac{3 \, g^2}{8 \pi^2} \left( T(A) -\frac{1}{3} T(R) \right) \left( \bar \lambda^a \bar \sigma^{\mu\nu} \right)_A F^a_{\mu\nu }\,, \\
\label{AnomalyT}
\eta_{\mu\nu} T^{\mu\nu} &=& -  \frac{3 \, g^2}{32 \pi^2} \left(T(A) - \frac{1}{3} T(R) \right) F^{a \, \mu\nu}  F^a_{\mu\nu} \,.
\eea
The first and the last equations are respectively extracted from the imaginary and the real part of the $\theta$ component of Eq.(\ref{HyperAnomaly}), while the gamma-trace of the supercurrent comes from the lowest component.

\section{The perturbative expansion in the component formalism}
In this section we will present the one-loop perturbative analysis of the one-particle irreducible correlators, built with a single current insertion contributing - at leading order in the gauge coupling constant - to the anomaly equations previously discussed. \\
We define the three correlation functions, $\Gamma_{(R)}$, $\Gamma_{(S)}$ and $\Gamma_{(T)}$ as
\bea
\label{RSTCorrelators}
\delta^{ab} \, \Gamma_{(R)}^{\mu\alpha\beta}(p,q) &\equiv& \langle R^{\mu}(k)\, A^{a \, \alpha}(p) \, A^{b \, \beta}(q) \rangle \qquad \langle RVV \rangle \,, \nn \\
\delta^{ab} \, \Gamma_{(S) \, A\dot B}^{\mu\alpha}(p,q) &\equiv& \langle S^{\mu}_A (k) \, A^{a \, \alpha}(p) \, \bar \lambda^b_{\dot B}(q) \rangle \qquad \langle SVF \rangle \,, \nn \\
\delta^{ab} \, \Gamma_{(T)}^{\mu\nu\alpha\beta}(p,q) &\equiv& \langle T^{\mu\nu}(k) \, A^{a \, \alpha}(p) \, A^{b \, \beta}(q)\rangle \qquad \langle TVV \rangle  \,,
\eea
with $k = p+q$ and where we have factorized, for the sake of simplicity, the Kronecker delta on the adjoint indices. These correlation functions have been computed at one-loop order in the dimensional reduction scheme (DRed) using the Feynman rules listed in Appendix \ref{AppFeynmanRules}. We recall that in this scheme the tensor and scalar loop integrals are computed in the analytically continued spacetime while the sigma algebra is restricted to four dimensions. \\
In order to provide more details, we will present the results for the matter chiral and gauge vector multiplets separately, for on-shell external gauge lines. The chiral contribution will be discussed first, and the result will be given with the inclusion of 
the corresponding mass corrections. \\
Notice that the matter chiral superfield belongs to a certain representation $R$ of the gauge group. If the representation is complex, for instance the fundamental of $SU(N)$, then the superfield must be accompanied by another superfield (eventually with the same mass) belonging to the complex-conjugate representation $\bar R$. In this case, the generator $\bar T^a$ of $\bar R$ are related to those of $R$ by the equation $\bar T^a = - (T^a)^T = - (T^a)^*$. For simplicity, in the following we will consider just the case of a single chiral superfield in a real representation of the gauge group. The extension to a complex representation amounts just to a factor of 2 in front of all the expressions which are generated for the chiral multiplet. Indeed these terms are all proportional to $T(R)$, which is equal to $T(\bar R)$.

The one-particle irreducible correlation functions of the Ferrara-Zumino multiplet are ultraviolet (UV) divergent, as one can see from a direct computation, and we need a suitable renormalization procedure in order to get finite results. In particular we have explicitly checked that, at one-loop order, among the three correlators defined in Eq. (\ref{RSTCorrelators}), only those with $S^\mu_A$ and $T^{\mu\nu}$ require a UV counterterm. 
The renormalization of the correlation functions is guaranteed by replacing the bare operators in Eq. (\ref{Scurrent}) and Eq. (\ref{EMT}) with their renormalized counterparts. This introduces the renormalized parameters and wave-function renormalization constants which are fixed by some conditions that specify the renormalization scheme. In particular, for the correlation functions we are interested in, the bare $S^\mu_A$ and $T^{\mu\nu}$ current become
\bea
\label{RenormalizedST}
S^\mu_A &=&  i Z_\lambda^{1/2} Z_V^{1/2}(\sigma^{\nu \rho} \sigma^\mu \bar \lambda^a_R)_A F^a_{R \, \nu\rho} + \ldots \,, \nn \\
T^{\mu\nu} &=&   Z_V \left( - F_R^{a \, \mu \rho} {F^{a \, \nu}}_{R \, \rho} + \frac{1}{4} \eta^{\mu\nu}  F_R^{a \, \rho \sigma} F^a_{R \, \rho \sigma}  \right)  + \ldots \,, 
\eea 
where the suffix $R$ denotes renormalized quantities. $Z_V$ and $Z_\lambda$ are the wave-function renormalization constants of the gauge and gaugino field respectively, while the ellipses stand for all the remaining operators. In the previous equations we have explicitly shown only the contributions from which, at one-loop order, we can extract the counterterms needed to renormalize our correlation functions. All the other terms, not shown, play a role at higher perturbative orders.\\ Expanding the wave-function renormalization constants at one-loop as $Z = 1 + \delta Z$ we obtain the vertices of the counterterms
\bea
\label{counterterms}
\delta[S^{\mu}_A(k) A^{a \, \alpha}(p) \bar \lambda^b_{\dot B}(q)] &=&  \left( \delta Z_V + \delta Z_\lambda \right) \,  p_\rho \, \left( \sigma^{\alpha \rho} \sigma^{\mu} \right)_{A \dot B} \,, \nn \\
\delta[T^{\mu\nu}(k)A^{a \, \alpha}(p) A^{b \, \beta}(q)] &=& \delta Z_V \, \delta^{ab} \left\{ p \cdot q \, C^{\mu\nu\alpha\beta} + D^{\mu\nu\alpha\beta}(p,q) \right\} \,,
\eea
with $p$ and $q$ outgoing momenta and where the two tensor structures $C^{\mu\nu\alpha\beta}$ and $D^{\mu\nu\alpha\beta}(p,q)$ are given in Appendix \ref{AppFeynmanRules}. The $\delta Z$ counterterms can be defined, for instance, by requiring a unit residue of the full two-point functions on the physical particle poles. This implies that 
\bea
\delta Z_V = - \frac{\partial}{\partial p^2} \Sigma^{(VV)}(p^2) \bigg|_{p^2 = 0}  \qquad \mbox{and} \qquad \delta Z_\lambda = - \Sigma^{(\lambda \bar \lambda)}(0) \,, 
\eea
where the one-loop corrections to the gauge and gaugino two-point functions are defined as
\bea
\Gamma^{(VV)}_{\mu\nu}(p) &=& - i \delta^{ab} \left( \eta_{\mu\nu} - \frac{p_\mu p_\nu}{p^2} \right) \Sigma^{(VV)}(p^2) \,, \\
\Gamma^{(\lambda \bar \lambda)}_{A \dot B}(p) &=& i \delta^{ab} \, p_{\mu} \sigma^{\mu}_{A \dot B} \, \Sigma^{(\lambda \bar \lambda)}(p^2) \,,
\eea 
with
\bea
\Sigma^{(VV)}(p^2) &=& \frac{g^2}{16 \pi^2} p^2 \left\{ T(R) \, \mathcal B_0(p^2,m^2)  - T(A) \, \mathcal B_0(p^2,0) \right\}   \,, \\
\Sigma^{(\lambda \bar \lambda)}(p^2) &=&  \frac{g^2}{16 \pi^2} \left\{ T(R) \, \mathcal B_0(p^2,m^2)  + T(A) \, \mathcal B_0(p^2,0)\right\} \,.
\eea
Using the previous expressions we can easily compute the wave-function renormalization constants
\bea
\delta Z_V &=& -  \frac{g^2}{16 \pi^2}  \left\{ T(R) \, \mathcal B_0(0,m^2)  - T(A) \, \mathcal B_0(0,0) \right\}   \,, \nn \\
\delta Z_\lambda &=& - \frac{g^2}{16 \pi^2}  \left\{ T(R) \, \mathcal B_0(0,m^2)  + T(A) \, \mathcal B_0(0,0)\right\} \,,
\eea
and therefore obtain the one-loop counterterms needed to renormalize our correlators. In the following we will always present results for the renormalized correlation functions. \\ It is interesting to observe that, accordingly to Eq. (\ref{counterterms}), the one-loop counterterm to the supercurrent correlation function is identically zero for the vector gauge multiplet, due to a cancellation between $\delta Z_V$ and $\delta Z_\lambda$. Therefore we expect a finite result for the vector supermultiplet contribution to the $\Gamma^{\mu\alpha}_{(S)}$. Indeed this is the case as we will show below.\\
The correctness of our computations is secured by the check of some Ward identities. These arise from gauge invariance, from the conservation of the energy-momentum tensor and of the supercurrent. In particular, for the three point correlators defined above, we have
\bea
\label{VectorWI}
&& p_\alpha \, \Gamma_{(R)}^{\mu\alpha\beta}(p, q) = 0 \,, \qquad \qquad q_\beta \, \Gamma_{(R)}^{\mu\alpha\beta}(p, q) = 0 \,, \nn \\
&& p_\alpha \, \Gamma_{(S)}^{\mu\alpha}(p, q) = 0 \,, \nn \\
&& p_\alpha \, \Gamma_{(T)}^{\mu\nu\alpha\beta}(p, q) = 0 \,, \qquad \qquad q_\beta \, \Gamma_{(T)}^{\mu\nu\alpha\beta}(p, q) = 0 \,
\eea
from the conservation of the vector current, and 
\bea
\label{TensorWI}
i \, k_\mu \, \Gamma_{(S)}^{\mu\alpha}(p, q) &=& - 2 p_\mu \, \sigma^{\mu \alpha} \hat \Gamma_{(\lambda \bar \lambda)}(q) - i \sigma_\mu \hat \Gamma^{\mu \alpha}_{(VV)}(p)\,, \nn \\
i \, k_\mu \, \Gamma_{(T)}^{\mu\nu\alpha\beta}(p, q) &=& q_\mu \hat \Gamma^{\alpha \mu}_{(VV)} \eta^{\beta \nu}(p ) + p_\mu \hat \Gamma^{\beta \mu}_{(VV)}(q) \eta^{\alpha \nu} - q^\nu \hat \Gamma^{\alpha \beta}_{(VV)}(p ) - p^\nu \hat \Gamma^{\alpha \beta}_{(VV)}(q) \,, 
\eea
for the conservation of the supercurrent and of the EMT respectively, where $\hat \Gamma_{(VV)}$ and $\hat \Gamma_{(\lambda \bar \lambda)}$ are the renormalized self-energies. Their derivation follows closely the analysis presented in \cite{Armillis:2010qk}.
Notice that, for on-shell gauge and gaugino external lines, the two identities in Eq. (\ref{TensorWI}) simplify considerably because their right-hand sides vanish identically. 

\section{The supercorrelator in the on-shell and massless case} 
In this section we discuss the explicit results of the computation of supercorrelator when the components of the external vector supercurrents are on-shell and the superpotential of the chiral multiplet is absent. We will consider first the contributions due to the exchange of the chiral multiplet, followed by a subsection in which we address the exchange of a virtual vector multiplet. 

\subsection{The chiral multiplet contribution}
We start from the chiral multiplet, presenting the result of the computation for massless fields and on shell gauge and gaugino external lines. 
\begin{figure}[t]
\centering
\subfigure[]{\includegraphics[scale=0.6]{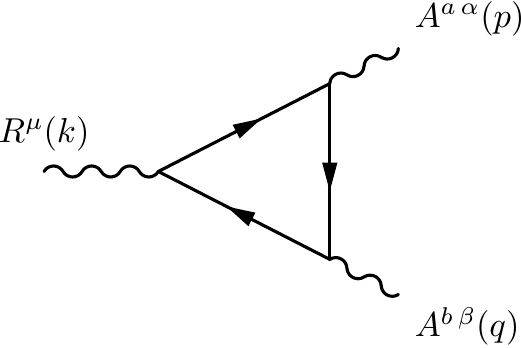}} \hspace{.5cm}
\subfigure[]{\includegraphics[scale=0.6]{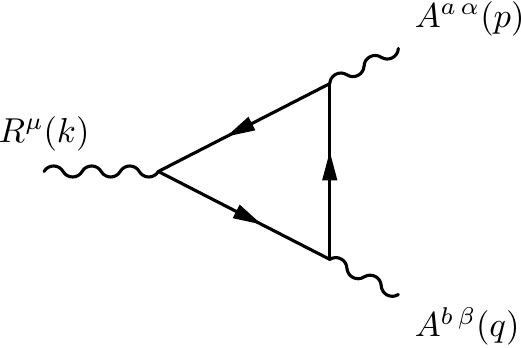}} \hspace{.5cm}
\subfigure[]{\includegraphics[scale=0.6]{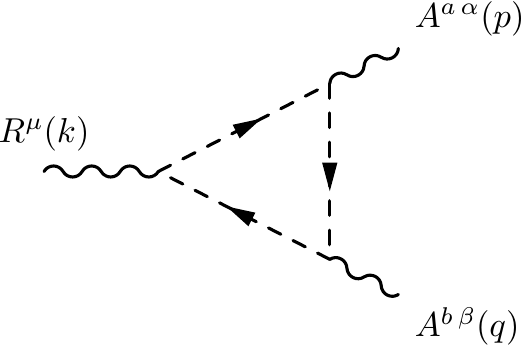}} \hspace{.5cm}
\subfigure[]{\includegraphics[scale=0.6]{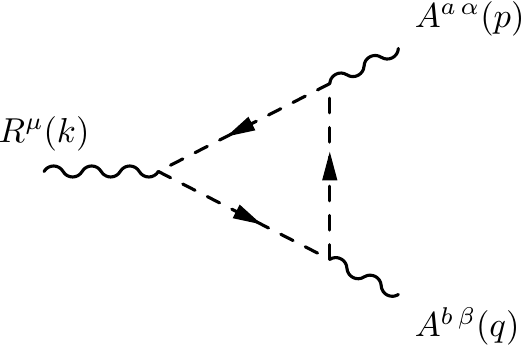}} \\
\subfigure[]{\includegraphics[scale=0.6]{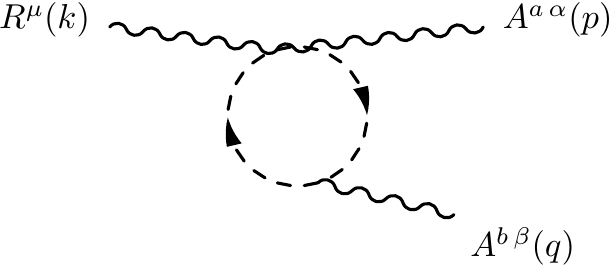}} \hspace{.5cm}
\subfigure[]{\includegraphics[scale=0.6]{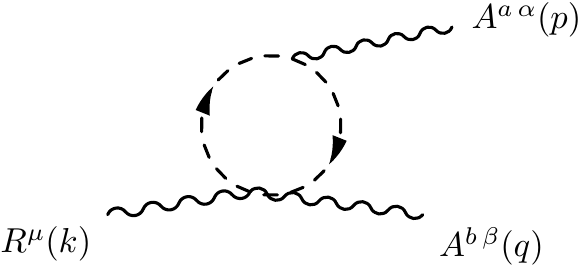}} \hspace{.5cm}
\subfigure[]{\includegraphics[scale=0.6]{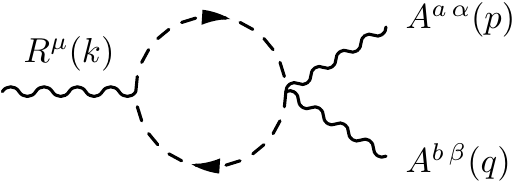}} 
\caption{The one-loop perturbative expansion of the $\langle RVV \rangle$ correlator with a massless chiral multiplet running in the loops. \label{Fig.Rchiral}}
\end{figure}
\begin{figure}[t]
\centering
\subfigure[]{\includegraphics[scale=0.6]{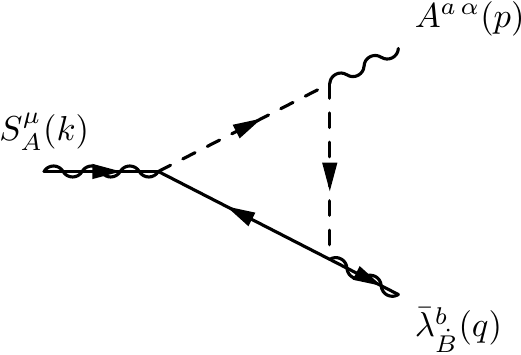}} \hspace{.5cm}
\subfigure[]{\includegraphics[scale=0.6]{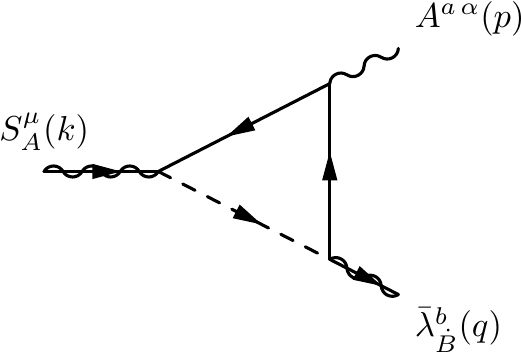}} \hspace{.5cm}
\subfigure[]{\includegraphics[scale=0.6]{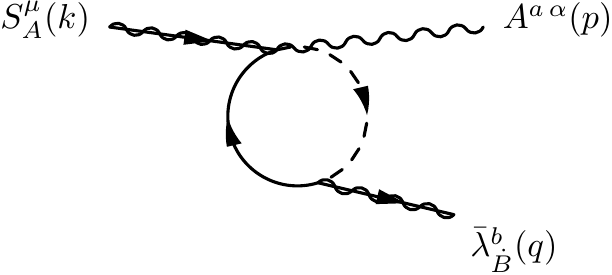}} \hspace{.5cm}
\subfigure[]{\includegraphics[scale=0.6]{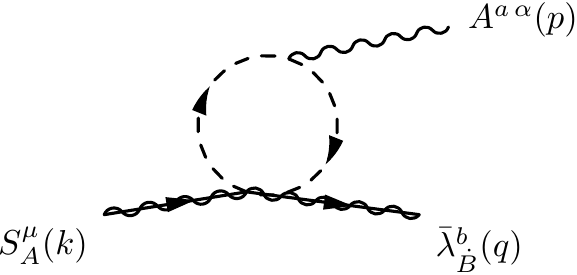}}
\caption{The one-loop perturbative expansion of the $\langle SVF \rangle$ correlator with a massless chiral multiplet running in the loops. \label{Fig.Schiral}}
\end{figure}
\begin{figure}[t]
\centering
\subfigure[]{\includegraphics[scale=0.6]{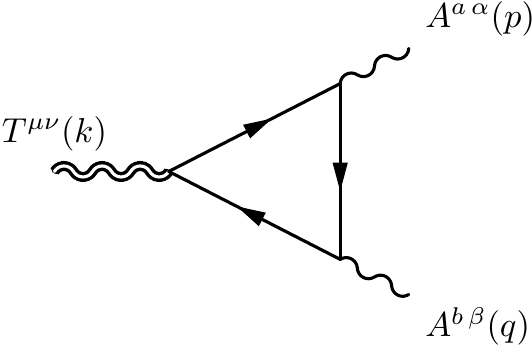}} \hspace{.5cm}
\subfigure[]{\includegraphics[scale=0.6]{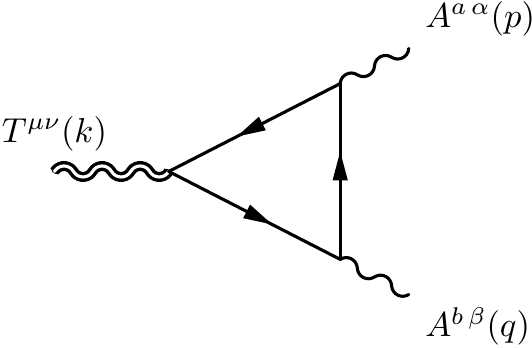}} \hspace{.5cm}
\subfigure[]{\includegraphics[scale=0.6]{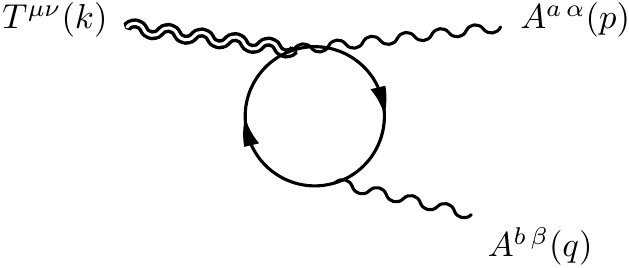}} \hspace{.5cm}
\subfigure[]{\includegraphics[scale=0.6]{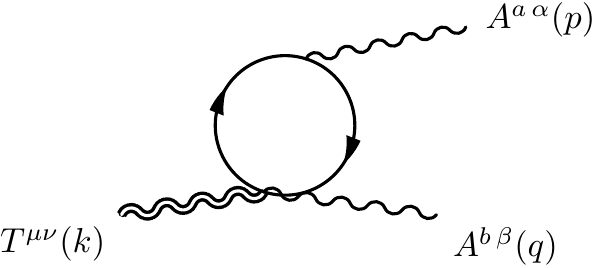}} \\
\subfigure[]{\includegraphics[scale=0.6]{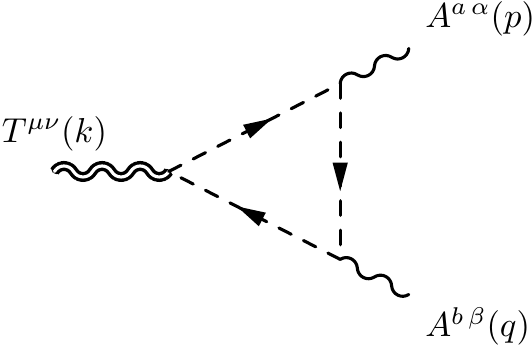}} \hspace{.5cm}
\subfigure[]{\includegraphics[scale=0.6]{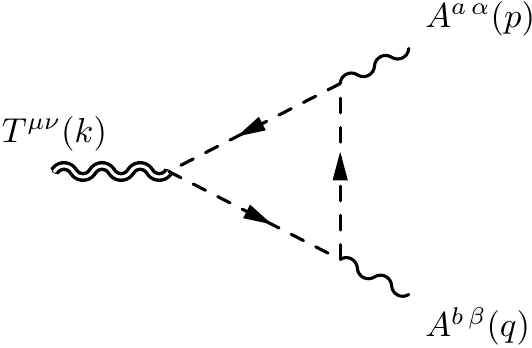}} \hspace{.5cm}
\subfigure[]{\includegraphics[scale=0.6]{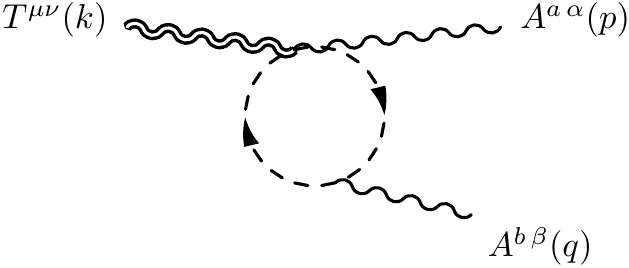}} \hspace{.5cm}
\subfigure[]{\includegraphics[scale=0.6]{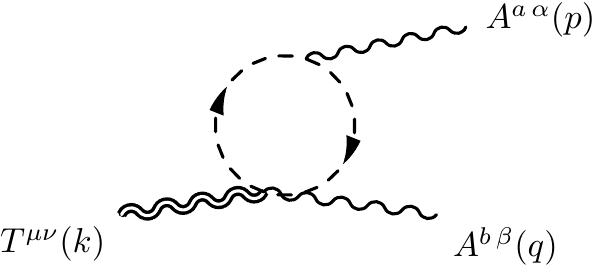}} \\
\subfigure[]{\includegraphics[scale=0.6]{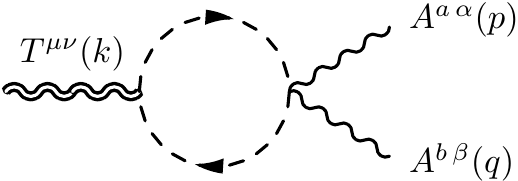}} \hspace{.5cm}
\subfigure[]{\includegraphics[scale=0.6]{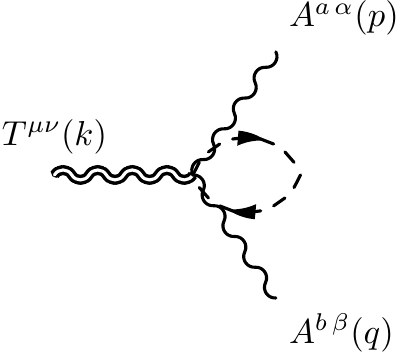}} \hspace{.5cm}
\caption{The one-loop perturbative expansion of the $\langle TVV \rangle$ correlator with a massless chiral multiplet running in the loops. The last diagram, being a massless tadpole, is identically zero in dimensional regularization. \label{Fig.Tchiral}}
\end{figure}

{\bf -Three-point function of the $R^\mu$ current}\\
The diagrams defining the one-loop expansion of the $\Gamma_{(R)}$ correlator are shown in Fig.~(\ref{Fig.Rchiral}). They consist of triangle and bubble topologies with fermions, since the scalars do not contribute. The explicit result for a massless chiral multiplet with on-shell external gauge bosons is given by
\bea
\label{RChiralOSMassless}
\Gamma_{(R)}^{\mu\alpha\beta}(p,q) = - i \frac{g^2 \, T(R)}{12 \pi^2} \frac{k^\mu}{k^2} \eps[p, q, \alpha ,\beta] \,,
\eea
The correlator in Eq.(\ref{RChiralOSMassless}) satisfies the vector current conservation constraints given in Eq.(\ref{VectorWI}) and the anomalous equation of Eq.(\ref{AnomalyR}) 
\bea
\label{AnomalyRmom}
i k_\mu \, \Gamma_{(R)}^{\mu\alpha\beta}(p,q) = \frac{g^2 \, T(R)}{12 \pi^2} \, \eps[p, q, \alpha ,\beta] \,.
\eea
There is no much surprise, obviously, for the anomalous structure of Eq. (\ref{RChiralOSMassless}) which is characterized by a pole $1/k^2$ term, since in the on-shell case and for massless fermions (which are the only fields contributing to the $\langle RVV \rangle$ at this perturbative order), we recover the usual structure of the $\langle AVV \rangle$ diagram.

{\bf -Three-point function of the $S^\mu_A$ current}\\
The perturbative expansion of the $\Gamma^{\mu\alpha}_{(S) \, A \dot B}$ correlation function is depicted in Fig.~(\ref{Fig.Schiral}). For simplicity we will remove, from now on, the spinorial indices from the corresponding expressions. The explicit result for a massless chiral supermultiplet with on-shell external gauge and gaugino lines is then given by
\bea
\label{SChiralOSMassless}
\Gamma^{\mu\alpha}_{(S)}(p,q) = - i \frac{g^2 T(R)}{6 \pi^2 \, k^2} s_1^{\mu\alpha}
+ i \frac{g^2 T(R)}{64 \pi^2} \Phi_2(k^2,0) \, s_2^{\mu\alpha} \,,
\eea
where the form factor $\Phi_2(k^2,0)$ is defined as
\bea
\label{Phi2massless}
\Phi_2(k^2,0) = 1 - \mathcal B_0(0,0) + \mathcal B_0(k^2,0) \,,
\eea
and the two tensor structures are 
\bea
s_1^{\mu\alpha} &=&   \sigma^{\mu \nu} k_\nu \, \sigma^\rho k_\rho \,  \bar \sigma^{\alpha \beta} p_\beta \,,\nn \\ 
s_2^{\mu\alpha} &=&  2 p_\beta \, \sigma^{\alpha \beta} \sigma^\mu \,.
\eea 
The $\mathcal B_0$ function appearing in Eq.(\ref{Phi2massless}) is a two-point scalar integral defined in Appendix \ref{AppScalarIntegrals}. Notice that the form factor multiplying the second tensor structure $s_2$ is ultraviolet finite, due to the renormalization procedure, but has an infrared singularity inherited by the counterterms in Eq.~(\ref{counterterms}).  \\
It is important to observe that the only pole contribution comes from the anomalous structure $s_1^{\mu\alpha}$, which shows that 
the origin of the anomaly has to be attributed to a unique fermionic pole ($\sigma^\rho k_\rho / k^2$) in the correlator, in the form factor multiplying $s_1^{\mu\alpha}$. 
It is easy to show that Eq.~(\ref{SChiralOSMassless}) satisfies the vector current and EMT conservation equations. Moreover, the anomalous equation reads as
\bea
\label{AnomalySmom}
\bar \sigma_{\mu} \, \Gamma^{\mu\alpha}_{(S)}(p,q) =  \frac{g^2 T(R)}{ 4 \pi^2}  \bar \sigma^{\alpha \beta} p_\beta \,,
\eea
where only the first tensor structure contributes to the $\sigma$-trace of the correlator. This result is clearly in agreement with Eq.(\ref{AnomalyS}) after Fourier transform $(\mathcal{F.T.})$, owing to
\bea
\mathcal{F.T.} \left\{ \frac{i}{2} \frac{\delta^2  F_{\mu\nu} \bar \sigma^{\mu\nu} \bar \lambda }{\delta A_\alpha(x) \delta \bar \lambda (y)} \right\} =  \bar \sigma^{\alpha \beta}  p_\beta \,.
\eea
Notice also that
\bea
\mathcal{F.T.} \left\{ \frac{\delta^2 S^\mu}{\delta A_\alpha(x) \delta \bar \lambda (y)} \right\} = s_{2}^{\mu\alpha} \,.
\eea

{\bf -Three-point function of the energy-momentum tensor $T^{\mu \nu}$}\\
The diagrams appearing in the perturbative expansions of the $\Gamma_{(T)}$ are depicted in Fig.(\ref{Fig.Tchiral}). They consist of triangle and bubble topologies. There is also a tadpole-like contribution, Fig.(\ref{Fig.Tchiral}j), which is non-zero only in the massive case. \\
The explicit expression of the  $\Gamma_{(T)}$ correlator for a massless chiral supermultiplet and on-shell gauge lines is given by
\bea
\label{TChiralOSMassless}
\Gamma_{(T)}^{\mu\nu\alpha\beta}(p,q) = - \frac{g^2 \, T(R)}{24 \pi^2 \, k^2} t_{1S}^{\mu\nu\alpha\beta}(p,q)  + \frac{g^2 \, T(R)}{16 \pi^2} \Phi_2(k^2,0) \, t_{2S}^{\mu\nu\alpha\beta}(p,q) \,,
\eea
where the $\Phi_2$ is defined in Eq.(\ref{Phi2massless}) and
\bea
t_{1S}^{\mu\nu\alpha\beta}(p,q) &\equiv& \phi_1^{\mu\nu\alpha\beta}(p,q) = (\eta^{\mu\nu} k^2 - k^\mu k^\nu) u^{\alpha\beta}(p,q)\,, \\
t_{2S}^{\mu\nu\alpha\beta}(p,q) &\equiv& \phi_3^{\mu\nu\alpha\beta}(p,q) = (p^\mu q^\nu + p^\nu q^\mu) \eta^{\alpha\beta} + p \cdot q (\eta^{\alpha\nu} \eta^{\beta\mu} + \eta^{\alpha\mu} \eta^{\beta\nu}) - \eta^{\mu\nu} u^{\alpha \beta}(p,q) \nn \\
&-&  (\eta^{\beta\nu}p^\mu + \eta^{\beta\mu}p^\nu)q^\alpha - (\eta^{\alpha\nu}q^\mu + \eta^{\alpha\mu}q^\nu)p^\beta\,,
\eea
where $\phi_1^{\mu\nu\alpha\beta}, \phi_3^{\mu\nu\alpha\beta}$ and $u^{\alpha\beta}$ are given in Eqs. (\ref{phitensors}) and (\ref{utensor}).
As in the previous cases we have explicitly checked all the Ward identities originating from gauge invariance and conservation of the energy-momentum tensor. As one can easily verify by inspection, only the first one of the two tensor structures is traceful and contributes to the anomaly equation of the $\Gamma_{(T)}$ correlator
\bea
\label{AnomalyTmom}
\eta_{\mu\nu} \, \Gamma_{(T)}^{\mu\nu\alpha\beta}(p,q) = - \frac{g^2 \, T(R)}{8 \pi^2} u^{\alpha\beta}(p,q) \,.
\eea 
The comparison of Eq.(\ref{AnomalyTmom}) to Eq.(\ref{AnomalyT}) is evident if one recognizes that
\bea
\mathcal{F.T.} \left\{ - \frac{1}{4} \frac{\delta^2 F_{\mu\nu}F^{\mu\nu}}{\delta A_\alpha(x) \delta A_\beta(y)} \right\} = u^{\alpha\beta}(p,q) \,.
\eea
For completeness we give also the inverse Fourier transform of $t_{2S}^{\mu\nu\alpha\beta}(p,q)$ which is obtained from
\bea
\mathcal{F.T.} \left\{ \frac{\delta^2 T^{\mu\nu}_{gauge}}{\delta A_\alpha(x) \delta A_\beta(y)} \right\} = t_{2S}^{\mu\nu\alpha\beta}(p,q) \,,
\eea 
where $T^{\mu\nu}_{gauge}$ is the pure gauge part of the energy-momentum tensor. Notice that $t_{2S}$ is nothing else than the tree-level vertex with two onshell gauge fields on the external lines. 

As in the previous subsection, concerning the supersymmetric current $S^\mu_A$, also in the case of this correlator there is only one structure containing a pole term, which appears in the only form factor (which multiplies $t_{1S}$) with a nonvanishing trace. Differently from the non supersymmetric case, such as in QED and QCD, with fermions or scalars running in the loops, as shown in Eqs.~(\ref{RVectorOS}), (\ref{SVectorOS}), and (\ref{TVectorOS}), there are {\em no extra poles} in the traceless structures of the decomposition of the correlators. This shows that in a  supersymmetric theory the signature of all the anomalies in the 
$\langle \mathcal{J} \mathcal{V} \mathcal{V} \rangle$ correlator are only due to anomaly poles in each channel.
 
\subsection{The vector multiplet contribution}
Finally, we come to a discussion of the perturbative results for the vector (gauge) multiplet to the three anomalous correlation functions presented in the previous sections. Notice that due to the quantization of the gauge field, gauge fixing and ghost terms must be taken into account both, increasing the complexity of the computation. This technical problem is completely circumvented with on-shell gauge boson and gaugino, which is the case analyzed in this work. \\
Concerning the diagrammatic expansion, the topologies of the various contributions defining the three correlators is analogous to those illustrated in massless chiral case. The explicit results are given by
\bea
\label{RVectorOS}
\Gamma_{(R)}^{\mu\alpha\beta}(p,q) &=&  i \frac{g^2 \, T(A)}{4 \pi^2} \frac{k^\mu}{k^2} \eps[p, q, \alpha ,\beta] \,,  \\
\label{SVectorOS}
\Gamma_{(S)}^{\mu\alpha}(p,q) &=&   i \frac{g^2 T(A)}{2 \pi^2 \, k^2} s_1^{\mu\alpha} + i \frac{g^2 T(A)}{64 \pi^2} V(k^2) \, s_2^{\mu\alpha} \,, \\
\label{TVectorOS}
\Gamma_{(T)}^{\mu\nu\alpha\beta}(p,q) &=&  \frac{g^2 \, T(A)}{8 \pi^2 \, k^2} t_1^{\mu\nu\alpha\beta}(p,q)  + \frac{g^2 \, T(A)}{16 \pi^2} V(k^2) \, t_{2}^{\mu\nu\alpha\beta}(p,q) \,,
\eea
where
\bea
V(k^2) = -3 + 3 \, \mathcal B_0(0,0) - 3 \, \mathcal B_0(k^2,0) - 2 k^2 \, \mathcal C_0(k^2,0) \,.
\eea
The tensor expansion of the correlators is the same as in the previous cases. The only differences are in the form factors. In particular, the first in each of them is the only one responsible for the anomaly and is multiplied, respect to the chiral case, by a factor $-3$ and by a different group factor. The result reproduces exactly the anomaly Eqs (\ref{AnomalyR},\ref{AnomalyS},\ref{AnomalyT}). Concerning the ultraviolet divergences of these correlators, the explicit computation shows that the vector multiplet contribution to $\Gamma_{(S)}^{\mu\nu}$ is indeed finite at one-loop order before any renormalization. This confirms a result obtained in the analysis of the renormalization properties of these correlators presented in a previous section, where we have shown the vanishing of the counterterm of $\Gamma^{\mu\alpha}_{(S)}$ for the vector multiplet.

Also for the vector multiplet, the result is similar, since the only anomaly poles present in the three correlators 
(\ref{RVectorOS}), (\ref{SVectorOS}) and (\ref{TVectorOS}) are those belonging to anomalous structures. We conclude that in all the cases discussed so far, anomaly poles are the signature of an anomaly in a superconformal theory.

\section{ The supercorrelator in the on-shell and massive case}
We now extend our previous analysis to the case of a massive chiral multiplet. This will turn out to be extremely useful in order to discuss the general behaviour of the spectral densities away from the conformal point.

\begin{figure}[t]
\centering
\subfigure[]{\includegraphics[scale=0.6]{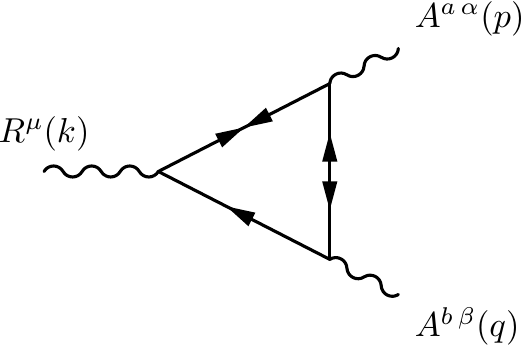}} \hspace{.5cm}
\subfigure[]{\includegraphics[scale=0.6]{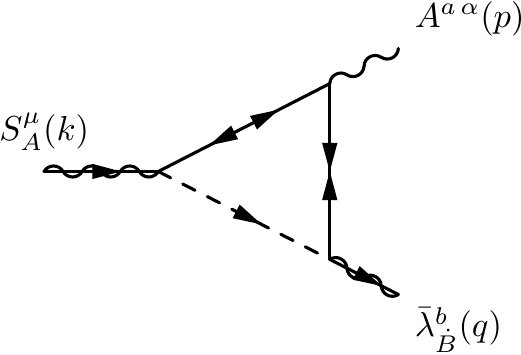}} \hspace{.5cm}
\subfigure[]{\includegraphics[scale=0.6]{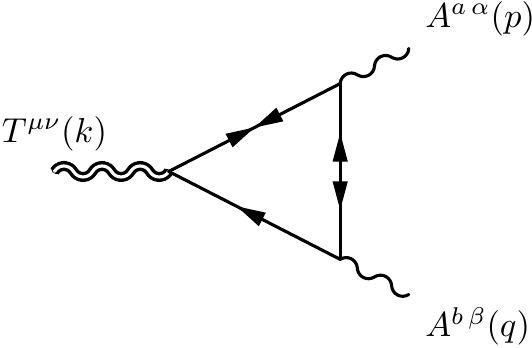}} 
\caption{A sample of diagrams, for a massive chiral multiplet, mass insertions in the fermion propagators. \label{Fig.Massivechiral}}
\end{figure}
The diagrammatic expansion of the three correlators for a massive chiral multiplet in the loops grows larger, with a bigger set of contributions. These are characterized by mass insertions on the $S^\mu_A$ and $T^{\mu\nu}$ vertices and on the propagators of the Weyl fermions. A sample of them is shown in Fig. (\ref{Fig.Massivechiral}). An explicit computation, in this case, gives
\bea
\label{RChiralOSMassive}
\Gamma_{(R)}^{\mu\alpha\beta}(p,q) &=&  i \frac{g^2 \, T(R)}{12 \pi^2} \, \Phi_1(k^2,m^2) \, \frac{k^\mu}{k^2} \eps[p, q, \alpha ,\beta]  \,, \\
\label{SChiralOSMassive}
\Gamma^{\mu\alpha}_{(S)}(p,q) &=&  i \frac{g^2 T(R)}{6 \pi^2 \, k^2} \, \Phi_1(k^2,m^2) \, s_1^{\mu\alpha}
+ i \frac{g^2 T(R)}{64 \pi^2}  \, \Phi_2(k^2,m^2) \, s_2^{\mu\alpha} \,, \\
\label{TChiralOSMassive}
\Gamma_{(T)}^{\mu\nu\alpha\beta}(p,q) &=& \frac{g^2 \, T(R)}{24 \pi^2 \, k^2} \, \Phi_1(k^2,m^2) \, t_{1S}^{\mu\nu\alpha\beta}(p,q) + \frac{g^2 \, T(R)}{16 \pi^2} \, \Phi_2(k^2,m^2) \, t_{2S}^{\mu\nu\alpha\beta}(p,q) \,, 
\eea
with
\bea
\Phi_1(k^2,m^2) &=& - 1 - 2\, m^2 \, \mathcal C_0(k^2,m^2) \,, \nn \\
\Phi_2(k^2,m^2) &=& 1 - \mathcal B_0(0,m^2) + \mathcal B_0(k^2,m^2) + 2 m^2  \mathcal C_0(k^2,m^2) \,.
\label{exp1}
\eea
The expressions above show that the only modifications introduced by the mass corrections are in the form factors, while the tensor structures remain unchanged. \\
As we have previously discussed, if the superpotential is quadratic in the chiral superfield, the conservation equation of the hypercurrent is non homogeneous. Its four-divergence equals a classical (non-anomalous) contribution due to the explicit breaking of the conformal symmetry. Therefore, in this case, the anomaly equations (\ref{AnomalyRmom}),(\ref{AnomalySmom}), and (\ref{AnomalyTmom}) must be modified in order to account for the mass dependence. The new conservation equations for a massive chiral supermultiplet become
\bea
i k_\mu \, \Gamma^{\mu\alpha\beta}_{(R)}(p,q) &=& -\frac{g^2 T(R)}{12\pi^2} \Phi_1(k^2,m^2) \eps[p,q,\alpha,\beta] \,, \\
\bar \sigma_{\mu} \, \Gamma^{\mu\alpha}_{(S)}(p,q) &=& - \frac{g^2 T(R)}{ 4 \pi^2}  \Phi_1(k^2,m^2) \bar \sigma^{\alpha \beta} p_\beta \,, \\
\eta_{\mu\nu} \, \Gamma^{\mu\nu\alpha\beta}_{(T)}(p,q) &=&  \frac{g^2 T(R)}{8\pi^2} \Phi_1(k^2,m^2) u^{\alpha\beta}(p,q) \,.
\eea
It is interesting to observe that supersymmetry prevents the appearance of new structures in the conservation equations, at least for these correlation functions, being the explicit classical breaking terms just a correction to the anomaly coefficient. This does not occur in non-supersymmetric theories  \cite{Giannotti:2008cv, Armillis:2009pq}.

\section{The flavor chiral symmetries and the Konishi anomaly}
If the superpotential $\mathcal W(\Phi)$ is absent, the action in Eq.(\ref{SUSYactionSF}) is also invariant under a phase rotation of the chiral superfield alone. This transformation, differently from the $R$ transformation, does not affect the $\theta, \bar\theta$ coordinates. If the theory contains $N_f$ flavor chiral superfields $\Phi^f$, then we can construct $N_f$ chiral currents associated to the each of the independent $U(1)$ flavor rotations. In the superfield formalism these are given by
\bea
\mathcal J^f_{A \dot A} = - \frac{1}{2} [D_A, \bar D_{\dot A}] \mathcal J^f
\eea
where $\mathcal J^f$ is the Konishi operator defined as
\bea
\mathcal J^f = \bar \Phi^f e^V \Phi^f \,.
\eea
In the component formalism the chiral currents are extracted from the $\theta \bar \theta$ component of the Konishi operator and are given by
\bea
\label{Jfcurrent}
J^f_\mu =  \bar \chi^f \bar \sigma_\mu \chi^f + i \, \phi^{f \, \dag} (\mathcal D_\mu \phi^f) - i \, (\mathcal D_\mu \phi^f)^\dag \phi^f \,.
\eea
Differently from the $R$ current, which belongs to a supermultiplet together with the supercurrent and the energy-momentum tensor, the $U(1)$ chiral currents discussed here are the only non-trivially conserved components of the Konishi operator.\\
As for non-supersymmetric theories, these $U(1)$ chiral symmetries suffer from an anomaly whose equation in the superfield formalism is given by
\bea
\bar D^2 \mathcal J^f = \frac{T(R_f)}{2 \pi^2} \textrm{Tr} W^2 \,, \qquad \mbox{or} \qquad \bar \sigma_\mu^{\dot A A}\partial^\mu \mathcal J^f_{A \dot A} = i \frac{T(R_f)}{16 \pi^2} D^2 \textrm{Tr} W^2 + h.c. \,, 
\eea
or, equivalently, in components as
\bea
\partial^\mu  J^f_\mu = \frac{g^2 \, T(R_f)}{16 \pi^2} F^{a \, \mu\nu} \tilde F^a_{\mu\nu} \,.
\eea
The one-loop perturbative computation for the three-point function, $\Gamma^{\mu\alpha\beta}_{(J^f)}$, with a $J^f_\mu$ current insertion and two on-shell gauge fields on the external lines can be easily recovered from the previous computations. Indeed, due to its chiral nature, the $J^f$ current is quite similar to the $R$ current. Taking into account the fact that the scalar part of $J^f_\mu$ in Eq.(\ref{Jfcurrent}) does not contribute to the one-loop correlator, the result for $\Gamma^{\mu\alpha\beta}_{(J^f)}$ is obtained from Eq.(\ref{RChiralOSMassless}) with a multiplicative factor $-3$, or from the vector multiplet contribution of Eq.(\ref{RVectorOS}) with a different group theoretical factor. Therefore, for massless chiral multiplets we have
\bea
\Gamma^{\mu\alpha\beta}_{(J^f)}(p,q) = i \frac{g^2 \, T(R_f)}{4 \pi^2} \frac{k^\mu}{k^2} \eps[p, q, \alpha ,\beta] \,,
\eea  
which manifests, also in this case, an anomaly pole. \\
We conclude this section by giving the expression of the correlator responsible for the Konishi anomaly in the massive case
\bea
\label{konishi}
\Gamma^{\mu\alpha\beta}_{(J^f)}(p,q) = - i \frac{g^2 \, T(R_f)}{4 \pi^2}\Phi_1(k^2,m^2) \frac{k^\mu}{k^2} \eps[p, q, \alpha ,\beta] \,,
\eea  
with $\Phi_1(k^2,m^2)$ given in Eq. (\ref{exp1}), in full analogy with the result for the correlator of the $R$ current. \\
In the next section we investigate the sum rule and 
the spectral density flows associated with these correlators, showing the universality of their behaviour.

\section{Mass deformations and the spectral densities flow } 
In this and in the following section, we turn to a detailed discussion of the dispersive structure of the form factors of the correlators computed above, since their spectral densities carry significant information on the anomaly. As before, we will be setting the momenta $p,q$ on-shell, and choose the incoming momentum $k$ to be either spacelike, timelike or null. Being interested in the analysis of the spectral density of the anomalous form factor $\Phi_1(k^2,m^2)$, it is convenient first to describe the analytic properties of the three-point scalar integral $\mathcal C_0(k^2, m^2)$ which enters in the definition of $\Phi_1$, as clear from Eq. (\ref{exp1}).\\ 
We start by introducing the spectral density $\rho(k^2)$, which is the discontinuity of $\mathcal C_0$ along the cut $(k^2>4 m^2)$, as 
\beq
\rho(k^2,m^2)=\frac{1}{2 i} \textrm{Disc}\, \mathcal C_0(k^2, m^2) \,,
\label{spect}
\eeq
with the usual $i\epsilon$ prescription ($\epsilon >0$)
\beq
\label{discdef}
 \textrm{Disc} \, \mathcal C_0(k^2, m^2)\equiv \mathcal C_0(k^2 +i \epsilon,m^2) - \mathcal C_0(k^2 -i \epsilon,m^2).
\eeq
\begin{figure}[t]
\centering
\subfigure{\includegraphics[scale=1.3]{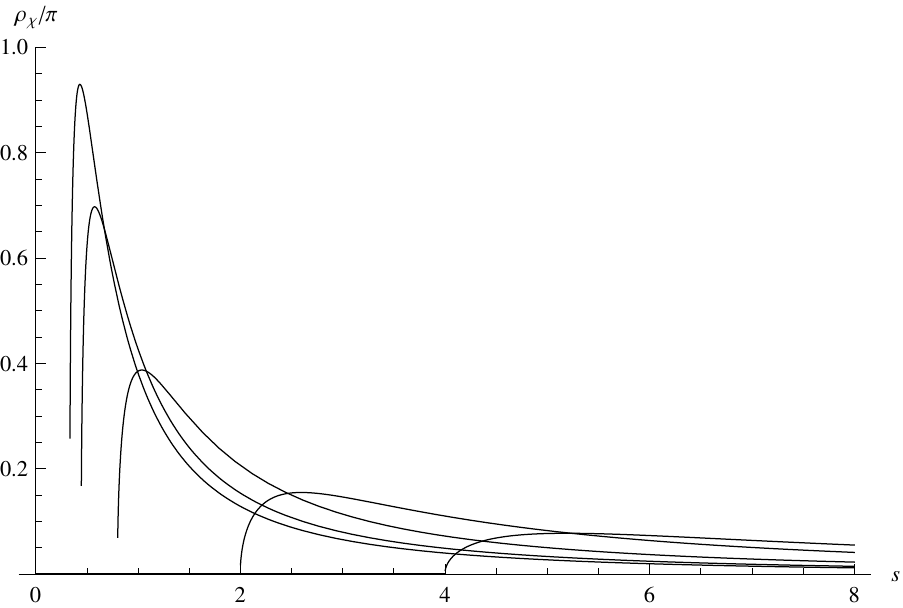}} \hspace{.5cm}
\caption{Representatives of the family of spectral densities $\frac{{{\rho}_\chi}^{(n)}}{\pi}(s)$ plotted versus $ s $ in units of $m^2$. The family "flows"  towards the $s=0$ region becoming a $\delta(s)$ function as $m^2$ goes to zero. }
\label{seq}
\end{figure}
To determine the discontinuity above the two-particle cut we can proceed in two different ways. We can use the unitarity cutting rules and therefore compute the integral
\bea
\label{disco}
\textrm{Disc} \, \mathcal C_0(k^2, m^2)&=& \frac{1}{i \pi^2} 
 \int d^4 l \frac{2 \pi i \delta_+(l^2 - m^2) 2 \pi i \delta_+((l - k)^2- m^2)}{(l-p)^2 - m^2 + i \epsilon} \nn\\
 &=&  \frac{2 \pi}{i k^2}\log\left(\frac{1 + \sqrt{\tau(k^2 ,m^2)}}{1-\sqrt{\tau(k^2 ,m^2)}}\right)\theta(k^2 - 4 m^2) \,,
\label{cut}
 \eea
where $\tau(k^2,m^2) = \sqrt{1 -4m^2/k^2}$.
 The integral has been computed by sitting in the rest frame of the off-shell line of momentum $k$.
Alternatively, we can exploit directly the analytic continuation of the explicit expression of the $\mathcal C_0(k^2,m^2)$ integral in the various regions. This is given by
\bea  
\label{ti}
 \mathcal C_0(k^2\pm i\epsilon,m^2) = \left\{ 
\begin{array}{ll} 
\frac{1}{2 k^2}\log^2 \frac{\sqrt{\tau(k^2,m^2)}+1}{\sqrt{\tau(k^2,m^2)}-1} & \mbox{for} \quad k^2 < 0 \,, \\
- \frac{2}{k^2} \arctan^2{\frac{1}{\sqrt{-\tau(k^2,m^2)}}} & \mbox{for} \quad 0 < k^2 < 4 m^2 \,, \\
\frac{1}{2 k^2} \left( \log \frac{1 + \sqrt{\tau(k^2,m^2)}}{1 - \sqrt{\tau(k^2,m^2)} } \mp i \, \pi\right)^2 & \mbox{for} \quad k^2 > 4 m^2 \,.  
\end{array} 
\right.
\eea
From the two branches encountered with the $\pm i \epsilon$ prescriptions, the discontinuity is then present only for $k^2> 4m^2$, as expected from unitarity arguments, and the result for the discontinuity, obtained using the definition in Eq. (\ref{discdef}), 
clearly agrees with Eq. (\ref{disco}), computed instead by the cutting rules. \\
The dispersive representation of $\mathcal C_0(k^2,m^2)$ in this case is written as
\beq
\mathcal C_0(k^2,m^2)=\frac{1}{\pi} \int_{4 m^2}^{\infty} d s \frac{\rho(s, m^2)}{s - k^2 }, 
\eeq
which, for $k^2 < 0$ gives the identity  
\beq
\label{loop}
 \int_{4 m^2}^{\infty} \frac{d s} {(s - k^2) s}\log\left(\frac{1 + \sqrt{\tau(s,m^2)}}{1-\sqrt{\tau(s,m^2)}}\right)=- \frac{1}{2 k^2}\log^2 \frac{\sqrt{\tau(k^2,m^2)}+1}{\sqrt{\tau(k^2,m^2)}-1},
 \eeq
with $\rho(s,m^2)$ given by Eqs. (\ref{spect}) and (\ref{cut}). The identity in Eq. (\ref{loop}) allows to reconstruct the scalar integral 
$\mathcal C_0(k^2,m^2)$ from its dispersive part. 

Having determined the spectral function of the scalar integral $\mathcal C_0(k^2,m^2)$, we can extract the spectral density associated with the anomaly form factors in Eqs. (\ref{RChiralOSMassive}), (\ref{SChiralOSMassive}), (\ref{TChiralOSMassive}) and (\ref{konishi}), which is given by
\beq
 \chi(k^2, m^2)\equiv \Phi_1(k^2,m^2)/k^2, 
 \eeq
and which can be computed as
\bea
\textrm{Disc}\, \chi(k^2, m^2)= \chi(k^2+i\epsilon,m^2)-\chi(k^2-i\epsilon, m^2) = - \textrm{Disc}\left( \frac{1}{k^2} \right) - 2 m^2 \textrm{Disc}\left(\frac{\mathcal C_0(k^2,m^2)}{k^2}\right).
\label{cancel1}
\eea
Using the principal value prescription 
\beq
\frac{1}{x\pm i\epsilon}=P\left(\frac{1}{x} \right) \mp i\pi \delta(x),
\eeq
we obtain
\bea
&& \textrm{Disc} \left( \frac{1}{k^2} \right) = - 2 i\pi\delta(k^2) \nn\\
&& \textrm{Disc}\left( \frac{\mathcal C_0(k^2,m^2)}{k^2}\right) = P\left(\frac{1}{k^2}\right)\textrm{Disc}\,\mathcal C_0(k^2,m^2) - i\pi \delta(k^2) A(0) \,,
\eea 
where we have defined  
\beq
\label{As}
A(k^2)\equiv C_0(k^2+i\epsilon,m^2)+C_0(k^2-i\epsilon,m^2),
\eeq
and
\bea
A(0) &=& \lim_{k^2\to 0} A(k^2) =  -\frac{1}{m^2}. 
\eea
This gives, together with the discontinuity of $\mathcal C_0(k^2,m^2)$ which we have computed previously in Eq. (\ref{cut}),
\beq
\label{discC0}
\textrm{Disc}\left( \frac{\mathcal C_0(k^2,m^2)}{k^2}\right)=-2 i\frac {\pi}{(k^2)^2} \log \frac{1 + \sqrt{\tau(k^2,m^2)}}{1 - \sqrt{\tau(k^2,m^2)}}\theta(k^2-4 m^2)
+i\frac{\pi}{m^2}\delta(k^2).
\eeq
The discontinuity of the anomalous form factor $\chi(k^2,m^2)$ is then given by
\beq
\textrm{Disc} \, \chi(k^2,m^2)=4 i \pi \frac{m^2}{ (k^2)^2}\log \frac{1 + \sqrt{\tau(k^2,m^2)}}{1 - \sqrt{\tau(k^2,m^2)} }\theta(k^2- 4 m^2).
 \eeq 
The total discontinuity of $\chi(k^2,m^2)$, as seen from the result above, is characterized just by a single cut for $k^2> 4 m^2$, since the $\delta(k^2)$ (massless resonance) contributions cancel between the first and the second term of Eq. (\ref{cancel1}). This result proves the {\em decoupling } of the anomaly pole at $k^2=0$ in the massive case due to the disappearance of the resonant state. \\
The function describing the anomaly form factor, $\chi(k^2,m^2)$, then admits a dispersive representation over a single branch cut
\beq
\chi(k^2,m^2)=\frac{1}{\pi}\int_{4 m^2}^{\infty}\frac{{\rho}_\chi(s,m^2)}{s- k^2 }ds
\eeq
corresponding to the ordinary threshold at $k^2=4 m^2$,
with 
\bea
\label{spectralrho}
{\rho}_\chi(s,m^2) = \frac{1}{2 i} \textrm{Disc} \, \chi(s,m^2)  
=\frac{2 \pi m^2}{s^2}\log\left(\frac{1 + \sqrt{\tau(s,m^2)}}{1-\sqrt{\tau(s,m^2)}}\right)\theta(s- 4 m^2). 
\eea
From the spectral function given above and from the corresponding integral representation one can extract a new nontrivial integral relation 
\beq
\int_{4 m^2}^{\infty} \frac{1}{s^2 (s - k^2)}\log\left(\frac{1 + \sqrt{\tau(s,m^2)}}{1-\sqrt{\tau(s,m^2)}}\right) ds 
=- \frac{1}{2 k^2 m^2} -\frac{1}{2 (k^2)^2} \log^2 \frac{\sqrt{\tau(k^2,m^2)}+1}{\sqrt{\tau(k^2,m^2)}-1},
\eeq
which is the analogue of Eq. (\ref{loop}). 

\begin{figure}[t]
\centering
\subfigure[]{\includegraphics[scale=0.8]{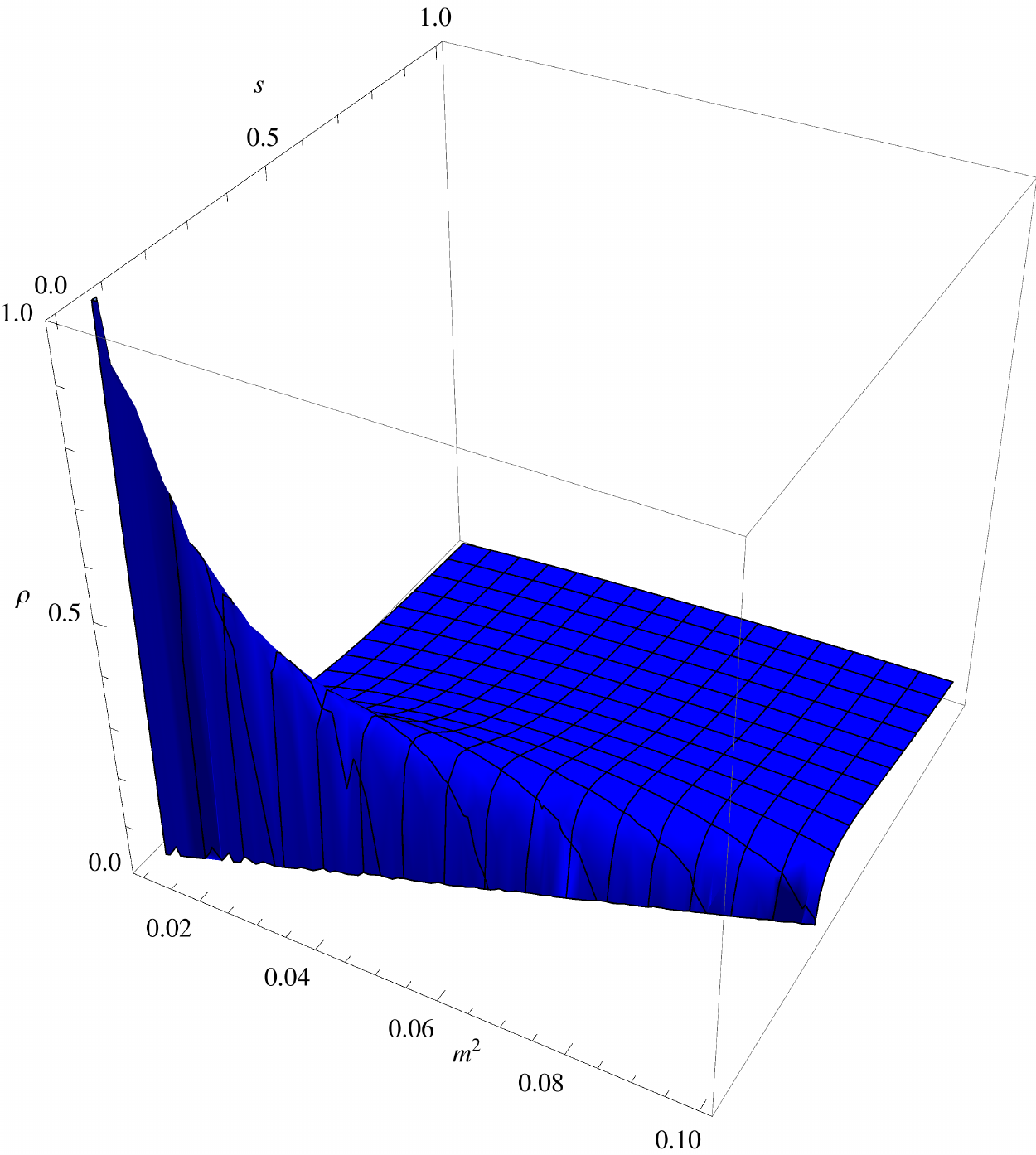}} \hspace{.5cm}
\caption{3-D Plot of the spectral density $\rho_\chi$ in the variables $s$ and $m^2$. }
\end{figure}
As we have anticipated above, a crucial feature of these spectral densities is the existence of a sum rule. In this case it is given by
\beq
\label{superc}
\frac{1}{\pi}\int_{4 m^2}^{\infty}ds{{\rho}_\chi(s,m^2)}=1.
\eeq
At this point, to show the convergence of the family of spectral densities to a resonant behaviour, it is convenient to extract a  
discrete sequence of functions, parameterized by an integer $n$ and let $n$ go to infinity.
\bea
\rho^{(n)}_\chi(s) &\equiv& \rho_\chi(s,m_n^2)  \qquad \mbox{with} \quad m_n^2 =\frac{4 m^2}{n}.
\eea
One can show that this sequence $\{\rho^{(n)}_\chi\}$ then converges to a Dirac delta function 
\bea
\lim_{m\to 0} \rho_\chi(s,m^2) = \lim_{m\to 0} \frac{2 \pi m^2}{s^2}\log\left(\frac{1 + \sqrt{\tau(s,m^2)}}{1-\sqrt{\tau(s,m^2)}}\right)\theta(s- 4 m^2) = \pi \delta(s)
\eea
in a distributional sense. We have shown in Fig. (\ref{seq}), on the left, the sequel of spectral densities which characterize the flow as we turn the mass parameter to zero. The area under each curve is fixed by the sum rule and is a characteristic of the entire flow. Clearly, the $\rho^{(n)}$ are normalized distributions for each given value of $m$. They describe, for each invariant mass value $s$, the absolute weight of the intermediate state - of that specific invariant mass - to a given anomaly form factor. Notice that the function $\chi(s,m^2)$ is a universal function, since it provides a full description of the flow for the 
anomaly form factors of all the components of the multiplet. \\
 One can see from the same figure how the density gets more and more peaked towards the lower end of the region of the interval $4 m_n^2 \leq s < \infty$ as $m_n^2$ tends to zero. In physical terms this means that the branch cut is replaced by a single massless anomaly pole. In Fig. (\ref{seq}), on the right, we have included a 3D plot of $\rho_{\chi}(s,m^2)$ in the $(s, m^2)$ plane, giving a visual perspective on the entire flow. 
\subsection{The analytic structure of $\Phi_2$}
Here we discuss the spectral representation of the second of the form factors appearing in the same $\Gamma_{T}$ and $\Gamma_{S}$ correlators, which is proportional to the renormalized function 
$\Phi_2$ 
\bea
\label{exp11}
\Phi_2(k^2,m^2)&= &  1 - \mathcal B_0(0,m^2) + \mathcal B_0(k^2,m^2) + 2 m^2 \mathcal C_0(k^2,m^2) 
\eea
which needs a subtraction for its integrability, due to the UV singularities of $\Phi_2$. 
Clearly, in this case, 
$\Phi_2$ does not admit a dispersive representation, due to its logarithmic divergence at large $k^2$, and, as we are going to show, it is 
characterized just by an ordinary cut for $k^2 > 4 m^2$, as in the previous case. We are now going to briefly illustrate this point. 

\begin{figure}[t]
\centering
\subfigure{\includegraphics[scale=1.2]{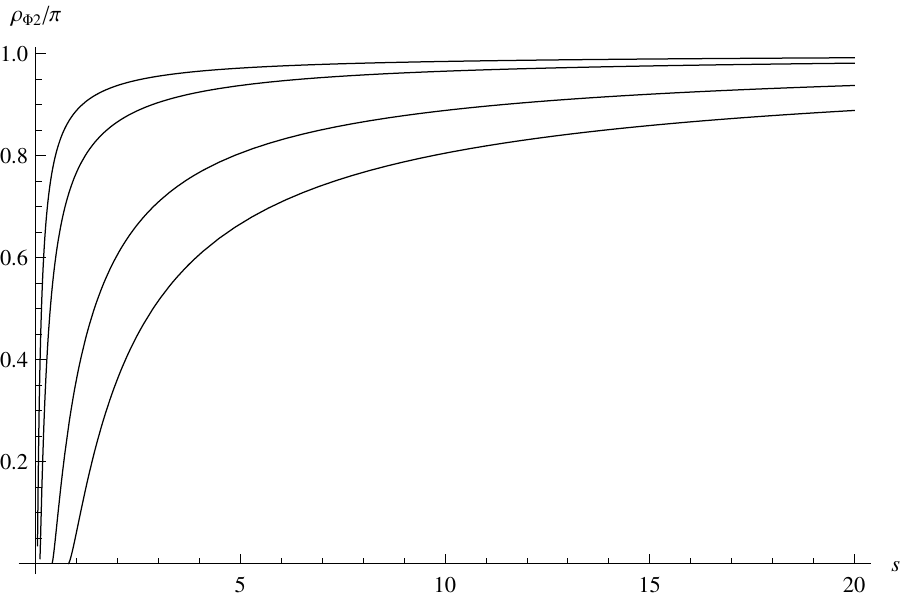}}
\caption{Spectral density flow of $\frac{{{\rho}_{\Phi_2}}}{\pi}(s,m^2)$ versus $s$. As $m^2$ decreases they turn to a unit step function $\theta(s)$. \label{phi2spectral}}
\end{figure}

As for $\mathcal C_0(k^2,m^2)$ also in this case we give the three branches of $\mathcal B_0(k^2,m^2)$ in the $k^2<0, 0<k^2<4 m^2$ and $k^2> 4 m^2$ regions
\bea  
 \mathcal B_0(k^2\pm i\epsilon,m^2) = \left\{ 
\begin{array}{ll} 
\frac{2}{\epsilon_{UV}} + 2 - \log\frac{m^2}{\mu^2} + \sqrt{\tau(k^2,m^2)} \log \frac{\sqrt{\tau(k^2,m^2)} -1}{\sqrt{\tau(k^2,m^2)} +1} & \mbox{for}  \quad k^2 < 0 \,, \\
\frac{2}{\epsilon_{UV}} + 2 - \log\frac{m^2}{\mu^2} - 2 \sqrt{- \tau(k^2,m^2)} \arctan{\frac{1}{\sqrt{- \tau(k^2,m^2)}}} & \mbox{for} \quad 0 < k^2 < 4 m^2 \,, \\
\frac{2}{\epsilon_{UV}} + 2 - \log\frac{m^2}{\mu^2} - \sqrt{ \tau(k^2,m^2)} \left( \log \frac{1 + \sqrt{\tau(k^2,m^2)}}{1 - \sqrt{\tau(k^2,m^2)}} \mp i \pi \right )& \mbox{for} \quad k^2 > 4 m^2 \,.  
\end{array} 
\right.
\eea
The discontinuity of the two-point scalar integral $\mathcal B_0(k^2,m^2)$ is then easily computed and it is given by
\bea
 \textrm{Disc}\, \mathcal B_0(k^2,m^2) = \mathcal B_0(k^2 + i\epsilon,m^2) - \mathcal B_0(k^2 - i\epsilon,m^2) = 2 i \pi \sqrt{\tau(k^2,m^2)} \, \theta(k^2 - 4 m^2) \,.
\eea
From the previous equation and from Eq. (\ref{cut}) we extract the discontinuity of $\Phi_2$ in the form 
 \beq
 \textrm{Disc}\, \Phi_2(k^2,m^2)=   2 i \pi \left( \sqrt{\tau(k^2,m^2)}- \frac{2 m^2}{k^2} \log\frac{1+ \sqrt{\tau(k^2,m^2}) }{1-\sqrt{\tau(k^2,m^2} )}\right)\theta(k^2- 4 m^2).
\eeq
This shows that both $\Phi_1/k^2$ and $\Phi_2$ are characterized by a single 2-particle cut for a nonzero mass deformation.  
It is important to observe that the spectral density of $\Phi_2$ tends to a uniform distribution
\beq
\frac{1}{\pi}\lim_{m\to 0}\rho_{\Phi_2}(k^2,m^2)= 1
\eeq
in the massless limit.
It is obvious, from this analysis, that the spectral density $\rho_{\Phi_2}$ of $\Phi_2$, which characterizes all the non anomalous form factors 
of the correlators that we have investigated, does not satisfy an unsubtracted dispersion relation. There is however a sort of duality between the spectral densities of the two form factors, since while $\rho_{\chi}$ becomes more and more localized 
at $k^2=0$ as $m\to 0$, the opposite is true for the spectral density of the non anomallous form factor $\rho_{\Phi_2}$, as clear from Fig. \ref{phi2spectral}. In this case, as $m$ goes to zero, the flow singles out - in the form factor which is relevant for the anomaly - {\em a single massless state}, while all the continuum region carries uniform weight in $\rho_{\Phi_2}$.
 \section{Constraining the flow: scaling behaviour and sum rules} 
The large momentum behaviour of the anomaly form factors, beside the sum rule, can be studied directly also from the explicit expressions of these.  
For this goal, we are going to investigate the behaviour of both $\Phi_1$ and $\Phi_2$ in the two opposite limits $k^2 \rightarrow 0$ and $k^2 \rightarrow - \infty$, which cover the light-cone as well as the deep euclidean regions of the correlators. 
For $k^2$ approaching zero we have
\bea
\label{smallk2limit}
\Phi_1(k^2,m^2) \sim  \frac{1}{12} \frac{k^2}{m^2} + O(k^4/m^4) \,, \qquad \qquad \Phi_2(k^2,m^2) \sim  \frac{1}{12} \frac{k^2}{m^2} + O(k^4/m^4) \,,
\eea
while for a large and negative $k^2$ we find
\bea
\label{largek2limit}
\Phi_1(k^2,m^2) &\sim& - 1 - \frac{m^2}{k^2} \log^2 \frac{-k^2}{m^2} + O(m^4/k^4) \,, \nn \\
\Phi_2(k^2,m^2) &\sim& 3 - \log \frac{-k^2}{m^2} + \frac{m^2}{k^2} \left( 2 + 2  \log \frac{-k^2}{m^2} +  \log^2 \frac{-k^2}{m^2}\right) + O(m^4/k^4) \,.
\eea
Because these form factors are characterized by only two mass scales, namely $m^2$ and $k^2$, performing the $k^2 \rightarrow - \infty$ limit is equivalent to taking the massless limit. Indeed it is easy to show that the leading order terms in Eq. (\ref{largek2limit}) reproduce the massless chiral contributions described in the previous sections. Notice also the presence of an infrared singularity, for $m^2 \rightarrow 0$, in the $\Phi_2$ form factor (see Eq. (\ref{largek2limit})). This is  due to the $\mathcal B_0(0,0)$ scalar integral appearing in Eq. (\ref{Phi2massless}). \\
The argument can be formally stated as follows. The anomaly form factor $\chi=\Phi_1/k^2$ satisfies the relation under rescaling with a constant $\lambda$
\beq
\chi(\lambda k^2,\lambda m^2)=\frac{1}{\lambda}\chi(k^2,m^2)
\eeq
being a homogeneous function. A similar property of homogeneity holds for the spectral density itself 
\beq
\rho_\chi(\lambda s,\lambda m^2)=\frac{1}{\lambda}\rho_\chi(s,m^2),
\label{scale}
\eeq
which under a partial rescaling, involving only the mass parameter $m$, with $m^2 \to m^2/\lambda$ and $\lambda$ large (which is the same as $m\to 0$) has the resonant behaviour
\beq
\label{scaling}
\lim_{\lambda\to\infty}\rho_\chi(s,\frac{m^2}{\lambda})= \pi \delta(s). 
\eeq
At this point, using Eq. (\ref{scale}) a large rescaling of the invariant mass $s$ gives 
\beq
\label{cone}
\rho_\chi(\lambda s, m^2)=\frac{1}{\lambda}\rho_\chi(s, \frac{m^2}{\lambda})\sim \frac {\pi}{\lambda}\delta(s)= \pi \delta(\lambda s),
\eeq
showing that the asymptotic behaviour of $\rho_\chi$ under a rescaling of $s$ with $\lambda$ identifies its support on the $s=0$ region. Notice that Eq. (\ref{cone}) should be interpreted as a light-cone dominance ($s\to 0$) of the asymptotic limit of the correlator as $\lambda$ goes to infinity. \\
On the other hand, the vanishing of the massive form factors in the $k^2 \rightarrow 0$ region, and the consequent disappearance of the $1/k^2$ pole in the anomalous correlators, may be understood as a consequence of decoupling of the massive states. \\
Scaling relations (a), combined with the sum rule (b) and the resonant behaviour of the densities for $m$ going to zero (c), provide some important constraints on 
the structure of the flow, although they are not exclusive to anomalous form factors.
We recall that as a consequence of the scaling relation, one has the constraint
\beq
\label{one1}
k^2 \frac{\partial \chi(k^2,m^2) }{\partial k^2} +m^2 \frac{\partial \chi(k^2,m^2) }{\partial m^2} + \chi(k^2,m^2)=0.
\eeq
Similar conditions are satisfied by the related spectral density $(\rho_\chi)$
\bea
s \frac{\partial \rho_\chi}{\partial s} + m^2 \frac{\partial \rho_\chi }{\partial m^2} +\rho_\chi&=& 0.
\eea
The combination of scaling behaviour and of the sum rule, together with the vanishing of $\rho_\chi(s,m^2)$ at the threshold (i.e. at $s=4 m^2$), induces further constraints on its functional form, for instance 
\beq
\label{eq22}
\frac{1}{\pi} \int_{4 m^2}^\infty \frac{\partial \rho_\chi(s,m^2)}{\partial s} ds =0, \qquad \frac{1}{\pi} \int_{4 m^2}^\infty \frac{\partial \rho_\chi(s,m^2)}{\partial m^2} ds =0,\qquad \frac{1}{\pi}\int_{4 m^2}^\infty s\frac{\partial \rho_\chi(s,m^2)}{\partial s} ds =-f.
\eeq
In the previous equation, and in the following ones, $f$ is a nonzero constant which normalizes the sum rule of the spectral density. For $\rho_\chi$ introduced in the previous section $f=1$. \\
Eq. (\ref{eq22}) can be generalized to give an infinite set of ordinary and superconvergent sum rules 
\bea
\frac{1}{\pi} \int_{4 m^2}^{\infty} ds \,(s - 4 m^2)^n \frac{\partial^n \rho_\chi}{\partial s^n} &=&(-1)^{n} n! f, \qquad n\geq 1\nn\\
\frac{1}{\pi} \int_{4 m^2}^{\infty} ds\,(s - 4 m^2)^n \frac{\partial^{n+1} \rho_\chi}{\partial s^{n+1}} &=& 0.
\eea
Additional constraints come from the scaling relation expanded to second order, 
\beq
s^2  \frac{\partial^2 \rho_\chi}{\partial s^2} + m^4  \frac{\partial^2 \rho_\chi}{\partial ({m^2})^2} + 2 s\, m^2 \frac{\partial^2 \rho_\chi}{\partial s\, \partial m^2}=2 \pi \, f.
\label{duesim}
\eeq
Using the information that the density has only a branch cut for nonzero $m$, integrating over the cut Eq. (\ref{duesim}) we get 
\beq
m^4 \int_{4m^2}^{\infty} ds    \frac{\partial^2 \rho_\chi}{\partial ({m^2})^2} =-2  m^2 \int_{4m^2}^{\infty} ds \, s\frac{\partial^2 \rho_\chi}{\partial s\, \partial m^2}.
\label{pm}
\eeq
At this point, the sign of the dispersive integrals above can be determined by exploiting the derivative form of the sum rule 
\beq
\frac{1}{\pi} \int_{4m^2}^{\infty} ds s \frac{\partial \rho_\chi}{\partial s}=-f,
\eeq
which is satisfied because of the convergence condition of the integral of $\rho_\chi$. Differentiated respect to $m^2$ the sum rule above gives
\beq
 \int_{4m^2}^{\infty} ds \frac{\partial^2 \rho_\chi}{\partial s\, \partial m^2}= 16 m^2 \frac{\partial \rho_\chi}{\partial s}\bigg|_{s=4m^2} \,, 
\eeq
which relates the integral of the mixed derivatives to the spectral density at the threshold. If the spectral density is properly normalized with a positive 
constant $f$ in the sum rule, then it will be always positive along the entire cut and, in particular, at threshold $t$. Notice that as $m$ goes to zero, 
the density is saturated by the pole behaviour, and it is then clear that it implies the local positivity relation 
\beq
 \frac{\partial^2 \rho_\chi}{\partial s\, \partial m^2} >0  \qquad m\sim 0,
 \eeq
 being the integral dominated just by the region around the threshold  $s\sim 4 m^2 $. Clearly this implies that 
\beq
\label{one2}
 \int_{4m^2}^{\infty} ds    \frac{\partial^2 \rho_\chi}{\partial ({m^2})^2} < 0,
 \eeq
 having used Eq. (\ref{pm}). Also in this case, in the $m\to 0$ limit, the inequality becomes a local condition  
 \beq
 \frac{\partial^2 \rho_\chi}{\partial ({m^2})^2}<0 
 \eeq
 which has to be satisfied by the flow. Notice that in the presence of multiple thresholds at specific masses $m_n$ the density jumps at every threshold by a positive or a negative amount. The jump is proportional to the contribution of the new threshold to the 
 $\beta$ function of the theory. This point can be easily illustrated by reintroducing the prefactor contribution of each massless state in front of the corresponding density. For this purpose we define the contributions of each field to the $\beta$ function of a theory at 1-loop, which for a Dirac fermion and a complex scalar in the representation $R_f$ and $R_s$ respectively, and for a spin 1 in the adjoint are 
 \beq
 \beta(g)=\sum_n \frac{g^3}{16 \pi^2} c^{(n)},
 \eeq
 with
 \beq
 c^{(D)} = \frac{4}{3} T(R_f) \qquad  c^{(A)} =- \frac{11}{3} T(A) \qquad c^{(\phi)} = \frac{1}{3} T(R_s)
 \eeq 
with $T(R_f)$, $T(A)$, $T(R_s)$ being the Dynkin indices of the respective representations. Real scalars and Weyl fermions contribute with an additional factor of $1/2$ respect to complex scalars and Dirac fermions. 
We recall that in a $SU(N)$ $\mathcal{N}=1$ theory, the vector multiplet contributes with
$-11/3 \,T(A)$ and $2/3\, T(A)$ for the gauge field and the gaugino respectively, while the chiral supermultiplet gives $2/3 \, T(R)$ and $1/3 \, T(R)$ for the Weyl fermion and the complex scalar.\\
We use the notation 
\beq
\rho (s,\{m_n^2\}) = \sum_n c^{(n)} \rho_\chi (s,m_n^2)
\eeq
to refer to the total spectral density of a certain theory, with intermediate thresholds at increasing mass values $\{m_n^2\}\equiv(m_1^2,m_2^2, \ldots , m_I^2)$ with
$(m_1< m_2< \ldots < m_I)$, where $I$ counts the total number of degrees of freedom. The corresponding anomaly form factor will be given by 
\beq
\label{th1}
F(Q^2,\{m_n^2\})=\frac{-2}{3 g} \frac{g^3}{16 \pi^2}\sum_n  c^{(n)}\frac{1}{\pi}\int_{4 m_n^2}^\infty ds \frac{\rho_\chi(s,m_n^2)}{s + Q^2}. 
 \eeq
 Notice that if $Q^2 \gg 4 m^2_n$, for a certain mass threshold $n$, then we can set $Q^2=4 m_n^2\lambda$, with $1/\lambda= 4 m_n^2/Q^2 \ll 1$.
Due to scaling, the $n_{th}$ threshold will then contribute to the total form factor with the amount
\beq
F_n(Q^2,m_n^2)=\frac{-2}{3 g} \frac{g^3}{16 \pi^2} c^{(n)} \frac{1}{\pi}\int_{4 m_n^2/\lambda}^\infty ds \frac{\rho_\chi(s,m_n^2/\lambda)}{s + 4 m_n^2 \lambda}, 
\eeq
which in the $1/\lambda \ll 1$ limit will give
\beq
\label{inter}
F_n(Q^2,m_n^2) \sim \frac{-2}{3 g} \frac{g^3}{16 \pi^2} c^{(n)} \int_{0}^\infty ds\frac{\delta(s)}{\lambda(s + 4 m_n^2)} 
= \frac{-2}{3 g}\beta^{(n)}(g)\frac{1}{Q^2}.
\eeq
Eq. (\ref{inter}) reduces to the anomaly pole contribution times the contribution of the state $(n)$ to the expression of the total $\beta$ function. 
 As $Q^2$ grows larger than any intermediate scale, the total spectral density $\rho$ in the dispersive integral is asymptotically given by the expression
 \bea
 \rho(s, \{m_n^2\}) \sim  \sum_n  c^{(n)}  \delta(s)  = \frac{16 \pi^2}{g^3} \beta(g) \pi \delta(s) \,,
 \eea
where we have used Eq. (\ref{scaling}).
 Notice that  $\rho(s, \{m_n^2\})$ satisfies a total sum rule to which contribute all the intermediate thresholds for $0<s<\infty$ 
 \bea
\frac{1}{\pi} \int_{0}^\infty ds \, \rho(s, \{m_n^2\})&=&\sum_n c^{(n)} \frac{1}{\pi} \int_{4 m_n^2}^\infty ds \, \rho_\chi(s,m_n^2) =  \frac{16 \pi^2}{g^3} \beta(g). 
\eea
 In supersymmetric theories this function is the only one which developes a resonant behaviour at the conformal point and satisfies a sum rule, as we have pointed out. The sum of the densities stripped of the gauge factors, integrated over the thresholds 
  \beq
  \label{th2}
\frac{1}{\pi}\sum_n \int_{4 m_n^2}^\infty ds \, \rho_\chi (s,m_n^2)= I  
\eeq
simply counts the number of degrees of freedom $(I)$.\\
Notice that the analysis of this section related to Eqs. (\ref{one1}-\ref{one2}) remains valid also for any form factor which is characterized by a finite (non superconvergent) sum rule. The asymptotic analysis discussed in Eqs. (\ref{th1}-\ref{th2}), can be also easily extended to cases unrelated to the anomaly, with coefficients $c^{(n)}$ replaced by some new coefficients, not related to the $\beta$ function.

\section{Comparing supersymmetric and non supersymmetric cases: sum rules and extra poles in the Standard Model}
In this section and in the following one, we compare the structure of the spectral densities in supersymmetric and in non supersymmetric theories in the presence of mass terms. In particular we will be looking for additional sum rules not directly related to the anomalies, which may be present in the $\langle TVV \rangle$ and $\langle AVV \rangle$ correlators. We anticipate that these are found in the $\langle TVV \rangle$ (hence in the non supersymmetric case) in all the gauge invariant sectors of the Standard Model. We start our analysis with the conformal anomaly action of QCD, described by the EMT-gluon-gluon vertex and then move to the EMT-$\gamma\gamma$  vertex of the complete electroweak theory. Obviously, the spectral densitites develope anomaly poles in the limit in which all the second scales of the vertices turn to zero. By this we refer to fermion masses, to the $W$ mass and to the external virtualities of the diagrams. Moreover, we are going to identify the explicit form of the sum rules satisfied by these correlators in perturbation theory. 
\subsection{The extra pole of QCD}
For definiteness we focus our attention on a specific gauge theory, QCD.   
We write the whole amplitude $\Gamma^{\mu\nu\a\b}(p,q)$ of the $\langle TVV \rangle$ diagram in QCD in the form
\bea
\Gamma^{\mu\nu\a\b}(p,q) = \Gamma_q^{\mu\nu\a\b}(p,q) + \Gamma_g^{\mu\nu\a\b}(p,q),
\eea
having separated the quark $(\Gamma_q)$ and the gluons/ghosts $(\Gamma_g)$ contributions. We have omitted the colour indices for simplicity, being the correlator diagonal in colour space.
As described before in Section \ref{TVVsection} in the massless case, also in the massive case the amplitude $\Gamma$ is expressed in terms of 3 tensor structures. In the $\overline{MS}$ scheme these are given by \cite{Armillis:2010qk}
\beq
\Gamma^{\mu\nu\alpha\beta}_{q/g}(p,q) =  \, \sum_{i=1}^{3} \Phi_{i\,q/g} (k^2,m^2)\, \phi_i^{\mu\nu\alpha\beta}(p,q)\,.
\label{Gamt}
\eeq
For on-shell and transverse gluons, only 3 invariant amplitudes contribute, which for the quark loop case are given by
\bea
\Phi_{1\, q} (k^2,m^2) &=&
\frac{g^2}{6 \pi^2  k^2} \bigg\{  - \frac{1}{6}   +  \frac{ m^2}{k^2}  - m^2 \mathcal C_0(k^2,m^2)
\bigg[\frac{1}{2 \,  }-\frac{2 m^2}{ k^2}\bigg] \bigg\} \,,  \\
\Phi_{2\, q} (k^2,m^2)  &=&
- \frac{g^2}{4 \pi^2 k^2} \bigg\{   \frac{1}{72} + \frac{m^2}{6  k^2}
+  \frac{ m^2}{2  k^2} \mathcal D (k^2,m^2) +  \frac{ m^2}{3 } \mathcal C_0(k^2,m^2 )\, \left[ \frac{1}{2} + \frac{m^2}{k^2}\right] \bigg\} \,,  \\
\Phi_{3\,q} (k^2,m^2) &=& 
\frac{g^2}{4 \pi^2} \bigg\{ \frac{11}{72}  +   \frac{ m^2}{2 k^2}
 +  m^2  \mathcal C_0(k^2,m^2) \,\left[ \frac{1}{2 } + \frac{m^2}{k^2}\right]  +  \frac{5  \, m^2}{6 k^2}  \mathcal D (k^2,m^2) + \frac{1}{6} \mathcal B_0^{\overline{MS}}(k^2, m^2) \bigg\},
\label{masslesslimit}
\eea
where the on-shell scalar integrals $\mathcal D (k^2,m^2)$, $\mathcal C_0(k^2, m^2)$ and $\mathcal B_0^{\overline{MS}}(k^2, m^2)$ are given in Appendix \ref{AppScalarIntegrals}. \\
Here we concentrate on the two form factors which are unaffected by renormalization, namely $\Phi_{1,2 q}$. Both admit convergent dispersive integrals of the form 
\bea
\Phi_{1,2 q}(k^2,m^2) &=& \frac{1}{\pi} \int_0^{\infty} ds \frac{\rho_{1,2 q}(s,m^2)}{s-k^2} \,,
\eea
in terms of spectral densities ${\rho_{1,2 q}(s,m^2)}$. From the explicit expressions of these two form factors, the corresponding spectral densities are obtained using the relations 
\bea
&& \textrm{Disc}\left( \frac{1}{s^2}\right) = 2i\pi \delta'(s),\nn\\
&& \textrm{Disc}\left( \frac{\mathcal C_0(s,m^2)}{s^2} \right)= - \frac{2i \pi}{s^3} \log \frac{1+\sqrt{\tau(s,m^2)}}{1-\sqrt{\tau(s,m^2)}}\theta(s-4 m^2) +   i\pi \delta'(s) A(s), \label{disc1}
\eea
where $A(s)$ is defined in Eq.(\ref{As}) and we have used the general relation 
\beq
\left( \frac{1}{x +i \epsilon}\right)^n - \left( \frac{1}{x -i \epsilon}\right)^n=(-1)^n \frac{2 \pi i}{(n-1)!}\delta^{(n-1)}(x) \,,
\eeq
with $\delta^{(n)}(x)$ the $n$-th derivative of the delta function. 
The contribution proportional to $\delta'(s)$ in Eq.(\ref{disc1}) can be rewritten in the form
\bea
\delta'(s) A(s) = -\delta(s) A'(0)+ \delta'(s) A(0), \qquad  \mbox{with} \quad A(0)=-\frac{1}{m^2} \,, \quad A'(0)=-\frac{1}{12 m^4} \,,
\eea
giving for the spectral densities
\bea
\label{rhoq}
\rho_{1q}(s,m^2) &=& \frac{g^2}{12 \pi} \frac{m^2}{s^2} \tau(s,m^2) \log \frac{1+\sqrt{\tau(s,m^2)}}{1-\sqrt{\tau(s,m^2)}} \theta(s-4m^2) \,, \nn \\
\rho_{2q}(s,m^2) &=& - \frac{g^2}{12 \pi} \left[ \frac{3 m^2}{2 s^2} \sqrt{\tau(s,m^2)} - \frac{m^2}{s} \left( \frac{1}{2 s} + \frac{m^2}{s^2}   \right) \log \frac{1+\sqrt{\tau(s,m^2)}}{1-\sqrt{\tau(s,m^2)}} \right] \theta(s-4m^2).
\eea
Both functions are characterized by a two particle cut starting at $4m^2$, with $m$ the quark mass.
Notice also that in this case there is a cancellation of the localized contributions related to the $\delta(s)$, showing that for nonzero mass there are no pole terms in the dispersive integral. The crucial difference, respect to the supersymmetric case discussed above, is that now we have two independent sum rules   
\bea
 \frac{1}{\pi} \int_{0}^{\infty} ds \, \rho_{1 q}(s,m^2) = \frac{g^2}{36 \pi^2} \,, \qquad \qquad
 \frac{1}{\pi} \int_0^{\infty} ds \, \rho_{2q}(s,m^2) = \frac{g^2}{288 \pi^2} \,,
\eea
one for each form factor, as it can be verified by a direct integration. We can normalize both densities as
\beq
\bar{\rho}_{1 q}(s,m^2)\equiv  \frac{36 \pi^2}{g^2}\rho_{1 q}(s,m^2) \qquad \bar{\rho}_{2 q}(s,m^2) \equiv  \frac{288 \pi^2}{g^2}\rho_{2 q}(s,m^2)
\eeq
in order to describe the two respective flows, which are homogeneuos, since both densities carry the same physical dimension and both converge to a $\delta(s)$ as the quark mass $m$ is sent to zero 
\beq
\lim_{m\to 0}\bar{ \rho}_{1 q}=\lim_{m\to 0}\bar{ \rho}_{2 q}=\delta(s). 
\eeq
Indeed at $m=0$, $\Phi_{1,2 q}$ are just given by pole terms, while $\Phi_{3q}$ is logarithmic in momentum 
\bea
\Phi_{1\,q} (k^2,0) &=& - \frac{g^2}{36 \pi^2  k^2}, \qquad
\Phi_{2\,q} (k^2,0) = - \frac{g^2}{288 \pi^2 \, k^2}, \\
\Phi_{3\,q} (k^2,0) &=& - \frac{g^2}{288 \pi^2} \, \left( 12 \log \left( -\frac{k^2}{\mu^2} \right ) - 35\right), \qquad \mbox{for} \quad k^2<0.
\eea
It is then clear, from this comparative analysis, that the supersymmetric and the non supersymmetric anomaly correlators can be easily differentiated with respect to their spectral behaviour. In the non supersymmetric case 
the spectral analysis of the $\langle TVV \rangle$ correlator shows the appearance of two flows, one of them being anomalous, the other not. 
A similar pattern is found in the gluon sector, which obviously is not affected by the mass term. 
In this case the on-shell and transverse condition on the external gluons brings to three very simple form factors  
whose expressions are
\bea
\Phi_{1\,g}(k^2) &=& \frac{11 \, g^2}{72 \pi^2 \, k^2} \, C_A \,, \qquad 
\Phi_{2\,g}(k^2) = \frac{g^2}{288 \pi^2 \, k^2} \, C_A \,, \\
\Phi_{3\,g}(k^2) &=& - \frac{g^2}{8 \pi^2}  C_A \bigg[ \frac{65}{36} + \frac{11}{6} \mathcal{ B}_0^{\overline{MS}}(k^2,0) - \mathcal {B}_0^{\overline{MS}}(0,0) +  k^2  \,\mathcal C_0(k^2,0) \bigg].
\label{gl2}
\eea
The $\overline{MS}$ renormalized scalar integrals can be found in Appendix \ref{AppScalarIntegrals}.
Also in this case, it is clear that the simple poles in $\Phi_{1\,g}$ and $\Phi_{2\,g}$, the two form factors which are not affected by the renormalization, are accounted for by two spectral densities which are proportional to $\delta(s)$. The anomaly pole in $\Phi_{1\,g}$ is accompanied by a second pole in the non anomalous form factor $\Phi_{2\,g}$. Notice that $\Phi_{3 g}$ is affected by renormalization, and as such it is not considered relevant in the spectral analysis. 

\subsection{$\langle TVV \rangle$ and the two spectral flows of the electroweak theory} 
\begin{figure}[t]
\centering
\includegraphics[scale=0.8]{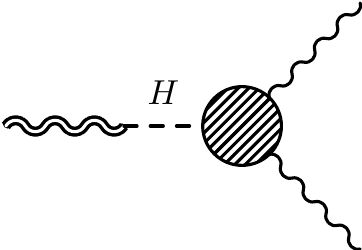}
\caption{Amplitude with the graviton - Higgs mixing vertex generated by the term of improvement. The blob represents the SM Higgs -VV' vertex at one-loop.}
 \label{HVVImpr}
\end{figure}

The point illustrated above can be extended to the entire electroweak theory by looking at some typical diagrams which manifest a trace anomaly. The simplest case is the $\langle TVV \rangle$ in the full electroweak theory, where $V$, in this case, denotes on-shell photons. At one loop level it is given by the vertex $\Gamma^{\mu\nu\alpha\beta}$ and expanded onto two terms
 \beq
\label{TVVew}
 \Gamma^{\mu\nu\alpha\beta}(p,q)=\Sigma^{\mu\nu\alpha\beta}(p,q) +\Delta^{\mu\nu\alpha\beta}(p,q)\,,
 \eeq
where $\Sigma^{\mu\nu\alpha\beta}(p,q)$ is a full irreducible contribution derived from the set of diagrams given in the appendices and depicted in Figs.(\ref{ewtops}), corresponding to topologies of triangles, bubbles and tadpoles. %
In this case $\Sigma^{\mu\nu\alpha\beta}(p,q)$ is given by the expression%
 \cite{Coriano:2011ti,Coriano:2011zk,Coriano:2012nm}
\bea
\Sigma ^{\mu\nu\a\b}(p,q) = \Sigma_{F}^{\mu\nu\a\b}(p,q) + \Sigma_{B}^{\mu\nu\a\b}(p,q) + \Sigma_{I}^{\mu\nu\a\b}(p,q),
\eea
corresponding to the exchange of fermions ($\Sigma_{F}$), gauge bosons ($\Sigma_{B}$) and to a term of improvement $(\Sigma_{I})$. The latter is generated by an EMT of the form 
\bea
T^I_{\mu\nu} = - \frac{1}{3} \bigg[ \partial_{\mu} \partial_{\nu} - \eta_{\mu\nu} \, \Box \bigg] \mathcal H^\dag \mathcal H = - \frac{1}{3} \bigg[ \partial_{\mu} \partial_{\nu} - \eta_{\mu\nu} \, \Box \bigg] \bigg( \frac{H^2}{2} + \frac{\phi^2}{2} + \phi^{+}\phi^{-} + v \, H \bigg).
\eea
and is responsible for a bilinear mixing between the EMT and the Higgs field. \\
The term $\Delta^{\mu\nu\alpha\beta}(p,q)$ in Eq.(\ref{TVVew}) comes from the insertion of the EMT of improvement given above on the Standard Model $H\gamma\gamma$ vertex. The relevant diagram is reported in Fig. (\ref{HVVImpr}). The inclusion of this term 
is necessary in order to guarantee consistent Ward identities, as discussed in \cite{Coriano:2011zk}. \\
They complete irreducible contributions are expanded as 
\bea
\Sigma ^{\mu\nu\alpha\beta}_{F}(p,q) &=&  \, \sum_{i=1}^{3} \Phi_{i\,F} (s,0, 0,m_f^2) \, \phi_i^{\mu\nu\alpha\beta}(p,q)\,, \\
\Sigma ^{\mu\nu\alpha\beta}_{B}(p,q) &=&  \, \sum_{i=1}^{3} \Phi_{i\,B} (s,0, 0,M_W^2) \, \phi_i^{\mu\nu\alpha\beta}(p,q)\,, \\
\Sigma ^{\mu\nu\alpha\beta}_{I}(p,q) &=&  \Phi_{1\,I} (s,0, 0,M_W^2) \, \phi_1^{\mu\nu\alpha\beta}(p,q) + \Phi_{4\,I} (s,0, 0,M_W^2) \, \phi_4^{\mu\nu\alpha\beta}(p,q) \,.
\eea
with $s=k^2=(p+q)^2$, $\phi_i^{\mu\nu\alpha\beta}(p,q)$ given in Eq. (\ref{phitensors}) and 
\beq 
\phi_4^{\mu\nu\alpha\beta}(p,q) = (s \, \eta^{\mu\nu} - k^{\mu}k^{\nu}) \, \eta^{\alpha\beta},
\eeq
while the $\Delta$ term reads as
\bea
\Delta^{\mu\nu\alpha\beta}(p,q) &=& \Delta^{\mu\nu\alpha\beta}_I (p,q) \nn \\
&=&  \Psi_{1\, I} (s,0, 0,m_f^2,M_W^2,M_H^2) \, \phi_1^{\mu\nu\alpha\beta}(p,q) + \Psi_{4 \, I} (s,0, 0,M_W^2)  \, \phi_4^{\mu\nu\alpha\beta}(p,q)\, .
\label{DAA}
\eea
This is obtained by combining the tree level vertex for EMT/Higgs mixing, coming from the improved EMT, and the Standard Model $H\gamma\gamma$ correlator at one-loop. For convenience, we have included in Appendix \ref{apptvv} the explicit expression of these form factors, from which we extract the corresponding spectral densities and sum rules.\\
The spectral densities of the fermion contributions, related to $\Sigma_F$  have structure similar to those computed above in Eq. (\ref{rhoq}), as one can easily deduce from the explicit expression of the form factor given in Eq.(\ref{ap1}), with $\rho_{\Phi_{1F}} \sim \rho_{1 q}(s)$ and $\rho_{\Phi_{2F}} \sim \rho_{2 q} (s)$. Therefore we have two sum rules and two spectral flows also in this case, 
following the pattern discussed before for the spectral densities in Eq. (\ref{rhoq}).\\
A similar analysis on the two form factors $\Phi_B$ in the gauge boson sector gives 
\beq
\rho_{\phi_{1 B}}(s)=\frac{ 2 M_W^2}{s^3} (2 M_W^2 - s) \alpha \log\left(\frac{1+ \sqrt{\tau(s,M_W^2}) }{1-\sqrt{\tau(s,M_W^2} )}\right)\theta(s- 4 M_W^2)
\eeq
while $\rho_{\phi_{2 B}}$ has the same functional form of $\rho_{\phi_{2 F}}$, modulo an overall factor, with $m$, the fermion mass, replaced by the $W$ mass $M_W$. Notice that both $\rho_{\phi_{1 B}}$ and $\rho_{\phi_{2 B}}$, as well as $\rho_{\phi_{1 F}}$ and $\rho_{\phi_{2 F}}$ are deprived of resonant contributions, being the diagrams massive. \\
Coming to the form factors in $\Sigma_I$, whose explicit expressions are given in Eq.(\ref{imp}), one immediately realizes that the spectral density of $\Phi_{1I}$ 
shares the same functional form of $\rho_\chi$, extracted from Eq. (\ref{spectralrho}), and there is clearly a sum rule associated to it. Also in this case, this result is accompanied by the $1/k^2$ behaviour of the corresponding form factor, due to the anomaly. \\
Finally, for the case of $\psi_{1I}$, one can also show that the spectral density finds support only above the two particle cuts. The cuts are linked to $2 m$ and $2 M_W$. In this case there is no sum rule and the contribution is not affected by an anomaly pole, as expected, being the virtual loop connected with the $H\gamma\gamma$ vertex (see Fig. \ref{HVVImpr}). The explicit expression of this density is given in Appendix \ref{apptvv}.

\subsection{The non-transverse $\langle AVV \rangle$ correlator } 
Before closing the analysis of the spectral densitites for non supersymmetric theories, we pause for a few comments 
on the structure of the $\langle AVV \rangle$ diagram. This correlator, as we are going to show, is affected by a single flow even if we do not impose the transversality condition on the two photons. As a clarification of this point we consider once more the anomaly vertex as parameterized in 
Eq. (\ref{anom1}), and consider the second form factor $A_{4+6}\equiv A_4 + A_6$, which contributes to the anomaly loop for non transverse (but on-shell) photons. The expression of $A_6$, the anomalous form factor, has been given in 
Eq. (\ref{a6}), while $A_4$ is given by 
\beq
A_4(k^2,m^2)=-\frac{1}{2 \pi^2 k^2}\left[ 2 - \sqrt{\tau(k^2,m^2)}\log \frac{\sqrt{\tau(k^2,m^2)}+1}{\sqrt{\tau(k^2,m^2)}-1} \right], \qquad k^2<0
\label{a4}
\eeq
and $A_{4+6}$ takes the form 
\beq
A_{4+6}(k^2,m^2)=\frac{1}{2\pi^2 k^2}\left[ -1 + \sqrt{\tau(k^2,m^2)} \log\frac{\sqrt{\tau(k^2,m^2)} + 1}{\sqrt{\tau(k^2,m^2)} - 1}  + \frac{m^2}{k^2} \log^2 \frac{\sqrt{\tau(k^2,m^2)} + 1}{\sqrt{\tau(k^2,m^2)} - 1} \right].
\eeq
Its discontinuity is given by 
\beq
\textrm{Disc}\,A_{4+6}(k^2,m^2)= - 2 i \pi\left[ \frac{\sqrt{\tau(k^2,m^2)}}{k^2} + \frac{2 m^2}{(k^2)^2}\log \frac{\sqrt{\tau(k^2,m^2)}+1}{\sqrt{\tau(k^2,m^2)}-1} \right] \theta(k^2 - 4m^2).
\eeq
Notice that in this case there is no sum rule satisfied by this spectral density, being non-integrable along the cut.  
Coming to the spectral density for the anomaly coefficient $A_6$, this is proportional to the density of 
$\chi(s,m^2)$ given in Eq. (\ref{spectralrho})
and shares the same behaviour found for $\rho_{\chi}(s,m^2)$, as expected. This analysis shows that in the $\langle AVV \rangle$ case one encounters a single sum rule and a single massive flow which degenerates into a $\delta(s)$ behaviour, as in the supersymmetric case. This condition remains valid also for non-transverse vector currents. It is then clear that the crucial difference between the non supersymmetric case and the supersymmetric one manifests in the $\langle TVV \rangle$ diagram, due to the extra sum rule discussed above.

\subsection{Cancellations in the supersymmetric case}
In order to clarify even more how the cancellation of the extra poles occurs in the supersymmetric $\langle TVV \rangle$, we consider the non-anomalous form factor $f_2$ in a general theory (given in Eqs. (\ref{FFfermions},\ref{FFscalars},\ref{FFgauge})), with $N_f$ Weyl fermions, $N_s$ complex scalars and $N_A$ gauge fields. We work, for simplicity, in the massless limit. In this case the non anomalous form factor $f_2$, which is affected by pole terms, after combining scalar, fermions and gauge contributions can be written in the form
\bea
\label{F2total}
f_2(k^2) &=& \frac{N_f}{2} f_2^{(f)}(k^2) + N_s \, f_2^{(s)}(k^2) + N_A \,  f_2^{(A)}(k^2) \nn \\
&=& \frac{g^2}{144 \pi^2 \, k^2} \left[ -  \frac{N_f}{2} T(R_f) + N_s \, \frac{T(R_s)}{2} + N_A \, \frac{T(A)}{2} \right] \,, 
\eea
where the fermions give a negative contribution with respect to scalar and gauge fields. 
If we turn to a $\mathcal N=1$ Yang-Mills gauge theory, which is the theory that we are addressing, we need to consider in the anomaly diagrams the virtual exchanges both of a chiral and of a vector supermultiplet. In the first case the multiplet is built out of one Weyl fermion and one complex scalar, therefore in Eq.(\ref{F2total}) we have $N_f = 1, N_s = 1, N_A = 0$ with $T(R_f) = T(R_s)$. With this matter content, the form factor is set to vanish. \\
For a vector multiplet, on the othe other end, we have one vector field and one Weyl fermion, all belonging to the adjoint representation and then we obtain $N_f=1, N_s=0, N_A=1$ with $T(R_f) = T(A)$. Even in this case all the contributions in the $f_2$ form factor sum up to zero. It is then clear that the cancellation of the extra poles in the $\langle TVV \rangle$ is a specific tract of supersymmetric Yang Mills theories, due to their matter content, not shared by an ordinary gauge theory. A corollary of this is that in a supersymmetric theory we have just one spectral flow driven by the deformation parameter $m$, accompanied by one sum rule for the entire deformation.

\section{The anomaly effective action and the pole cancellations for $\mathcal{N}=4$ }
\begin{figure}[t]
\centering
\subfigure{\includegraphics[scale=0.6]{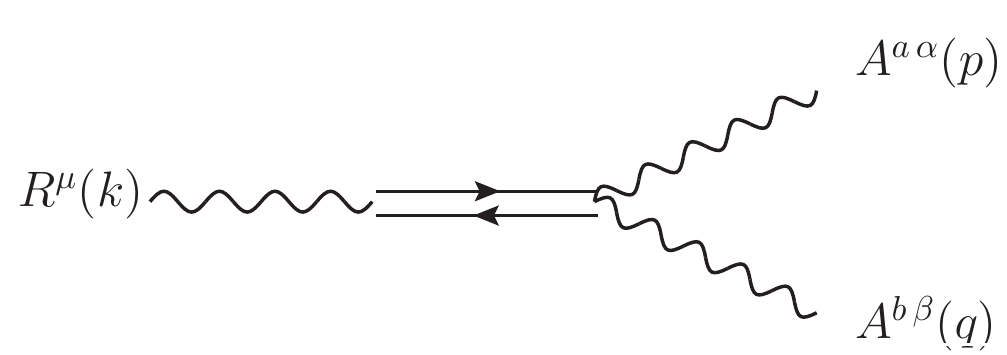}} \hspace{.5cm}
\subfigure{\includegraphics[scale=0.6]{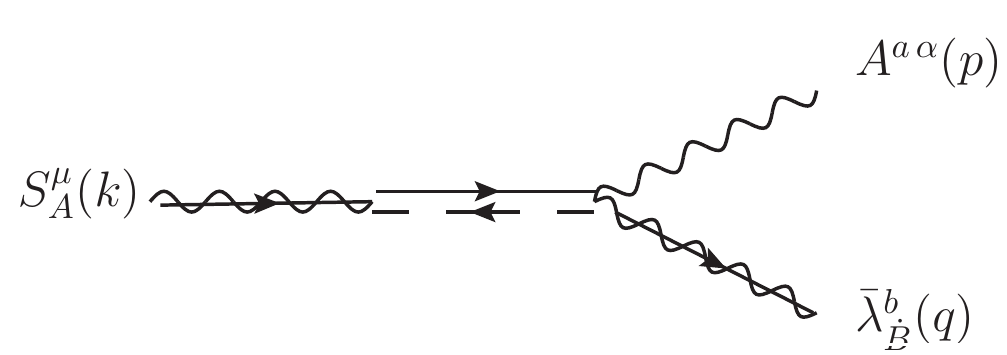}} 
\subfigure{\includegraphics[scale=0.6]{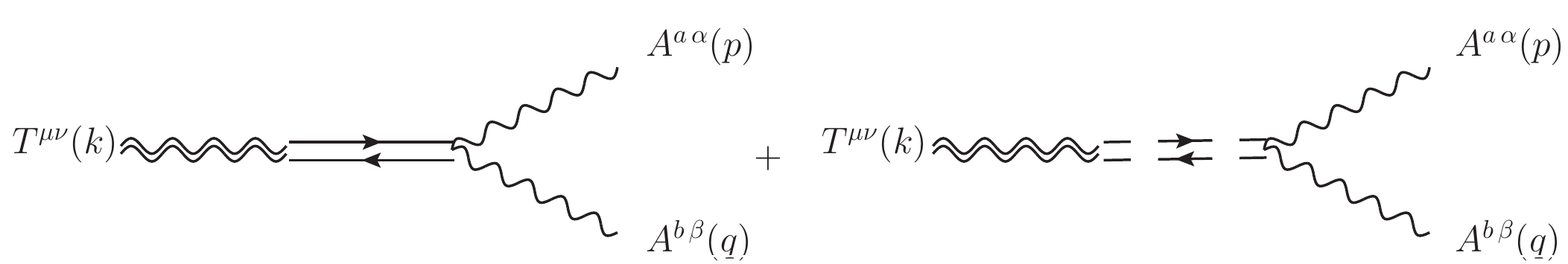}} 
\caption{The collinear diagrams corresponding to the exchange of a composite axion (top right), a dilatino (top left) and the two sectors of an intermediate dilaton (bottom). Dashed lines denote intermediate scalars.}
\label{RST}
\end{figure}

The appearance of poles in an effective action is associated, in general, either with the intermediate exchange of particles related to the fundamental fields in the defining Lagrangian or with the exchange of intermediate bound states. For convenience, this point has been briefly reviewed by us, in the case of three-point correlators, in Appendix \ref{polology}, to which we refer for further detais. Here, instead, we just present the expression of the quantum effective action obtained from the three-point correlation functions that we have previously discussed.\\
 We consider the massless case for the chiral supermultiplet and on-shell external gauge bosons and gauginos. 
The anomalous part is given by the three terms 
\beq
S_{\textrm{anom}}= S_{\textrm{axion}} +  S_{\textrm{dilatino}} + S_{\textrm{dilaton}}
\eeq
which are given, respectively, by
\bea
S_{\textrm{axion}}&=& - \frac{g^2}{4 \pi^2} \left(  T(A) - \frac{T(R)}{3} \right)  \int d^4 z \, d^4 x \,  \partial^\mu B_\mu(z) \, \frac{1}{\Box_{zx}} \, \frac{1}{4} F_{\alpha\beta}(x)\tilde F^{\alpha\beta}(x) \\
S_{\textrm{dilatino}} &=&  \frac{g^2}{2 \pi^2} \left(T(A) - \frac{T(R)}{3} \right)  \int d^4 z \, d^4 x \bigg[  \partial_\nu \Psi_\mu(z) \sigma^{\mu\nu} \sigma^\rho \frac{\stackrel{\leftarrow}{\partial_\rho}}{\Box_{zx}} \,  \bar \sigma^{\alpha\beta} \bar \lambda(x) \frac{1}{2} F_{\alpha\beta}(x) + h.c. \bigg]  \\
S_{\textrm{dilaton}}&=&- \frac{g^2}{8 \pi^2} \left(T(A)  - \frac{T(R)}{3} \right) \int d^4 z \, d^4 x \, \left(\Box h(z) - \partial^\mu \partial^\nu h_{\mu\nu}(z) \right) \,  \frac{1}{\Box_{zx}} \,\frac{1}{4} F_{\alpha\beta}(x) F^{\alpha\beta}(x). 
\eea
We show in Figs.\,\ref{RST} the three types of intermediate states which interpolate between the Ferrara-Zumino hypercurrent and the gauge $(A)$ and the gaugino ($\lambda$) of the final state. The axion is identified by the collinear exchange of a bound fermion/antifermion pair in a pseudoscalar state, generated in the $\langle RVV \rangle$ correlator. In the case of the $\langle SVF \rangle$ correlator, the intermediate state is a collinear scalar/fermion pair, interpreted as a dilatino. In the $\langle TVV \rangle$ case, the collinear exchange is a linear combination of a fermion/antifermion and scalar/scalar pairs.  

The non-anomalous contribution is associated with the extra term $S_0$ which is given by
\bea
S_0 &=&  \frac{g^2}{16 \pi^2}  \int d^4 z \, d^4 x \,  h_{\mu\nu}(z) \left( T(R) \, \tilde \Phi_2(z-x) + T(A) \, \tilde V(z-x) \right) T^{\mu\nu}_{gauge}(x) \nn \\
&+&   \frac{g^2}{64 \pi^2}  \int d^4 z \, d^4 x \bigg[ i \, \Psi_\mu(z) \left( T(R) \, \tilde \Phi_2(z-x) + T(A) \, \tilde V(z-x) \right) S^{\mu}_{gauge}(x)  + h.c. \bigg] \,,
\eea
where $\tilde \Phi_2(z-x)$ and $\tilde V(z-x)$ are the Fourier transforms of $\Phi_2(k^2,0)$ and $V(k^2)$ respectively. 
Their contributions in position space correspond to nonlocal logarithmic terms.

The relation between anomaly poles, spectral density flows and sum rules appear to be a significant feature of supersymmetric theories affected by anomalies. It is then clear that supersymmetric anomaly-free theories should be free of such contributions in the anomaly effective action. In this respect, it natural to turn to the $\mathcal N = 4$ theory, which is free of anomalies, in order to verify and validate this reasoning. Indeed the $\beta$ function of the gauge coupling constant in this theory has been shown to vanish up to three loops \cite{Grisaru:1980nk,Caswell:1980yi,Avdeev:1981ew}, and there are several arguments about its vanishing to all the perturbative orders. As a consequence, the anomaly coefficient in the trace of the energy-momentum tensor, being proportional to the $\beta$ function, must vanish identically and the same occurs for the other anomalous component, related to the $R$ and to the $S$ currents in the Ferrara-Zumino supermultiplet. \\
We recall that in the $\mathcal{N}=4$ theory the spectrum contains a gauge field $A^\mu$, four complex fermions $\lambda^i$ ($i=1,2,3,4$) and six real scalars $\phi_{ij} = - \phi_{ji}$ $(i,j=1,2,3,4)$. All fields are in the adjoint representation of the gauge group. 

From the point of view of the $\mathcal N=1$ SYM, this theory can be interpreted as describing a vector and three massless chiral supermultiplets, all in the adjoint representation. Therefore the $\langle TVV \rangle$ correlator in $\mathcal N=4$ can be easily computed from the general expressions in Eqs (\ref{TChiralOSMassless}) and (\ref{TVectorOS}) which give
\bea
\label{residualV}
\Gamma^{\mu\nu\alpha\beta}_{(T)}(p,q) = \frac{g^2 \, T(A)}{16 \pi^2} \left[ V(k^2) + 3 \Phi_2(k^2,0)\right] t_{2S}^{\mu\nu\alpha\beta}(p,q) 
= - \frac{g^2 \, T(A)}{8 \pi^2} k^2 \, \mathcal C_0(k^2,0) \, t_{2S}^{\mu\nu\alpha\beta}(p,q) \,.
\label{gamma}
\eea
One can immediately observe from the expression above the vanishing of the anomalous form factor proportional to the tracefull tensor structure 
$t_{1S}^{\mu\nu\alpha\beta}$. The partial contributions to the same form factor, which can be computed using Eqs. (\ref{TChiralOSMassless}) and (\ref{TVectorOS}) for the various components, are all affected by pole terms, but they add up to give a form factor whose residue at the pole is proportional to the $\beta$ function of the $\mathcal N=4$ theory. It is then clear that the vanishing of the conformal anomaly, via a vanishing $\beta$ function, is equivalent to the cancellation of the anomaly pole for the entire multiplet. \\
Notice also that the only surviving contribution in Eq. (\ref{gamma}), proportional to the traceless tensor structure 
$t_{2S}^{\mu\nu\alpha\beta}$, is finite. This is due to the various cancellations between the UV singular terms from $V(k^2)$ and $\Phi_2(k^2,0)$ which give a finite correlator without the necessity of any regularization. 

We recall that the cancellation of infinities and the renormalization procedure, as we have already seen in the $\mathcal N=1$ case, involves only the form factor of tensor $t_{2S}^{\mu\nu\alpha\beta}$, which gets renormalized with a counterterm 
 proportional to that of the two-point function $\langle AA \rangle$, and hence to the gauge coupling. 
 For this reason the finiteness of the second form factor and then of the entire $\langle TVV \rangle$ in $\mathcal N=4$ is directly connected to the vanishing of the anomalous term, because its non-renormalization naturally requires that the $\beta$ function has to vanish.

\section{Conclusions and Perspectives}
Our analysis and results show the consistency of a conjecture about the perturbative structure of the anomalies in supersymmetric theories, formulated by us in previous works \cite{Coriano:2011zk,Coriano:2011ti}. We have presented  additional evidence that anomaly poles are the signature of the anomalies in the perturbative anomaly action of these theories, extending former studies \cite{Giannotti:2008cv,Armillis:2010qk,Armillis:2009im}. For global anomalies it is expected that the massless states identified by the pole contributions can be promoted to new composite degrees of freedom by some non perturbative dynamics, as for the chiral anomaly and the pion.\\
In the QCD case \cite{Armillis:2010qk}, for instance, the breaking of classical scale invariance - in this perturbative picture - should manifest in the emergence of a dilaton, if gluons were asymptotic states. We have noticed, though, that the $\langle TVV \rangle$ vertex in QCD, as we have shown, has one extra pole and and one extra flow related to a non anomalous form factor which is both IR and UV safe, which should be interpreted as an extra interpolating state.  A similar perturbative pattern emerges in the Standard Model \cite{Coriano:2011ti, Coriano:2012nm}, as clear from the analysis of the conformal anomaly in the electroweak sector \cite{Coriano:2011ti}. \\
However, by turning to supersymmetry, we have shown that here the connection between anomalies, poles and sum rules for anomaly vertices are one to one.
The $1/ k^2$ feature of the anomaly form factors has been investigated in connection with the scaling properties of their spectral densities and with the finite (non zero) sum rule which it satisfies, in agreement with a previous analysis by Giannotti and Mottola  \cite{Giannotti:2008cv}. We have seen that the anomalous behaviour emerges from the $s\sim 0$ region of the spectral density of a given form factor and covers, therefore, 
the entire light-cone surface. The resonant behaviour at $s=0$ is present, as we have shown, also at very high momentum.\\
In supersymmetry we have focused our attention on the perturbative correlators which are responsible for the generation of the superconformal anomaly, and shown that the Ferrara-Zumino multiplet, as well as the Konishi currents, allow to identify some composite 
states in the effective action, interpolating between the currents and the on-shell final states. They correspond to a dilaton, a dilatino and an axion, plus 
a number of pseudoscalar states, one for each fermion flavour.
The description of these effective degrees of freedom not as anomaly poles but as asymptotic states of the S-matrix remains, obviously, an open issue, 
which goes beyond the simple perturbative picture discussed here, as demonstrated by the complex pattern of chiral dynamics in QCD. In particular would be interesting to compare this result with the anomaly action obtained in \cite{Schwimmer:2010za} in the superconformal case, which is of Wess-Zumino type, which is local. We expect both actions to share the same physical content. 

Following this pattern, it is then natural to ask if global anomalies are always connected to the generation of effective degrees of freedom, and hence to compositeness, as indicated by the poles of the effective action. These results are valid for all the anomalies characterized by a single flow, in particular for all the chiral currents affected by global anomalies. From this perspective, also the Peccei-Quinn current should induce as an interpolating  state a composite axion rather an elementary one, being our argument generic to anomalous global currents. 

We stress once again, that all our results are limited to perturbation theory. Obviously, nonperturbative effects may change drastically this picture, as in the case of the $\eta'$ in QCD. In general, indeed, one expects the appearance of massless poles in the spontaneous breaking of global symmetries and not in those driven by radiative effects, as in the case of anomalies. For this reason mass corrections related to non perturbative effects should modify this picture by shifting the position of these poles which could become massive.   

There are also some drastic implications of our analysis, at least in the supersymmetric case, whenever the symmetries of the hypercurrent are gauged, 
which concern the way anomalies should cancel when a theory affected by a superconformal anomaly is coupled to gravity. We have seen that the anomaly is entirely given by the 
$\beta/k^2$ term, in terms of the $\beta$ function of the theory, and it appears obvious that the coupling to gravity has necessarily to provide an extra massless sector in order to remove such contribution. This could only take place if the gravitational sector can contribute by an equal and opposite amount to the pole residue, at the cost, otherwise, of being left with an inconsistency in the total theory. It is important to remark, as we have already pointed out, that the cancellation of the pole may not be an identical cancellation of the anomaly vertex.
We have in fact explicitly shown that in a $\mathcal{N}=4$ theory, for instance, by setting the 
$\beta$ function to zero, one indeed is canceling the pole contributions, and hence the anomaly, but not the entire anomaly vertex, as clear from Eq. (\ref{residualV}). This situation is new compared to the case of anomalous abelian symmetries, where anomaly cancellation by charge assignments on the massless matter spectrum forces the entire $\langle AVV \rangle/\langle AAA \rangle$ vertices to vanish.\\
It is important to stress, at this point, that there are subtle issues related to the definition of the anomaly supermultiplet in general theories, of which the Ferrara-Zumino choice is only one realization. For instance, in the presence of Fayet-Iliopoulos terms the multiplet is not gauge invariant and requires an appropriate redefinition. Similar issues appear in theories with a K\"ahler form that is not exact, as discussed in recent works \cite{Dienes:2009td}, \cite{Komargodski:2009pc}, \cite{Komargodski:2010rb}.These issues have particular relevance in the investigation of the coupling of these theories to supergravity. In \cite{Komargodski:2010rb}, for instance, it is shown that it is always possible to construct a new supermultiplet which generalizes the FZ-multiplet. 
However, being our analysis limited to a non-abelian gauge theory with a simple K\"ahler potential, the pathologies described above are not present. In our case the Ferrara-Zumino supermultiplet is a good operator of the theory. Of course, it would be interesting to extend our results to the perturbative analysis of the supercurrent introduced in \cite{Komargodski:2010rb}. These issues are deferred to future studies.
  
\centerline{\bf Acknowledgments} 
We thank Emil Mottola for discussions on the issues addressed in this work and on his work with Maurizio Giannotti. 
We thank Pietro Colangelo and Nikos Irges for discussions and for comments on the manuscript and Andreas Wipf for discussions and for hospitality at Jena University. 
Finally, C.C. thanks Prof. Dino Vaira of the University of Bologna for his support and the theory division at CERN for hospitality during the elaboration of this work.

\appendix

\section{Appendix. Scalar integrals}
\label{AppScalarIntegrals}
One-, two- and three- point functions are denoted respectively as $\mathcal A_0$, $\mathcal B_0$ and $\mathcal C_0$ with
\bea
\mathcal A_0(m^2) &=& \frac{1}{i \pi^2} \int d^n l \frac{1}{l^2 - m^2} \,, \nn \\
\mathcal B_0(p_1^2, m_0^2,m_1^2) &=& \frac{1}{i \pi^2} \int d^n l \frac{1}{(l^2 -m_0^2)((l+p_1)^2 -m_1^2)} \,, \nn \\ 
\mathcal C_0((p+q)^2, p^2, q^2,m_0^2,m_1^2,m_2^2) &=&  \frac{1}{i \pi^2} \int d^n l \frac{1}{(l^2 -m_0^2)((l-p)^2 -m_1^2)((l-p-q)^2 -m_2^2)} \,.
\eea
Moreover, for equal internal masses and for $p^2=q^2=0$ we have used the more compact notation
\bea
\mathcal B_0(p_1^2,m^2) \equiv \mathcal B_0(p_1^2,m^2,m^2)\,, \qquad \mathcal C_0((p+q)^2,m^2) \equiv \mathcal C_0((p+q)^2,0,0,m^2,m^2,m^2) \,.
\eea
In the spacelike region ($k^2 < 0$), using two regulators for the ultraviolet and infrared singularities ($n=4-\epsilon_{UV} = 4 + \epsilon_{IR}$), where $n$ denotes the spacetime dimensions, the relevant 2-point functions appearing in the computation are
\bea
\mathcal B_0(k^2, 0) &=& \frac{2}{\epsilon_{UV}} + 2 - \log\frac{-k^2}{\mu^2} \,, \\
\mathcal B_0(k^2, m^2) &=& \frac{2}{\epsilon_{UV}} + 2 - \log\frac{m^2}{\mu^2} + \sqrt{\tau(k^2,m^2)} \log \frac{\sqrt{\tau(k^2,m^2)} -1}{\sqrt{\tau(k^2,m^2)} +1} \,, 
\eea
with $\tau(k^2,m^2) = 1-4m^2/k^2$,
while for $k^2$ null we obtain
\bea
\mathcal B_0(0, 0) &=& \frac{2}{\epsilon_{UV}} + \frac{2}{\epsilon_{IR}} \,, \\
\mathcal B_0(0, m^2) &=& \frac{2}{\epsilon_{UV}} - \log\frac{m^2}{\mu^2}. \, 
\eea
In the QCD computations we have also used the following finite two-point scalar integrals
\bea
\mathcal D(k^2,m^2) = \mathcal B_0(k^2,m^2) - \mathcal B_0(0,m^2) \,,
\eea
and we have renormalized all the divergent $\mathcal B_0$ functions in the $\overline{MS}$ scheme in which the $1/\eps_{UV}$ divergences have been subtracted. \\
The massless scalar 3-point function, for $k^2<0$,  is given by
\bea
\mathcal C_0(k^2,0) &=& \frac{1}{k^2} \left[ \frac{4}{\epsilon_{IR}} + \frac{2}{\epsilon_{IR}} \log \frac{-k^2}{\mu^2} +\frac{1}{2} \log^2 \frac{-k^2}{\mu^2} - \frac{\pi^2}{12} \right]\,,
\eea
while the massive $\mathcal C_0(k^2,m^2)$ is given in Eq. (\ref{ti}).

\section{Appendix. Electroweak form factors for the $\langle TVV \rangle$} 
\label{apptvv}

\begin{figure}[t]
\centering
\subfigure[]{\includegraphics[scale=0.8]{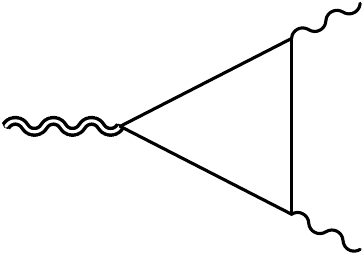}} \hspace{.5cm}
\subfigure[]{\includegraphics[scale=0.8]{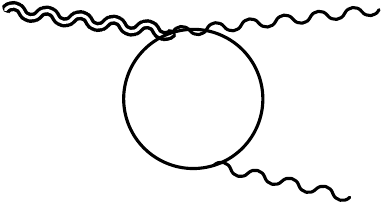}} \hspace{.5cm}
\subfigure[]{\includegraphics[scale=0.8]{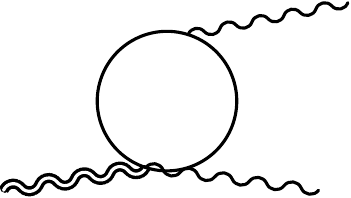}} \hspace{.5cm}
\subfigure[]{\includegraphics[scale=0.8]{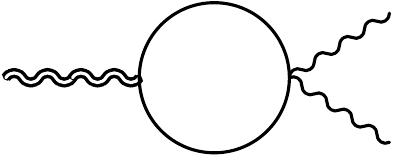}} \hspace{.5cm}
\subfigure[]{\includegraphics[scale=0.8]{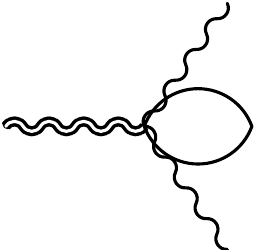}}
\caption{Different topologies for the $\langle TVV \rangle$ vertex. Internal lines can be fermions, $W$ bosons, goldstones and ghosts. \label{ewtops}}
\end{figure}

In the fermion sector the form factors are given by
\bea
\Phi_{1\, F} (s,\,0,\,0,\,m_f^2) &=& \, \frac{\alpha}{3 \pi \, s} \, Q_f^2 \bigg\{
- \frac{2}{3} + \frac{4\,m_f^2}{s} - 2\,m_f^2 \, \mathcal C_0 (s, 0, 0, m_f^2, m_f^2, m_f^2)
\bigg[1 - \frac{4 m_f^2}{s}\bigg] \bigg\}\, ,  \nn \\
\Phi_{2\, F} (s,\,0,\,0,\,m_f^2)  &=& \, \frac{\alpha}{3 \pi \, s} \, Q_f^2 \bigg\{
-\frac{1}{12} - \frac{m_f^2}{s} - \frac{3\,m_f^2}{s} \mathcal D_0 (s, 0, 0, m_f^2, m_f^2)  \nn \\
&-&  m_f^2 \mathcal C_0(s, 0, 0, m_f^2, m_f^2, m_f^2 )\, \left[ 1 + \frac{2\,m_f^2}{s}\right] \bigg\}\, , \\
\Phi_{3\,F} (s,\,0,\,0,\,m_f^2) &=&  \, \frac{\alpha}{3 \pi \, s} \, Q_f^2 \bigg\{
\frac{11\,s}{12}+ 3 m_f^2 +  \mathcal D_0 (s, 0, 0, m_f^2, m_f^2)\left[5 m_f^2 + s \right]\nn\\
&+&  s \, \mathcal B_0 (0, m_f^2, m_f^2) + 3 \, m_f^2 \, \mathcal C_0(s, 0, 0, m_f^2 , m_f^2, m_f^2) \left[s + 2m_f^2 \right] \bigg\}\, .
\label{ap1}
\eea
The other gauge-invariant sector of the $\langle TVV \rangle$ vertex is the one mediated by the exchange of bosons and ghosts in the loop. In this
sector the form factors are given by
\bea
\Phi_{1\, B} (s,\,0,\,0,\,M_W^2) &=&\, \frac{\alpha}{\pi \, s} \bigg\{
\frac{5}{6} - \frac{2\,M_W^2}{s} + 2\,M_W^2 \, \mathcal C_0 (s, 0, 0, M_W^2, M_W^2, M_W^2)
\bigg[1 - \frac{2 M_W^2}{s}\bigg] \bigg\},  \nn \\
\Phi_{2\, B} (s,\,0,\,0,\,M_W^2)  &=&\, \frac{\alpha}{\pi \, s} \bigg\{
\frac{1}{24} + \frac{M_W^2}{2\,s} + \frac{3\,M_W^2}{2\,s} \mathcal D_0 (s, 0, 0, M_W^2, M_W^2)  \nn \\
&+&  \frac{M_W^2}{2} \mathcal C_0(s, 0, 0, M_W^2, M_W^2, M_W^2 )\, \left[ 1 + \frac{2\,M_W^2}{s}\right] \bigg\}\, , \\
\Phi_{3\,B} (s,\,0,\,0,\,M_W^2) &=&\, \frac{\alpha}{\pi \, s} \bigg\{
-\frac{15\,s}{8}-\frac{3\,M_W^2}{2} - \frac{1}{2}\, \mathcal D_0 (s, 0, 0, M_W^2, M_W^2)\left[5 M_W^2+7\,s \right]\nn\\
&-& \frac{3}{4}s\, \mathcal B_0 (0, M_W^2, M_W^2) - \mathcal C_0(s, 0, 0, M_W^2 , M_W^2, M_W^2) \left[s^2 + 4 M_W^2\,s + 3\,M_W^4\right]
\bigg\}.
\eea
The contributions coming from the term of improvement are characterized bythe form factors
\bea
\Phi_{1\, I} (s,\,0,\,0,\,M_W^2) &=&\frac{\alpha}{3 \pi \, s} \bigg\{ 1 + 2 M_W^2 \,C_0 (s, 0, 0, M_W^2, M_W^2, M_W^2)\bigg\} \,,\\
\Phi_{4\, I} (s,\,0,\,0,\,M_W^2) &=&  -\frac{\alpha}{6 \pi }  M_W^2 \,C_0 (s, 0, 0, M_W^2, M_W^2, M_W^2), \,
\label{imp}
\eea
\bea
\Psi_{1\, I} (s,\,0,\,0,\,m_f^2,M_W^2,M_H^2) &=&  \frac{\alpha}{3 \pi \, s (s - M_H^2)} \bigg\{ 2 m_f^2 \, Q_f^2 \bigg[ 2 + (4 m_f^2 -s) C_0 (s, 0, 0, m_f^2, m_f^2, m_f^2) \bigg] \nn \\
&& \hspace{-2cm} + M_H^2 + 6 M_W^2 + 2 M_W^2 (M_H^2 + 6 M_W^2 - 4 s) C_0(s,0,0, M_W^2,M_W^2,M_W^2) \bigg\} \,, \\
\Psi_{4\, I} (s,\,0,\,0,M_W^2) &=& - \Phi_{4\, I} (s,\,0,\,0,\,M_W^2)  \, .
\eea
Finally, the spectral density associated to the form factor $\psi_{1I}$ for a light fermion ($m$) running in the loop, takes the form 
\bea
\rho_{\psi_{1I}}&=& A_1(s,m^2, M_H^2)  \log\left(\frac{1 +\sqrt{\tau(s,m^2)}}{1 -\sqrt{\tau(s,m^2)}}\right)\theta(s - 4 m^2)  \nn \\
&+& 
A_2(s,m^2, M_H^2) \log\left(\frac{1 +\sqrt{\tau(s,M_W^2)}}{1 -\sqrt{\tau(s,M_W^2)}}\right)\theta(s - 4 M_W^2) +  A_3(s,m^2,M_H^2, M_W^2) \delta(s- M_H^2)
\eea
with 
\bea
A_1(s, m^2,M_H^2)&=&\frac{2}{3} \frac{m^2 Q_f^2}{M_H^2 s^2} \alpha \left(4 m^2 + \frac{M_H^2 - 4 m^2}{s- M_H^2}\right)\nn\\
A_2(s, M_H^2,M_W^2)&=&\frac{2}{3 M_H^2 s^2}\alpha M_W^2 \left( M_H^2 + 6 M_W^2 + 3
s \frac{M_H^2 - 2 M_W^2}{s - M_H^2} \right) \nn\\
A_3(s, m^2, M_H^2,M_W^2)&=&\frac{\alpha}{3 M_h^4}\left( - M_H^2(M_H^2 + 6 M_W^2) + 
m^2 (4 m^2 \pi^2 - M_H^2(4 + \pi^2)) Q_f^2 \right.\nn\\
&&\left. + 6 M_h^2 (M_H^2 - 2 M_W^2) M_W^2 C_0(M_H^2,M_W^2)\right) \,.
\eea

\section{Appendix. List of spectral discontinuities} 
We summarize here, for convenience, a list of the discontinuities of functions needed in the computation of the spectral densities of the correlators.
\bea
\textrm{Disc}\frac{\mathcal C_0(s,m^2)}{s^2}&=& -\frac{2\pi i}{s^3}\log\left(\frac{1 +\sqrt{\tau(s,m^2)}}{1 -\sqrt{\tau(s,m^2)}}\right)\theta(s-4 m^2) +\frac{i\pi}{12 m^4}\delta(s) -\frac{ i \pi}{m^2}\delta'(s) \,, \nn\\ 
\textrm{Disc}\frac{\mathcal C_0(s,m^2)}{s}&=& -\frac{2 i\pi}{s^2}\log\left(\frac{1 +\sqrt{\tau(s,m^2)}}{1 -\sqrt{\tau(s,m^2)}}\right)\theta(s- 4 m^2) +\frac{i\pi}{m^2}\delta(s) \,, \nn \\
\textrm{Disc}\, \mathcal D(s,m^2)&=& 2 i \pi \sqrt{\tau(s,m^2)} \,, \nn\\
\textrm{Disc}\frac{\mathcal D(s,m^2)}{s}&=& 2 i \pi \frac{\sqrt{\tau(s,m^2)}}{s}\theta(s- 4 m^2) \,, \nn\\
\textrm{Disc}\frac{\mathcal D(s,m^2)}{s^2}&=&2 i \pi \frac{\sqrt{\tau(s,m^2)}}{s^2}\theta(s- 4 m^2)-\frac{i \pi}{ 3 m^2}\delta(s) \,.
\eea
For the computation of the spectral densities of $\psi_{1I}$ we need also
\beq
\textrm{Disc}\frac{ \mathcal C_0(s,m^2)}{s - M_H^2}=- 2 i \pi \textrm{Re} \, \mathcal C_0(M_H^2,m^2)\delta(s- M_H^2) - 2 i \frac{\pi}{s(s- M_H^2)} \log\left(\frac{1 +\sqrt{\tau(s,m^2)}}{1 -\sqrt{\tau(s,m^2)}}\right)\theta(s- 4 m^2).
\eeq
where $m$ can be either the $W$ boson or the fermion mass.

\section{Appendix. Polology}
\label{polology}
\begin{figure}[t]
\begin{center}
\includegraphics[scale=0.5]{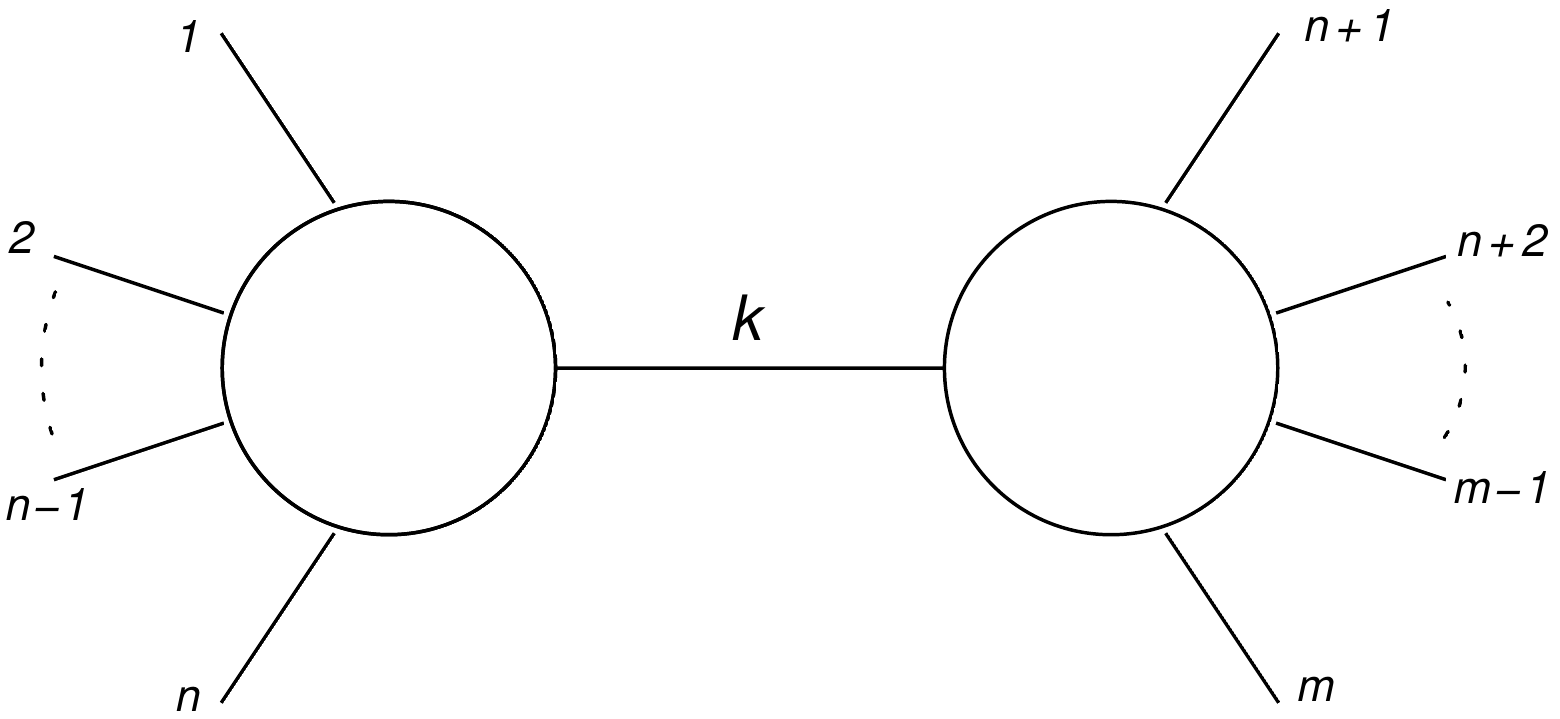}
\caption{\small A correlation function exhibits a pole exchange with momentum $k$ corresponding to an elementary particle which appears in the Lagrangian.}
\label{Polology.Elementary}
\end{center}
\begin{center}
\includegraphics[scale=0.5]{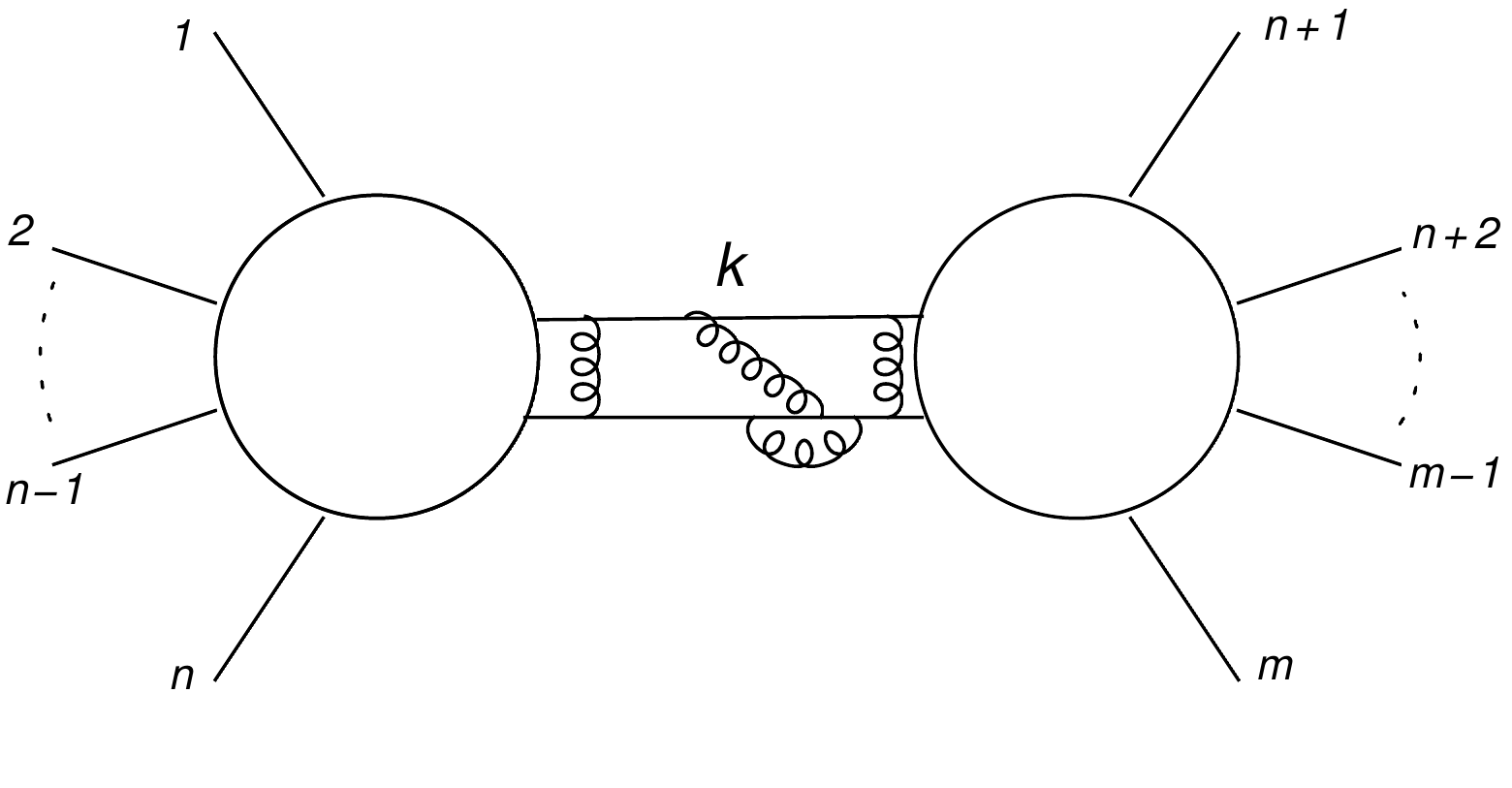}
\caption{\small A bound state interpolates between the two subamplitudes in a given correlation function. In this case the pole corresponds to a composite, a bound state of two elementary particles represented by straight lines, which interact together by the exchange of other elementary states (curly lines).}
\label{Polology.Composite}
\end{center}
\end{figure}
A pole in a correlation function may correspond either to the exchange of a fundamental particle in the defining Lagrangian, as shown in Fig. (\ref{Polology.Elementary}), or to the exchange of a composite particle, as in Fig.  (\ref{Polology.Composite}). We summarize the proof of the relation between the existence of poles in the S-matrix and the nature of the intermediate exchange for the specific case of the 3-point functions that we investigate. We consider a generic correlator $G$ as a function of the momenta of the external lines. In coordinate space it is given by
\bea
G(x_1, \ldots, x_n) = \langle 0 | T \left\{ A_1(x_1) \ldots A_n (x_n) \right\} | 0 \rangle 
\eea 
with the operatros $A$ denoting either fields appearing in the Lagrangian or even composite local operators.  We specialize to a simpler case with just three operators, say $\mathcal O(z), A_1(x_1)$ and $A_2(x_2)$. This is the situation encountered in our studies on chiral and conformal anomalies. Then we move to momentum space and consider the correlator
\bea
G(k, p_1, p_2) = \int d^4 z d^4 x_1 d^4 x_2 \, e^{- i k z - i p_1 x_1 - i p_2 x_2} \langle 0 | T \left\{ \mathcal O(z) A_1(x_1)  A_2 (x_2) \right\} | 0 \rangle
\eea
as a function of the virtuality of $\mathcal  O$, namely $k^2 = (- p_1 - p_2)^2$. Notice that the virtualities of the external momenta $p_1^2$ and $p_2^2$ are not fixed by any on-shellness condition and can be arbitrary. We isolate the operator $\mathcal O$ from the T product and retain only the term in which $\mathcal O$ appears to the far left
\bea
G(k, p_1, p_2) &=& \int d^4 z d^4 x_1 d^4 x_2 \, e^{- i k z - i p_1 x_1 - i p_2 x_2} \nn \\
&\times& \bigg\{ \theta(z^0 - \max\{x_1^0, x_2^0\}) \, \langle 0 | \mathcal O(z) T \left\{ A_1(x_1)  A_2 (x_2) \right\} | 0 \rangle + \ldots \bigg\} \,,
\eea
where the ellipsis stand for the other time ordering products which we have ignored. They do not contribute with any pole structure to the correlator.  
Now we insert a complete set of intermediate states between the operator $\mathcal O$ and the other ones, isolating only single particle states with a specific mass $m$. We discard the other single particle states with different masses (they will contribute with poles but at other kinematical positions) and multi particle states (which appear as branch cuts). We obtain
\bea
G(k, p_1, p_2) &=& \sum_\sigma \int d^4 z d^4 x_1 d^4 x_2 d^3 \vec{p} \, e^{- i k z - i p_1 x_1 - i p_2 x_2} \nn \\
&\times& \bigg\{ \theta(z^0 - \max\{x_1^0, x_2^0\}) \langle 0 | \mathcal O(z) | \vec{p}, \sigma \rangle \langle \vec{p}, \sigma | T \left\{ A_1(x_1)  A_2 (x_2) \right\} | 0 \rangle + \ldots \bigg\}
\eea
where $|\vec{p},\sigma \rangle$ is a single particle state with mass $m$ ($p^2  = m^2$) and with quantum numbers collectively identified by $\sigma$. We extract the $z$ and $x_1$ dependences from the matrix elements appearing in the previous equation, and introduce the new integration variable $y =  x_1 -  x_2$ in place of $x_2$. Finally we insert the integral representation of the step function $\theta(t)$ given by
\bea
\theta(t) = \frac{i}{2 \pi} \int_{- \infty}^{+ \infty} d \omega \frac{e^{- i \omega \, t}}{\omega + i \epsilon}
\eea
where $\epsilon$ is an infinitesimal and positive constant. We have 
\bea
G(k, p_1, p_2) &=& \frac{i}{2 \pi} \sum_\sigma \int d^4 z \, d^4 x_1 \, d^4 y \, d^3 \vec{p} \, \frac{d \omega}{\omega + i \epsilon}  \, e^{- i k z - i (p_1 + p_2) x_1 - i p_2 y } \nn \\
&& \hspace{-2cm} \times \, \, e^{ - i \omega (z^0 - x_1^0 - \max\{0, y^0\})} e^{i p z - i p x_1}  \langle 0 | \mathcal O(0) | \vec{p}, \sigma \rangle \langle \vec{p}, \sigma | T \left\{ A_1(0)  A_2 (y) \right\} | 0 \rangle + \ldots  \,,
\eea
where the integration over $z$ and $x_1$ is straightforward and gives only delta functions
\bea
&& \hspace{-0.7cm} G(k, p_1, p_2) = \frac{i}{2 \pi} \sum_\sigma \int  d^4 y \, d^3 \vec{p} \, \frac{d \omega}{\omega + i \epsilon}  \, e^{ - i p_2 y + i \omega \max\{0, y^0\} }  
\langle 0 | \mathcal O(0) | \vec{p}, \sigma \rangle \langle \vec{p}, \sigma | T \left\{ A_1(0)  A_2 (y) \right\} | 0 \rangle \nn \\
&& \hspace{-0.7cm} \times (2 \pi)^8 \delta^{(3)}(\vec{k} - \vec{p}) \, \delta(k^0 - \sqrt{\vec{p}^2 + m^2}  + \omega) \delta^{(3)}(\vec{p_1} + \vec{p_2} + \vec{p}) \, \delta(p_1^0 + p_2^0 + \sqrt{ \vec{p}^2 + m^2} - \omega) + \ldots \,. 
\eea
The integrations over the momenta $\vec{p}$ and $\omega$ are now trivial due to the delta functions and lead to
\bea
&& G(k, p_1, p_2) = (2 \pi)^4 \delta^{(4)}(k + p_1 + p_2) i \frac{(2 \pi)^3}{\sqrt{\vec{k}^2 + m^2} - k^0 + i \epsilon} \nn \\
&& \times \sum_\sigma \int d^4 y  \, e^{  i \left( \sqrt{\vec{k}^2 + m^2} - k^0  \right) \max\{0, y^0\}} e^{ - i p_2 y}
\langle 0 | \mathcal O(0) | \vec{k}, \sigma \rangle \langle \vec{k}, \sigma | T \left\{ A_1(0)  A_2 (y) \right\} | 0 \rangle \,.
\eea
The appearance of the pole in the limit $k^0 \rightarrow \sqrt{\vec{k}^2 + m^2}$ in the correlation function is now explicitly manifest and originates from the massless pole in $\omega$, which comes, in turn, from the integral parameterization of the step function. In order to make the pole structure more clear we notice that near the pole
\bea
\frac{1}{\sqrt{\vec{k}^2 + m^2} - k^0 + i \epsilon}  \sim  \frac{2 k^0  }{k^2 - m^2 - i \epsilon}
\eea
while the exponential function under integration goes to unity. This allows us to define the matrix elements
\bea
(2 \pi)^4 \delta^{(4)}(k - p) \, \mathcal M_{0 | (k,\sigma)}(k) &\equiv& \int d^4 z e^{-i p z} \langle 0 | \mathcal O(z) | \vec{k}, \sigma \rangle \, \\
(2 \pi)^4 \delta^{(4)}(k + p_1 + p_2) \, \mathcal M_{(k,\sigma) | 0}(k, p_1, p_2) &\equiv& \int d^4 x_1 d^4 x_2 e^{-i p_1 x_1 - i p_2 x_2} \langle \vec{k}, \sigma | T \left\{ A_1(x_1)  A_2 (x_2) \right\} | 0 \rangle \,.\nn \\
\eea
With these definitions and simplifications the pole behaviour of the correlator is now explicit and reads as
\bea
\label{polebehaviour}
G(k, p_1, p_2) & \stackrel{k^2 \rightarrow m^2}{\longrightarrow} & (2 \pi)^4 \delta^{(4)}(k + p_1 + p_2) \nn \\
&\times&
 \sum_\sigma   \sqrt{2 (2 \pi)^3 k^0} \mathcal M_{0 | (k,\sigma)}(k)      \frac{ i }{k^2 - m^2 - i \epsilon}    \sqrt{2 (2 \pi)^3 k^0}  \mathcal M_{(k,\sigma) | 0}(k, p_1, p_2) \,. 
\eea

\section{Appendix. Feynman rules}
\label{AppFeynmanRules}
We report the Feynman rules used for the massless computation. All momenta are incoming.

\begin{itemize}
\item{fermion - fermion - gauge boson vertex}
\\ \\
\begin{minipage}{98pt}
\includegraphics[scale=1]{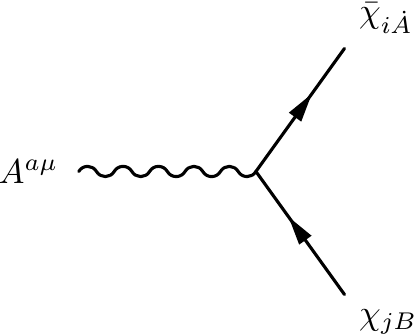}
\end{minipage}
\begin{minipage}{70pt}
\bea
\qquad\qquad=-ig\left(\bar{\sigma}^\mu\right)^{\dot{A}B}T^a_{ij}\ \ or\ \ ig\left(\sigma^\mu\right)_{B\dot{A}}T^a_{ij}
\nn
\eea
\end{minipage}
\item{gaugino - gaugino - gauge boson vertex}
\\ \\
\begin{minipage}{98pt}
\includegraphics[scale=1]{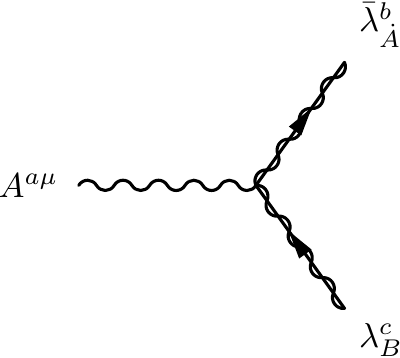}
\end{minipage}
\begin{minipage}{70pt}
\bea
\qquad\qquad=-g\left(\bar{\sigma}^\mu\right)^{\dot{A}B}t^{abc}\ \ or\ \ g\left(\sigma^\mu\right)_{B\dot{A}}t^{abc}
\nn
\eea
\end{minipage}
\item{scalar - scalar - gauge boson vertex}
\\ \\
\begin{minipage}{98pt}
\includegraphics[scale=1]{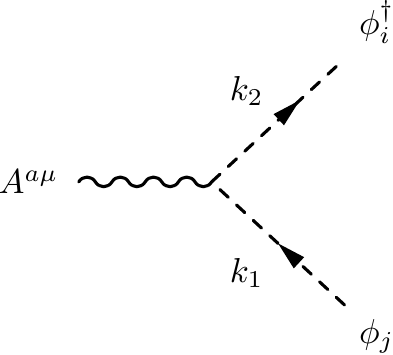}
\end{minipage}
\begin{minipage}{70pt}
\bea
\qquad\qquad=ig\left(k_2-k_1\right)^\mu T^a_{ij}
\nn
\eea
\end{minipage}
\item{scalar - scalar - gauge boson - gauge boson vertex}
\\ \\
\begin{minipage}{98pt}
\includegraphics[scale=1]{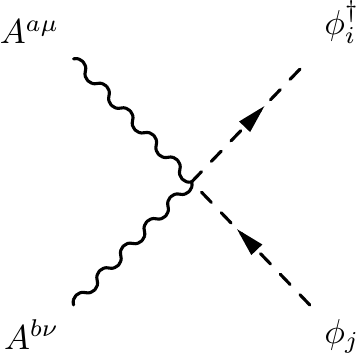}
\end{minipage}
\begin{minipage}{70pt}
\bea
\qquad\qquad=ig^2\eta^{\mu\nu}\left\{T^a,T^b\right\}_{ij}
\nn
\eea
\end{minipage}
\item{three gauge bosons vertex}
\\ \\
\begin{minipage}{98pt}
\includegraphics[scale=0.95]{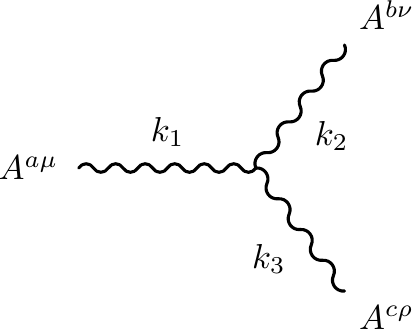}
\end{minipage}
\begin{minipage}{70pt}
\bea
\qquad&&=g\left[\eta^{\mu\nu}\left(k_1-k_2\right)^\rho+\eta^{\nu\rho}\left(k_2-k_3\right)^\mu+\eta^{\rho\mu}\left(k_3-k_1\right)^\nu\right]t^{abc}
\nn
\eea
\end{minipage}
\item{four gauge bosons vertex}
\\ \\
\begin{minipage}{98pt}
\includegraphics[scale=0.95]{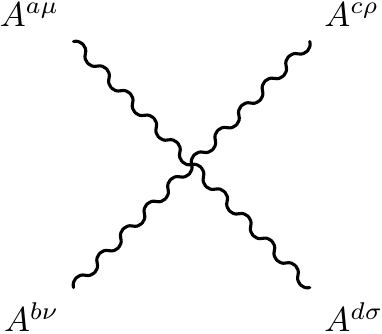}
\end{minipage}
\begin{minipage}{70pt}
\bea
\qquad&&=-ig^2\left[t^{abe}t^{cde}\left(\eta^{\mu\rho}\eta^{\nu\sigma}-\eta^{\mu\sigma}\eta^{\nu\rho}\right)+t^{ace}t^{bde}\left(\eta^{\mu\nu}\eta^{\rho\sigma}-\eta^{\mu\sigma}\eta^{\nu\rho}\right)\right.\nn\\
&&\left.\qquad\qquad+t^{ade}t^{bce}\left(\eta^{\mu\nu}\eta^{\rho\sigma}-\eta^{\mu\rho}\eta^{\nu\sigma}\right)\right]
\nn
\eea
\end{minipage}
\item{scalar - fermion - gaugino vertex}
\\ \\
\begin{minipage}{98pt}
\includegraphics[scale=0.95]{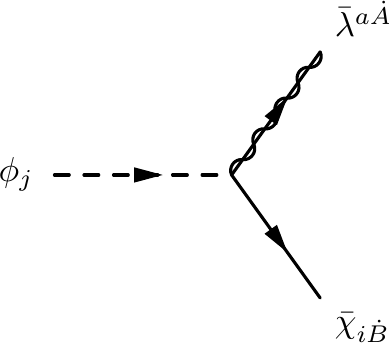}
\end{minipage}
\begin{minipage}{70pt}
\bea
\qquad\qquad=-i\sqrt{2}g\,T^a_{ij}\,\delta^{\dot{A}}_{\dot{B}}\ \ or\ \ -i\sqrt{2}g\,T^a_{ij}\,\delta^{\dot{B}}_{\dot{A}}
\nn
\eea
\end{minipage}
\item{scalar - fermion - gaugino vertex}
\\ \\
\begin{minipage}{98pt}
\includegraphics[scale=0.95]{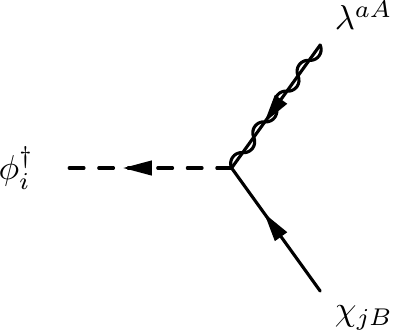}
\end{minipage}
\begin{minipage}{70pt}
\bea
\qquad\qquad=-i\sqrt{2}g\,T^a_{ij}\,\delta^A_B\ \ or\ \ -i\sqrt{2}g\,T^a_{ij}\,\delta^B_A
\nn
\eea
\end{minipage}
\item{R - gaugino - gaugino vertex}
\\ \\
\begin{minipage}{98pt}
\includegraphics[scale=0.95]{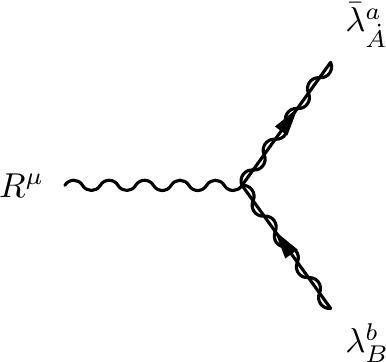}
\end{minipage}
\begin{minipage}{70pt}
\bea
\qquad\qquad=\left(\bar{\sigma}^\mu\right)^{\dot{A}B}\delta^{ab}\ \ or\ \ -\left(\sigma^\mu\right)_{B\dot{A}}\delta^{ab}
\nn
\eea
\end{minipage}
\item{R - fermion - fermion vertex}
\\ \\
\begin{minipage}{98pt}
\includegraphics[scale=0.9]{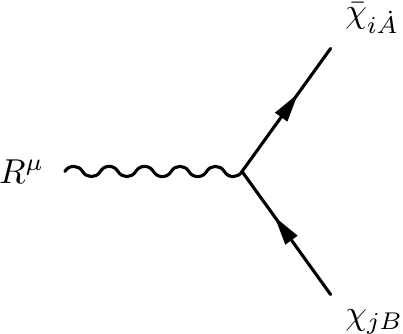}
\end{minipage}
\begin{minipage}{70pt}
\bea
\qquad\qquad=-\frac{1}{3}\left(\bar{\sigma}^\mu\right)^{\dot{A}B}\delta_{ij}\ \ or\ \ \frac{1}{3}\left(\sigma^\mu\right)_{B\dot{A}}\delta_{ij}
\nn
\eea
\end{minipage}
\item{R - scalar - scalar vertex}
\\ \\
\begin{minipage}{98pt}
\includegraphics[scale=0.9]{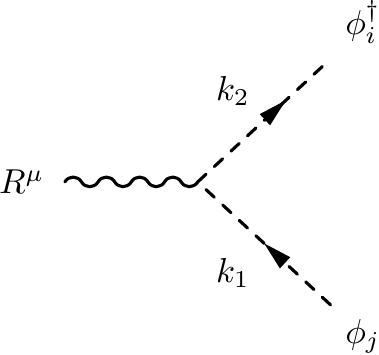}
\end{minipage}
\begin{minipage}{70pt}
\bea
\qquad\qquad=\frac{2}{3}\left(k_2-k_1\right)^\mu\delta_{ij}
\nn
\eea
\end{minipage}
\item{R - scalar - scalar - gauge boson vertex}
\\ \\
\begin{minipage}{98pt}
\includegraphics[scale=0.9]{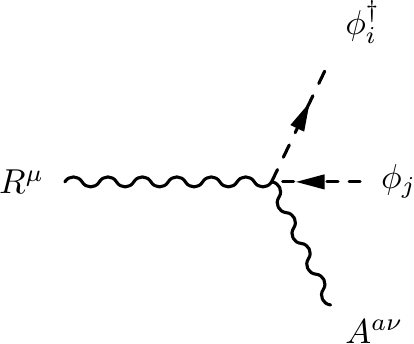}
\end{minipage}
\begin{minipage}{70pt}
\bea
\qquad\qquad=-\frac{4}{3}g\,\eta^{\mu\nu}\,T^a_{ij}
\nn
\eea
\end{minipage}
\item{S - scalar - fermion vertex}
\\ \\
\begin{minipage}{98pt}
\includegraphics[scale=0.9]{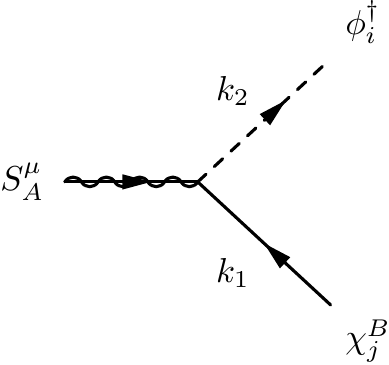}
\end{minipage}
\begin{minipage}{70pt}
\bea
\qquad\qquad=\sqrt{2}i\,k_2^\nu\left(\sigma_\nu\bar{\sigma}^\mu\right)_A^B\delta_{ij}-\frac{4}{3}\sqrt{2}i^2\left(k_1+k_2\right)_\nu\left(\sigma^{\mu\nu}\right)_A^B\delta_{ij}
\nn
\eea
\end{minipage}
\item{S - scalar - fermion - gauge boson vertex}
\\ \\
\begin{minipage}{98pt}
\includegraphics[scale=0.9]{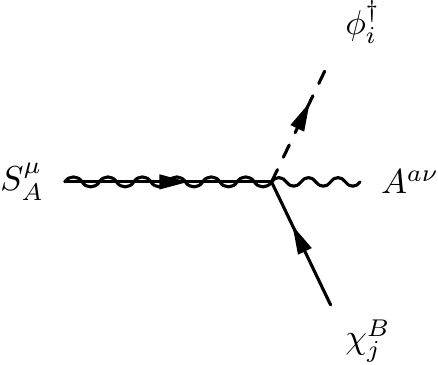}
\end{minipage}
\begin{minipage}{70pt}
\bea
\qquad\qquad=\sqrt{2}ig\left(\sigma^\nu\bar{\sigma}^\mu\right)_A^B T^a_{ij}
\nn
\eea
\end{minipage}
\item{S - scalar - scalar - gaugino vertex}
\\ \\
\begin{minipage}{98pt}
\includegraphics[scale=0.85]{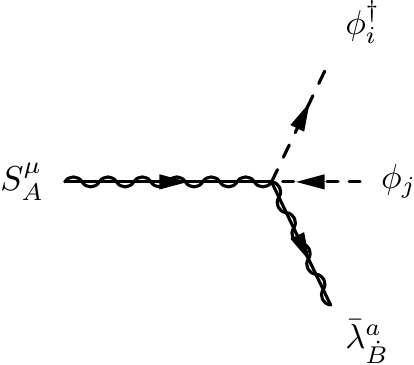}
\end{minipage}
\begin{minipage}{70pt}
\bea
\qquad\qquad=-ig\left(\sigma^\mu\right)_{A\dot{B}}T^a_{ij}
\nn
\eea
\end{minipage}
\item{S - gauge boson - gaugino vertex}
\\ \\
\begin{minipage}{98pt}
\includegraphics[scale=0.85]{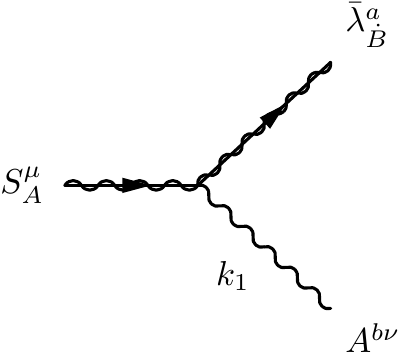}
\end{minipage}
\begin{minipage}{70pt}
\bea
\qquad\qquad=-2i^2 k_{1\rho}\left(\sigma^{\rho\nu}\sigma^\mu\right)_{A\dot{B}}\delta^{ab}
\nn
\eea
\end{minipage}
\item{S - gauge boson - gauge boson - gaugino vertex}
\\ \\
\begin{minipage}{98pt}
\includegraphics[scale=0.85]{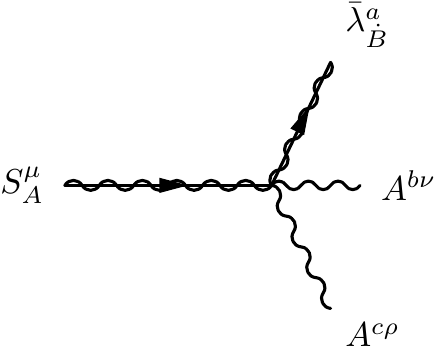}
\end{minipage}
\begin{minipage}{70pt}
\bea
\qquad\qquad=-2ig\left(\sigma^{\rho\nu}\sigma^\mu\right)_{A\dot{B}}t^{abc}
\nn
\eea
\end{minipage}
\item{T - scalar - scalar vertex}
\\ \\
\begin{minipage}{98pt}
\includegraphics[scale=0.85]{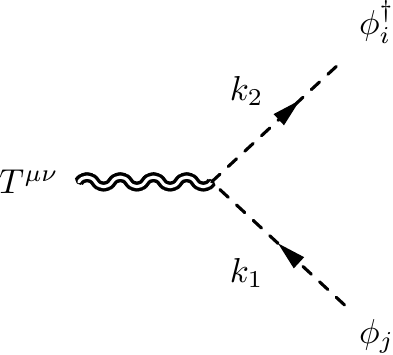}
\end{minipage}
\begin{minipage}{70pt}
\bea
\qquad\qquad=\left[-k_{2\rho}\,k_{1\sigma}\,C^{\mu\nu\rho\sigma}+\frac{1}{3}\left(\left(k_1+k_2\right)^\mu\,\left(k_1+k_2\right)^\nu-\eta^{\mu\nu}(k_1+k_2)^2\right)\right]\delta_{ij}
\nn
\eea
\end{minipage}
\item{T - scalar - scalar - gauge boson vertex}
\\ \\
\begin{minipage}{98pt}
\includegraphics[scale=0.85]{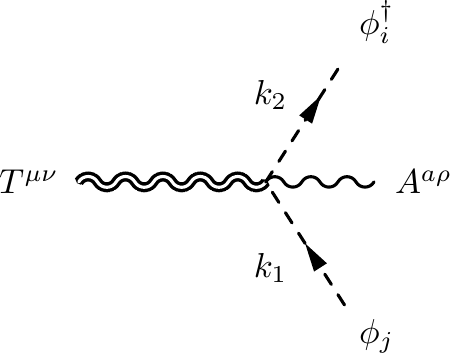}
\end{minipage}
\begin{minipage}{70pt}
\bea
\qquad\qquad=-g(k_2-k_1)_\sigma\,C^{\mu\nu\rho\sigma}T^a_{ij}
\nn
\eea
\end{minipage}
\item{T - scalar - scalar - gauge boson - gauge boson vertex}
\\ \\
\begin{minipage}{98pt}
\includegraphics[scale=0.9]{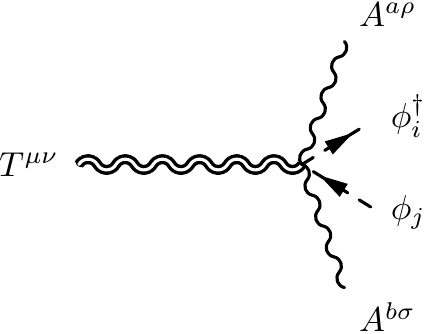}
\end{minipage}
\begin{minipage}{70pt}
\bea
\qquad\qquad=g^2 C^{\mu\nu\rho\sigma}\left\{T^a,T^b\right\}_{ij}
\nn
\eea
\end{minipage}
\item{T - fermion - fermion vertex}
\\ \\
\begin{minipage}{98pt}
\includegraphics[scale=0.9]{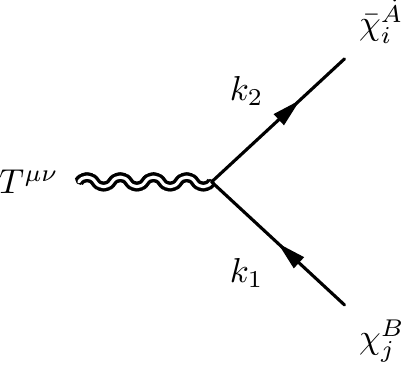}
\end{minipage}
\begin{minipage}{70pt}
\bea
\qquad\qquad=\frac{1}{4}\left(k_1-k_2\right)_\rho\left[\eta^{\rho\nu}\left(\bar{\sigma}^\mu\right)^{\dot{A}B}+\eta^{\rho\mu}\left(\bar{\sigma}^\nu\right)^{\dot{A}B}-2\eta^{\mu\nu}\left(\bar{\sigma}^\rho\right)^{\dot{A}B}\right]\delta_{ij}
\nn
\eea
\end{minipage}
\item{T - fermion - fermion - gauge boson vertex}
\\ \\
\begin{minipage}{98pt}
\includegraphics[scale=0.9]{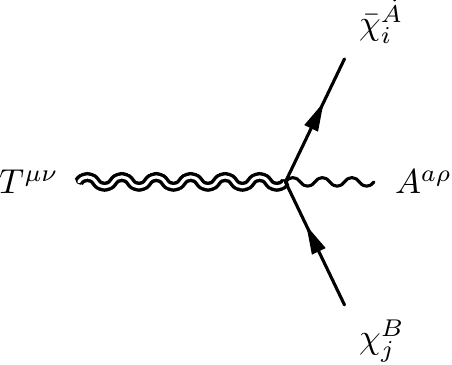}
\end{minipage}
\begin{minipage}{70pt}
\bea
\qquad\qquad=-\frac{g}{2}\left[\eta^{\rho\nu}\left(\bar{\sigma}^\mu\right)^{\dot{A}B}+\eta^{\rho\mu}\left(\bar{\sigma}^\nu\right)^{\dot{A}B}-2\eta^{\mu\nu}\left(\bar{\sigma}^\rho\right)^{\dot{A}B}\right]T^a_{ij}
\nn
\eea
\end{minipage}
\item{T - gaugino - gaugino vertex}
\\ \\
\begin{minipage}{98pt}
\includegraphics[scale=0.9]{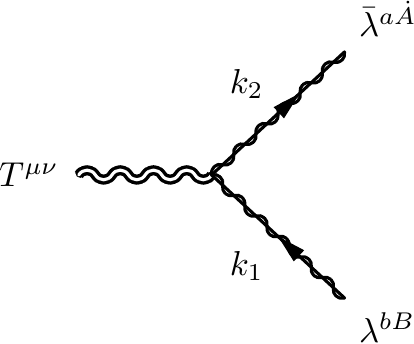}
\end{minipage}
\begin{minipage}{70pt}
\bea
\qquad\qquad=\frac{1}{4}\left(k_1-k_2\right)_\rho\left[\eta^{\rho\nu}\left(\bar{\sigma}^\mu\right)^{\dot{A}B}+\eta^{\rho\mu}\left(\bar{\sigma}^\nu\right)^{\dot{A}B}-2\eta^{\mu\nu}\left(\bar{\sigma}^\rho\right)^{\dot{A}B}\right]\delta^{ab}
\nn
\eea
\end{minipage}
\item{T - gaugino - gaugino - gauge boson vertex}
\\ \\
\begin{minipage}{98pt}
\includegraphics[scale=0.9]{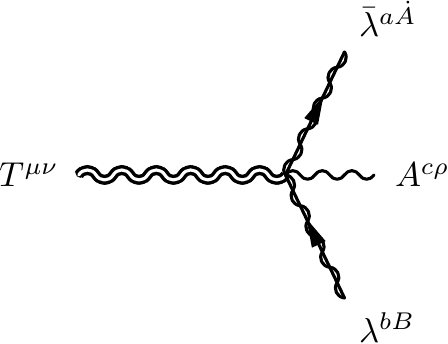}
\end{minipage}
\begin{minipage}{70pt}
\bea
\qquad\qquad=i\frac{g}{2}\left[\eta^{\rho\nu}\left(\bar{\sigma}^\mu\right)^{\dot{A}B}+\eta^{\rho\mu}\left(\bar{\sigma}^\nu\right)^{\dot{A}B}-2\eta^{\mu\nu}\left(\bar{\sigma}^\rho\right)^{\dot{A}B}\right]t^{abc}
\nn
\eea
\end{minipage}
\item{T - gauge boson - gauge boson vertex}
\\ \\
\begin{minipage}{98pt}
\includegraphics[scale=0.9]{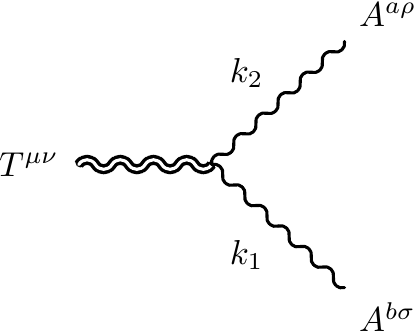}
\end{minipage}
\begin{minipage}{70pt}
\bea
\qquad\qquad=\left(k_1\cdot k_2\,C^{\mu\nu\rho\sigma}+D^{\mu\nu\rho\sigma}(k_1,k_2)+\frac{1}{\xi}E^{\mu\nu\rho\sigma}(k_1,k_2)\right)\delta^{ab}
\nn
\eea
\end{minipage}
\item{T - gauge boson - gauge boson - gauge boson vertex}
\\ \\
\begin{minipage}{98pt}
\includegraphics[scale=0.9]{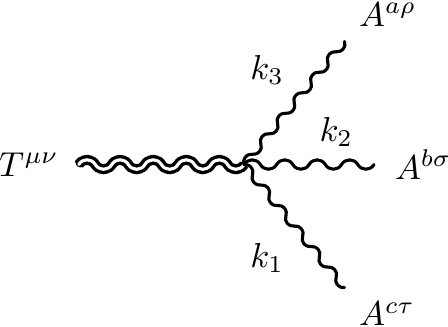}
\end{minipage}
\begin{minipage}{70pt}
\bea
\qquad\qquad&=&-ig\left(C^{\mu\nu\rho\sigma}(k_3-k_2)^\tau+C^{\mu\nu\rho\tau}(k_1-k_3)^\sigma\right.
\nn\\
&&\left. \quad + C^{\mu\nu\sigma\tau}(k_2-k_1)^\rho+F^{\mu\nu\rho\sigma\tau}(k_1,k_2,k_3)\right)t^{abc}
\nn
\eea
\end{minipage}
\end{itemize}

\bea
&& C_{\mu\nu\rho\sigma} = \eta_{\mu\rho}\, \eta_{\nu\sigma}
+\eta_{\mu\sigma} \, \eta_{\nu\rho}
-\eta_{\mu\nu} \, \eta_{\rho\sigma}\,
\\
&& D_{\mu\nu\rho\sigma} (k_1, k_2) =
\eta_{\mu\nu} \, k_{1 \, \sigma}\, k_{2 \, \rho}
- \biggl[\eta^{\mu\sigma} k_1^{\nu} k_2^{\rho}
  + \eta_{\mu\rho} \, k_{1 \, \sigma} \, k_{2 \, \nu}
  - \eta_{\rho\sigma} \, k_{1 \, \mu} \,r
 k_{2 \, \nu}
  + (\mu\leftrightarrow\nu)\biggr]\, \\
&& E_{\mu\nu\rho\sigma} (k_1, k_2) = \eta_{\mu\nu} \, (k_{1 \, \rho} \, k_{1 \, \sigma}
+k_{2 \, \rho} \, k_{2 \, \sigma} +k_{1 \, \rho} \, k_{2 \, \sigma})
-\biggl[\eta_{\nu\sigma} \, k_{1 \, \mu} \, k_{1 \, \rho}
+\eta_{\nu\rho} \, k_{2 \, \mu} \, k_{2 \, \sigma}
+(\mu\leftrightarrow\nu)\biggr]\,  \nn  \\ \\
&& F_{\mu\nu\rho\sigma\lambda} (k_1,k_2,k_3) =
g_{\mu\rho} \,  g_{\sigma\lambda} \, (k_2-k_3)_{\nu}
+g_{\mu\sigma} \, g_{\rho\lambda} \, (k_3-k_1)_{\nu}
+g_{\mu\lambda} \, g_{\rho\sigma}(k_1-k_2)_{\nu}
+ (\mu\leftrightarrow\nu) \nn \\
\eea


\begin{thebibliography}{10}

\bibitem{Gasperini:2007ar}
M.~Gasperini,
\newblock Lect.Notes Phys. {\bf 737}, 787 (2008), arXiv:hep-th/0702166.

\bibitem{LopesCardoso:1991zt}
G.~Lopes~Cardoso and B.~A. Ovrut,
\newblock Nucl.Phys. {\bf B369}, 351 (1992).

\bibitem{LopesCardoso:1992yd}
G.~Lopes~Cardoso and B.~A. Ovrut,
\newblock Nucl.Phys. {\bf B392}, 315 (1993), arXiv:hep-th/9205009.

\bibitem{Derendinger:1991kr}
J.-P. Derendinger, S.~Ferrara, C.~Kounnas, and F.~Zwirner,
\newblock Phys.Lett. {\bf B271}, 307 (1991).

\bibitem{Derendinger:1991hq}
J.~Derendinger, S.~Ferrara, C.~Kounnas, and F.~Zwirner,
\newblock Nucl.Phys. {\bf B372}, 145 (1992).

\bibitem{Derendinger:1985cv}
J.~Derendinger, L.~E. Ibanez, and H.~P. Nilles,
\newblock Nucl.Phys. {\bf B267}, 365 (1986).

\bibitem{Codello:2012sn}
A.~Codello, G.~D'Odorico, C.~Pagani, and R.~Percacci,
\newblock Class.Quant.Grav. {\bf 30}, 115015 (2013), arXiv:1210.3284.

\bibitem{Buchmuller:1988cj}
W.~Buchmuller and N.~Dragon,
\newblock Nucl.Phys. {\bf B321}, 207 (1989).

\bibitem{Coriano:2013nja}
C.~Corian\`o, L.~Delle~Rose, C.~Marzo, and M.~Serino,
\newblock (2013), arXiv:1311.1804.

\bibitem{Goldberger:2007zk}
W.~D. Goldberger, B.~Grinstein, and W.~Skiba,
\newblock Phys.Rev.Lett. {\bf 100}, 111802 (2008), arXiv:0708.1463.

\bibitem{Coriano:2012nm}
C.~Corian\`o, L.~Delle~Rose, A.~Quintavalle, and M.~Serino,
\newblock JHEP {\bf 1306}, 077 (2013), arXiv:1206.0590.

\bibitem{Coriano:2012dg}
C.~Corian\`o, L.~Delle~Rose, C.~Marzo, and M.~Serino,
\newblock Phys.Lett. {\bf B717}, 182 (2012), arXiv:1207.2930.

\bibitem{Schwimmer:2010za}
A.~Schwimmer and S.~Theisen,
\newblock Nucl.Phys. {\bf B847}, 590 (2011), arXiv:1011.0696.

\bibitem{Dudas:1993mm}
E.~Dudas,
\newblock Phys.Rev. {\bf D49}, 1109 (1994), arXiv:hep-ph/9307294.

\bibitem{Kors:2004ri}
B.~Kors and P.~Nath,
\newblock JHEP {\bf 12}, 005 (2004), arXiv:hep-ph/0406167.

\bibitem{Coriano:2008xa}
C.~Corian\`o, M.~Guzzi, A.~Mariano, and S.~Morelli,
\newblock Phys.Rev. {\bf D80}, 035006 (2009), arXiv:0811.3675.

\bibitem{Coriano:2008aw}
C.~Corian\`o, M.~Guzzi, N.~Irges, and A.~Mariano,
\newblock Phys.Lett. {\bf B671}, 87 (2009), arXiv:0811.0117.

\bibitem{Coriano:2010ws}
C.~Corian\`o, M.~Guzzi, and A.~Mariano,
\newblock Phys.Rev. {\bf D85}, 095008 (2012), arXiv:1010.2010.

\bibitem{Giannotti:2008cv}
M.~Giannotti and E.~Mottola,
\newblock Phys. Rev. {\bf D79}, 045014 (2009), arXiv:0812.0351.

\bibitem{Armillis:2009pq}
R.~Armillis, C.~Corian\`{o}, and L.~Delle~Rose,
\newblock Phys. Rev. {\bf D81}, 085001 (2010), arXiv:0910.3381.

\bibitem{Armillis:2010qk}
R.~Armillis, C.~Corian\`o, and L.~Delle~Rose,
\newblock Phys.Rev. {\bf D82}, 064023 (2010), arXiv:1005.4173.

\bibitem{Dolgov:1971ri}
A.~D. Dolgov and V.~I. Zakharov,
\newblock Nucl. Phys. {\bf B27}, 525 (1971).

\bibitem{Bertlmann:1981by}
R.~Bertlmann,
\newblock Acta Phys.Austriaca {\bf 53}, 305 (1981).

\bibitem{Horejsi:1985qu}
J.~Horejsi,
\newblock Phys.Rev. {\bf D32}, 1029 (1985).

\bibitem{Horejsi:1997yn}
J.~Horejsi and M.~Schnabl,
\newblock Z. Phys. {\bf C76}, 561 (1997), arXiv:hep-ph/9701397.

\bibitem{Horejsi:1994aj}
J.~Horejsi and O.~Teryaev,
\newblock Z. Phys. {\bf C65}, 691 (1995).

\bibitem{Bertlmann:2000da}
R.~Bertlmann and E.~Kohlprath,
\newblock Annals Phys. {\bf 288}, 137 (2001), arXiv:hep-th/0011067.

\bibitem{Freedman:1974ze}
D.~Z. Freedman and E.~J. Weinberg,
\newblock Ann. Phys. {\bf 87}, 354 (1974).

\bibitem{Adler:1976zt}
S.~L. Adler, J.~C. Collins, and A.~Duncan,
\newblock Phys. Rev. {\bf D15}, 1712 (1977).

\bibitem{Chanowitz:1972da}
M.~S. Chanowitz and J.~R. Ellis,
\newblock Phys.Rev. {\bf D7}, 2490 (1973).

\bibitem{Armillis:2009im}
R.~Armillis, C.~Corian\`o, and L.~Delle~Rose,
\newblock Phys. Lett. {\bf B682}, 322 (2009), arXiv:0909.4522.

\bibitem{Armillis:2009sm}
R.~Armillis, C.~Corian\`o, L.~Delle~Rose, and M.~Guzzi,
\newblock JHEP {\bf 0912}, 029 (2009), arXiv:0905.0865.

\bibitem{Capper:1975ig}
D.~Capper and M.~Duff,
\newblock Phys.Lett. {\bf A53}, 361 (1975).

\bibitem{Ferrara:1974pz}
S.~Ferrara and B.~Zumino,
\newblock Nucl.Phys. {\bf B87}, 207 (1975).

\bibitem{Coriano:2011zk}
C.~Corian\`{o}, L.~Delle~Rose, and M.~Serino,
\newblock Phys.Rev. {\bf D83}, 125028 (2011), arXiv:1102.4558.

\bibitem{Coriano:2011ti}
C.~Corian\`o, L.~Delle~Rose, A.~Quintavalle, and M.~Serino,
\newblock Phys.Lett. {\bf B700}, 29 (2011), arXiv:1101.1624.

\bibitem{Grisaru:1980nk}
M.~T. Grisaru, M.~Rocek, and W.~Siegel,
\newblock Phys.Rev.Lett. {\bf 45}, 1063 (1980).

\bibitem{Caswell:1980yi}
W.~E. Caswell and D.~Zanon,
\newblock Phys.Lett. {\bf B100}, 152 (1981).

\bibitem{Avdeev:1981ew}
L.~Avdeev and O.~Tarasov,
\newblock Phys.Lett. {\bf B112}, 356 (1982).

\bibitem{Dienes:2009td}
K.~R. Dienes and B.~Thomas,
\newblock Phys.Rev. {\bf D81}, 065023 (2010), arXiv:0911.0677.

\bibitem{Komargodski:2009pc}
Z.~Komargodski and N.~Seiberg,
\newblock JHEP {\bf 0906}, 007 (2009), arXiv:0904.1159.

\bibitem{Komargodski:2010rb}
Z.~Komargodski and N.~Seiberg,
\newblock JHEP {\bf 1007}, 017 (2010), arXiv:1002.2228.

\end{thebibliography}
\end{document}